\documentclass[10.5pt]{iopart}
\usepackage{graphicx,mathrsfs, color}
\usepackage{amssymb}
\graphicspath{{./images/}}
\usepackage{bm}
\usepackage[mathlines]{lineno}
\usepackage[normalem]{ulem}

\begin{document}
\title[Nuclear Fusion]{Helical Core Formation and MHD Stability in ITER-Scale Plasmas with Fusion-born Alpha Particles}

\author{P. Adulsiriswad$^1$, A. Bierwage$^{1,2}$, M. Yagi$^1$,}

\address{$^1$ National Institute for Quantum Science and Technology, Rokkasho Institute for Fusion Energy, Rokkasho, Aomori, Japan}
\address{$^2$ National Institute for Quantum Science and Technology, Naka Institute for Fusion Science and Technology, Naka, Ibaraki, Japan}

\vspace{10pt}
\begin{indented}

\item[] \today
\end{indented}

\begin{abstract}
The effect of fusion-born alpha particles on the helical core (HC), a long-lived ideal saturation state of the $m/n=1/1$ kink/quasi-interchange mode, is studied in the ITER-scale hybrid scenario where a core plasma has a low magnetic shear $q\gtrsim1$.  The HC state is determined by 3-D MHD force balance and all factors that contribute to it, such as plasma shaping, the safety factor profile, and the pressure profiles of all particle species. An incomplete but useful measure of the HC is the displacement of the magnetic axis, $\delta_\mathrm{HC}$. Using MHD-PIC simulations, we find	that $\delta_\mathrm{HC}$ is enhanced by increasing	alpha particle pressure $\beta_\mathrm{\alpha}$.  Within the ITER operating alpha pressure $\beta_\mathrm{\alpha}(0) \lesssim 1\%$, $\beta_\mathrm{\alpha}$ can be approximately treated as part of the total MHD pressure.  In this regime, there is no notable flattening of the pressure profile, indicating that the HC preserves the omnigenity of the plasma.  If one increases $\beta_\mathrm{\alpha}(0)$ beyond $1\%$, $\delta_\mathrm{HC}$ continues to increase with $\beta_\mathrm{\alpha}$ until it reaches an upper limit at $\beta_\mathrm{\alpha}(0)=3\%$ for our reference case.  At this limit, both the bulk and alpha pressure profiles are partially flattened, indicating a reduction in omnigenity.  After HC formation, a resistive pressure-driven MHD mode can become unstable, which seems to be triggered by the local steepening of the bulk plasma pressure gradient within the compressed magnetic flux region of the HC. This secondary mode consists of a broad spectrum of short-wavelength Fourier components that grow at identical rates and are thus part of a single coherent entity.  Our present simulation model is insufficient to adequately represent such a secondary mode; however, preliminary results suggest that it can facilitate magnetic chaos, which affects plasma confinement. We also discuss possible methods for suppressing this instability.

\end{abstract}

\noindent{\it Keywords}: Helical Core, Long-lived Mode, Hybrid Scenario, Alpha Particles, ITER

\maketitle

\ioptwocol

\section{\label{sec:level1}Introduction}
\begin{figure}[h]
\begin{center}
\includegraphics[width=1.00\linewidth]{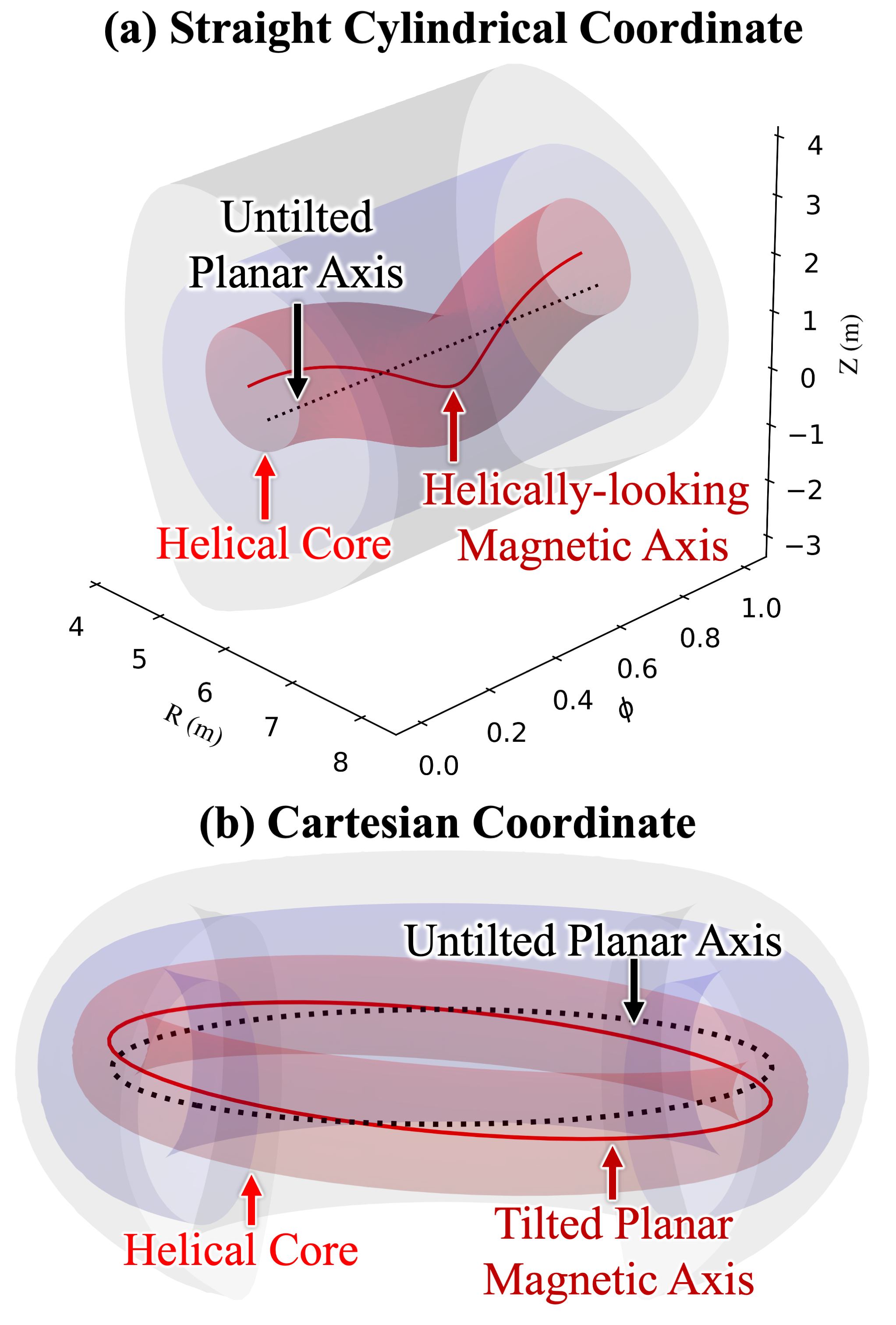}
\end{center} 
\caption{Illustrations of HC tokamak equilibrium in the (a) straight cylinder coordinate, and (b) Cartesian coordinate.  The red surface represents an arbitrary magnetic flux surface that resides within the HC, the blue surfaces represents a flux surface located outside of the HC, and the LCFS appears in gray.}
\label{fig:1HC}
\end{figure}

\quad
In a tokamak hybrid scenario where a core plasma has a weak magnetic shear at a safety factor slightly above unity ($q \gtrsim 1$), the non-resonant kink/quasi-interchange mode with poloidal ``$m$" and toroidal ``$n$" periodicities of unity can be unstable.  Since there is no $q=1$ surface, the $m/n = 1/1$ magnetic reconnection does not occur, and the mode instead saturates due to the stabilizing effect of field line bending\cite{chapman2010saturated} and potentially other nonlinear processes.  After the saturation, the plasma pressure profile remains mostly peaked and without any major radial mixing.  The $m/n=1/1$ structure does not decay to the original axisymmetric state but is maintained due to the establishment of a 3-D magnetohydrodynamic (MHD) equilibrium state known as helical core (HC).  The spatial structures of a HC in the straight cylindrical and Cartesian coordinates are illustrated in Figs.\ref{fig:1HC}(a) and (b), respectively.  The red surface represents an arbitrary magnetic flux surface that resides within the HC region ($q\gtrsim1$), while the blue and gray surfaces represent arbitrary magnetic flux surfaces located outside the HC region and at the last closed flux surface (LCFS), respectively.  In the straight cylindrical coordinate representation, the HC and its magnetic axis are helically twisted, as its name suggests. In contrast, the HC magnetic axis is a planar circle that is merely tilted with respect to the $z=0$ plane in the Cartesian frame. (Meanwhile, magnetic flux surfaces in the surrounding displaced domain of the HC do possess truly helical distortion.)  Prior studies have shown that a HC can have positive and negative consequences on plasma performance. For instance, the helical flow within the HC can contribute to the self-regulation of the plasma profiles in the $q \gtrsim 1$ region, enabling sawtooth mitigation with minimal to no external controls\cite{piovesan2017role, krebs2017magnetic, burckhart2023experimental}, which can be beneficial.  At the same time, the HC toroidal asymmetry can broaden the energetic particle (EP) spatial distribution \cite{pfefferle2014nbi,menard2005internal,yuan2019saturated} and plasma rotation profile \cite{menard2005internal,chapman2010saturated,yuan2019saturated}.  A small broadening of the EP distribution function can be beneficial because it can make plasma heating more uniform and reduce the likelihood of EP-gradient-driven instabilities.  However, excessive broadening can reduce the EP heating efficiency.  In terms of plasma diagnostics and control, a large displacement of the core plasma away from the diagnostic sight line introduces a spatiotemporal convolution with the HC phase. This can pose an additional challenge for measurements and for the plasma control systems that depend on those measurements.
s

Experimentally, the HC has been reported in several fusion devices, including MAST\cite{chapman2010saturated}, DIII-D\cite{piovesan2016impact}, Alcator C-Mod \cite{wingen2019helical}, EAST\cite{yuan2019saturated}, and JT-60U\cite{bando2019experimental}.  W.A. Cooper\cite{cooper2010tokamak,cooper2011jet,cooper2011helical,cooper2013bifurcated} found that the HC appears as an asymmetric solution of the MHD equilibrium, using the 3-D equilibrium solver ANIMEC\cite{cooper2009three}, an anisotropic pressure version of VMEC\cite{hirshman1983steepest}.  These codes calculate the MHD equilibrium by minimizing the total MHD energy functional.  The HC solution can then be found when its total MHD energy is lower than that of the axisymmetric state.  Using these codes, the dependencies of HC formation on normalized plasma pressure $\beta$ \cite{cooper2011helicaliter,wingen2018use,nakamura2020influence}, $q$ profile \cite{cooper2011helicaliter,kawagoe2022two}, and plasma shaping \cite{nakamura2020influence} have been investigated.  In addition to the steady-state solution obtained from these equilibrium solvers, initial value nonlinear MHD codes such as XTOR\cite{graves2012magnetohydrodynamic,brunetti2014ideal} and M3D-C$^1$\cite{krebs2017magnetic} have also been used to study the HC.  In ITER, an important milestone of magnetically confined fusion research, the studies by W.A. Cooper et al.\cite{cooper2011helicaliter} and A. Wingen et al.\cite{wingen2018use} predict that the $m/n=1/1$ kink/quasi-interchange mode in the hybrid scenario is linearly unstable and far above the bifurcation threshold; therefore, the spontaneous HC formation is likely to occur, at least within the ideal MHD regime.  

Since ITER aims to provide the physics basis and proof-of-principle for a controlled burning tokamak plasma, a significant alpha particle pressure $\beta_\mathrm{\alpha}$ is expected in the core plasma.  Both theoretical and experimental studies reported that EPs such as fusion-born alpha particles can affect MHD stability, including the stabilization of the $m/n=1$ kink mode by trapped energetic particles\cite{porcelli1991fast,fu2006global}, the destabilization of fishbone modes\cite{mcguire1983study,chen1984excitation,coppi1986theoretical}, and the excitation of Alfv{\'e}n eigenmodes (AEs)\cite{cheng1985low}. Nonlinear dynamics induced by EPs, such as mode chirping\cite{heidbrink1995beam} and bursting, have also been reported. Given that the HC can be viewed as an ideally saturated state of the $m/n=1/1$ kink/quasi-interchange mode, the following three questions arise concerning its interplay with fusion-born alpha particles: (1) Can a HC form in a plasma with ITER-relevant $\beta_\mathrm{\alpha}$? (2) If a HC can form, how is it affected by alpha particles? (3) How well are alpha particles confined after HC formation?

The present numerical study aims to answer these three key questions using mainly the code MEGA\cite{todo1998linear}, which describes the bulk plasma by the nonlinear single-fluid MHD model, and alpha particles by the drift-kinetic particle-in-cell (PIC) model. 
In the actual plasmas, HC formation, alpha particle dynamics, and plasma profiles are interconnected.  Modifying one element will alter the others.  Such a self-consistent simulation can be very expensive, and multiple effects can obscure the essence of the alpha particle effects on HC.  For this reason, MEGA is used to simulate a simplified situation, specifically the role of alpha particles during HC onset with a prescribed axisymmetric plasma and alpha profiles. 
The $5$ T/$13$ MA ITER hybrid scenario is used as a reference for our parameter scans, where we vary the profile of the safety factor $q$, the density of the alpha particles, and the electric resistivity of the plasma. This should be regarded as a physical study rather than a predictive one so that we will speak of a ``ITER-scale'' plasma.  The contents of this paper are organized as follows: Section \ref{sec:level2} presents the simulation model, assumptions, and numerical method.  Section \ref{sec:level3} presents the details of our simulation scenarios and parameters, including the MHD equilibrium, the grid resolution, and the alpha particle distribution function.  Section \ref{sec:level4} presents the results for the case without alpha particles, along with the benchmarking of the HC solution calculated with MEGA against those calculated with VMEC.  In our equilibrium scans, we found cases where a resistive pressure-driven MHD mode becomes unstable after HC formation.  The properties of this mode within the limits of our simulation model will also be reported.  Sections \ref{sec:level5}-\ref{sec:level6} present the simulation results for the case with alpha particles, where Section \ref{sec:level5} will focus on an equilibrium where the secondary MHD mode is unstable after HC formation, while Section \ref{sec:level6} will focus on a case with a stable HC. Lastly, Section \ref{sec:level7} contains a summary and concluding discussions.

\section{\label{sec:level2}Simulation Models and Setups}
\quad

\subsection{\label{sec:level2a}MEGA Code}
\quad

MEGA\cite{todo1998linear}, a global nonlinear MHD-PIC simulation code, is employed in this study.  This code solves the nonlinear resistive MHD equations coupled with EPs through the current coupling scheme\cite{park1992three} as an initial value problem.  The resistive MHD equations solved by MEGA are

\begin{eqnarray} \label{eq:mhd1}
\frac{\partial \rho_M} {\partial t}  = - \nabla \cdot (\rho_M \vec v) + \nu_n \nabla^2 (\rho_M), \ 
\end{eqnarray}
\begin{eqnarray} \label{eq:mhd2}
\rho_M \frac{\partial  \vec v}{\partial t}  = & -\rho_M (\vec v \cdot \nabla)\vec v - \nabla P + (\vec J - \vec J_\alpha) \times \vec B  \nonumber\\ & - \nabla \times (\nu \rho_M (\nabla \times \vec{v}))  + \frac{4}{3} \nabla (\nu \rho_M \nabla \cdot \vec v), \ 
\end{eqnarray}
\begin{eqnarray} \label{eq:mhd3}
\frac{\partial P} {\partial t}  = & - \nabla \cdot (P\vec v) - (\Gamma - 1)P \nabla \cdot \vec v \nonumber\\ & + (\Gamma - 1) \times [\nu \rho_M (\nabla \times \vec{v}) ^2 + \frac{4}{3} \nu \rho_M (\nabla \cdot \vec v)^2  \nonumber\\ & + \eta \vec J \cdot (\vec J - \vec J_{eq})] + \chi_\parallel \nabla^2_\parallel P \nonumber\\ & + \chi_\perp \nabla^2_\perp P,
\end{eqnarray}
\begin{eqnarray} \label{eq:mhd4}
\frac{\partial \vec B} {\partial t}  = - \nabla \times \vec E, \ 
\end{eqnarray}
\begin{eqnarray} \label{eq:mhd5}
\mu_0\vec J  = \nabla \times \vec B, \ 
\end{eqnarray}
\begin{eqnarray} \label{eq:mhd6}
\vec E = - \vec v \times \vec B + \eta (\vec J - \vec J_{eq}). \ 
\end{eqnarray}
\noindent

\begin{figure*}[t]
\begin{center}
\includegraphics[width=0.90\linewidth]{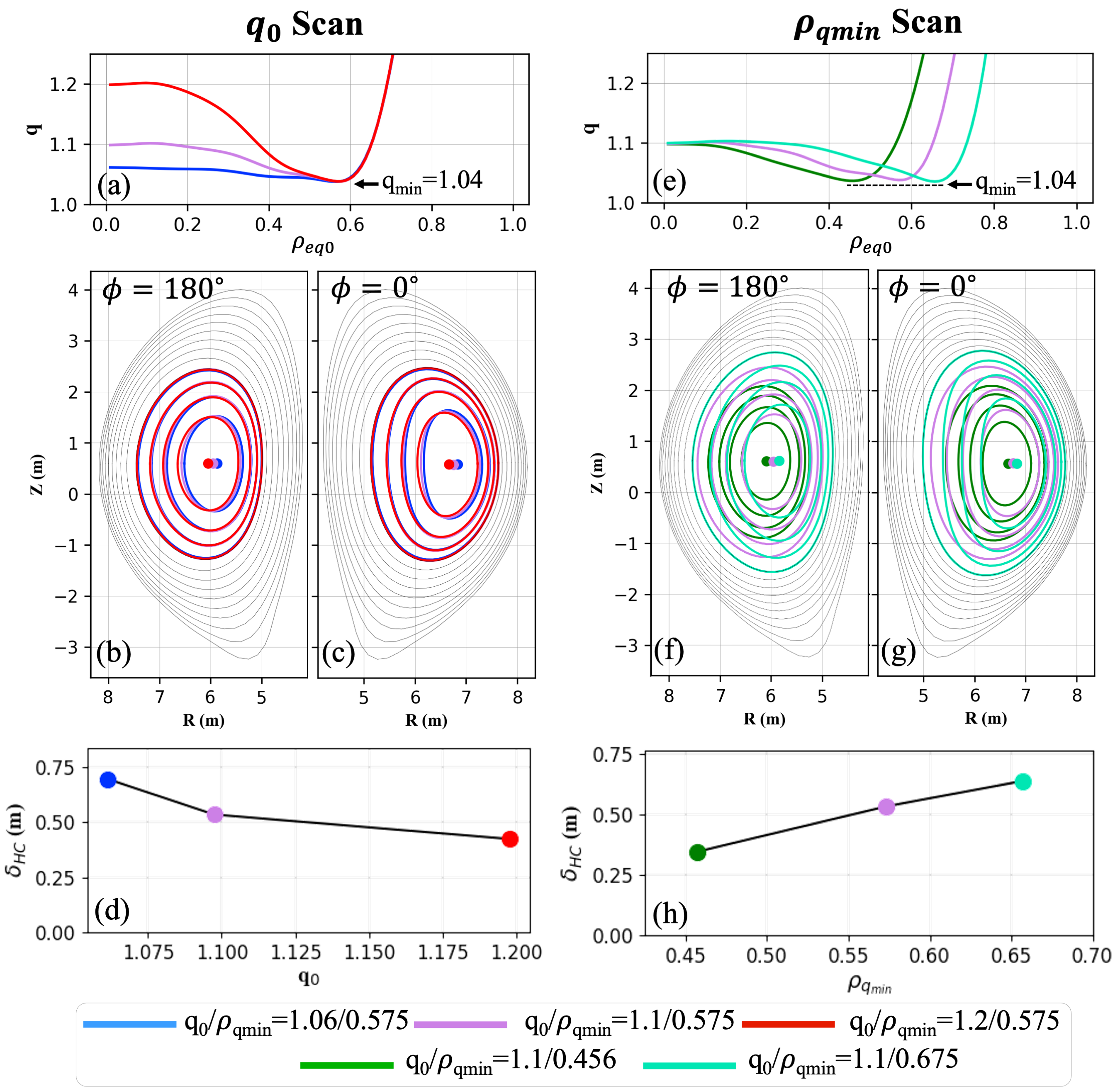}
\end{center} 
\caption{ITER-scale HC MHD equilibria calculated with VMEC while varying (a-d) $q_\mathrm{0}$ and (e-h) $\rho_\mathrm{qmin}$: (a,e) prescribed $q$ profiles;  magnetic Poincar{\'e} plots at (b,f) $\phi=180^\circ$ and (c,g)  $\phi=0^\circ$. The radial coordinate at $\phi = 180^\circ$ is plotted with an inverted horizontal axis.;  (d,h) $m/n=1/1$ radial displacement of the magnetic axis $\delta_\mathrm{HC}$.}
\label{fig:3mhdeq}
\end{figure*}

The variables appearing in these equations are defined as follows: bulk plasma mass density $\rho_{M}$, bulk plasma pressure $P$, total plasma current density $J$, alpha current density $J_\alpha$, equilibrium plasma current density $J_{eq}$, MHD velocity $\vec{v}$, magnetic field $\vec{B}$, electric field $\vec{E}$, and specific heat ratio $\Gamma$.  The EPs contribute via their current density $J_\alpha$, which appears as the $3^{rd}$ term on the right-hand side of the MHD momentum Eq.\ref{eq:mhd2}. The EP current density can be written as

\begin{eqnarray} \label{eq:jalpha}
\vec J_\alpha= & J_{\alpha\parallel}\hat{b} + \frac{1}{B}\left(P_{\alpha\parallel}\nabla\times\hat{b} -\frac{P_{\alpha\perp}}{B}\nabla B\times\hat{b}\right) \nonumber\\ &-\nabla\times\left(\frac{P_{\alpha \perp}}{B}\hat{b}\right). \ 
\end{eqnarray}
\noindent
The first term in Eq.\ref{eq:jalpha} represents the alpha particle current density along the magnetic field line.  Here, its contribution vanishes after taking the cross product with $\vec{B}$ in Eq.\ref{eq:mhd2}.  The second and third terms in Eq.\ref{eq:jalpha} represent the current density from the curvature and grad-$B$ drifts, respectively.  The last term is referred to as ``magnetization current"\footnote{Specifically, this is the component of the magnetization current that is associated with the particles' rapid gyration, after it has been isolated from the slow drifts.}.  The dissipation terms are viscosity $\nu$, particle diffusivity $\nu_n$, parallel heat conductivity $\chi_\parallel$, perpendicular heat conductivity $\chi_\perp$, and electric resistivity $\eta$.  Concerning the numerical methods, the spatial differentiation is performed using the $4^{th}$-order finite differences in right-handed cylindrical coordinates (R,$\phi$,Z), while the explicit $4^{th}$-order Runge-Kutta method is used for the time integration.

Alpha particles are represented by markers following guiding center trajectories. Gyroaveraging is not performed, as the focus of this study is on the long-wavelength $m/n=1/1$ mode. The time evolution of marker weights is computed using the full-$f$ PIC method, which is preferred over the conventional $\delta f$ approach because the HC can induce large and long-lived deformations in the alpha particle distribution, such that $\delta f_\alpha \sim \mathcal{O}(f_{\alpha0})$, where $f_{\alpha0}$ is the initial distribution of the axisymmetric equilibrium state.

\subsection{\label{sec:level2b}Simulation Setup and Dissipation Terms}
\quad

In this study, the particle diffusivity $\nu_n$, and viscosity $\nu$ are fixed at $10^{-6}v_AR_0$, while the normalized resistivity $\hat{\eta} = \eta / (\mu_0 v_A R_0)$ is scanned in the range $10^{-7} \leq \hat{\eta} \leq 10^{-4}$.  The resistivity scan allows us to examine trends related to the non-ideal bulk plasma effects.  In addition, the dissipation terms on the right-hand side of Eq.\ref{eq:mhd3} help to maintain numerical stability, particularly during the nonlinear phase by converting short-wavelength structures generated by the mode-mode coupling and shear Alfv{\'e}n continuum damping into thermal MHD energy.  The impact of these dissipation coefficients on the long-wavelength $m/n=1/1$ ideal kink/quasi-interchange mode is small because these viscous and resistive terms involve second-order spatial derivatives, which are weighted towards shorter wavelengths.

The parallel $\chi_\parallel$ and perpendicular $\chi_\perp$ heat conductivities are fixed at $10^{-4}v_AR_0$ and $10^{-6}v_AR_0$, respectively.  The value of $\chi_\parallel$ influences the simulation time step via the Courant–Friedrichs–Lewy condition, so our choice is constrained by computational cost. Although the ratio $\chi_\parallel / \chi_\perp=10^2$ used in this work accounts for the fact that plasma nonuniformities should relax relatively quickly along the magnetic field lines, this relaxation process is expected to be slower in our simulation than in reality. This is exacerbated by the fact that the continuity equation for the MHD density $\rho_M$ makes no distinction between the parallel and perpendicular diffusivity. This has the consequence that our MEGA simulations, which will cover a few milliseconds of physical time, will usually not yield a perfectly steady HC state. Meanwhile, extending the simulation time is not considered helpful because it would require more realistic modeling of sources and causes continued accumulation of numerical errors that can affect the nonlinear dynamics. Consequently, our MEGA simulations should be expected to yield oscillatory HC behavior that converges to a quasi-steady HC state.

Using a low $\chi_\parallel$ value is also expected to exaggerate ballooning modes at short wavelengths.  For instabilities in axisymmetric tokamak equilibria, MEGA addresses this issue with a low-pass filter that acts along the geometrical toroidal angle $\phi$ using the Fourier basis $\exp(in\phi)$. Given a maximal toroidal mode number $n_{\rm max}$, this filter removes all toroidal Fourier components with $n > n_{\rm max}$. However, this filter will not have the intended effect on modes that reside in a non-axisymmetric configuration like a HC. Since the geometric toroidal direction at any given $(R,Z)$ position will then intersect a range of flux surfaces, short-wavelength modes in the HC are not fully eliminated, and longer-wavelength modes are subject to unintended additional damping.

In this study, we discuss primarily results from simulations that contain toroidal mode numbers in the range $0\leq n\leq1$ or $0\leq n\leq8$. By considering these two ranges, we can discern the effects of the HC's dominant $n=1$ component and $n>1$ modes. In addition, results from a simulation containing $0\leq n\leq16$ will be briefly discussed in Section \ref{sec:level4c1} to clarify the poloidally and toroidally localized structure of the secondary MHD instability that emerged after HC formation. A specially prepared case that was simulated without the low-pass filter is discussed in \ref{sec:apen1}.

The cylindrical grid resolutions ($N_R,N_Z,N_\phi$) used in the $0\leq n\leq1$ and $0\leq n\leq8$ simulations are ($200,200,32$) and ($200,200,128$), and the respective numbers of PIC markers are $4.096\times10^7$ and $1.638\times10^8$. For convergence tests, particularly at lower resistivity $\hat{\eta}=10^{-7}$, we increase the poloidal resolution to  ($N_R, N_Z$)=($400,400$).
 
Our simulation domain is a cylindrical box with dimensions $4.04$ m $\leq R \leq$ $8.4$ m and $-3.32$ m $\leq Z \leq$ $4.12$ m.  The initial location of the magnetic axis is around $R_0 \approx 6.4$ m, $Z_0 \approx 0.6$ m The MHD fluctuations are constrained by a non-slip boundary condition at the LCFS, but alpha particles are free to travel through the entire simulation box. The absence of a realistic first wall may lead to an underestimation of alpha particle losses, but due to the relatively small magnetic drifts in the present ITER scenario ($5$ T/$13$ MA), this simplification is expected to have a negligible effect on HC dynamics.

\begin{figure}[h]
\begin{center}
\includegraphics[width=0.85\linewidth]{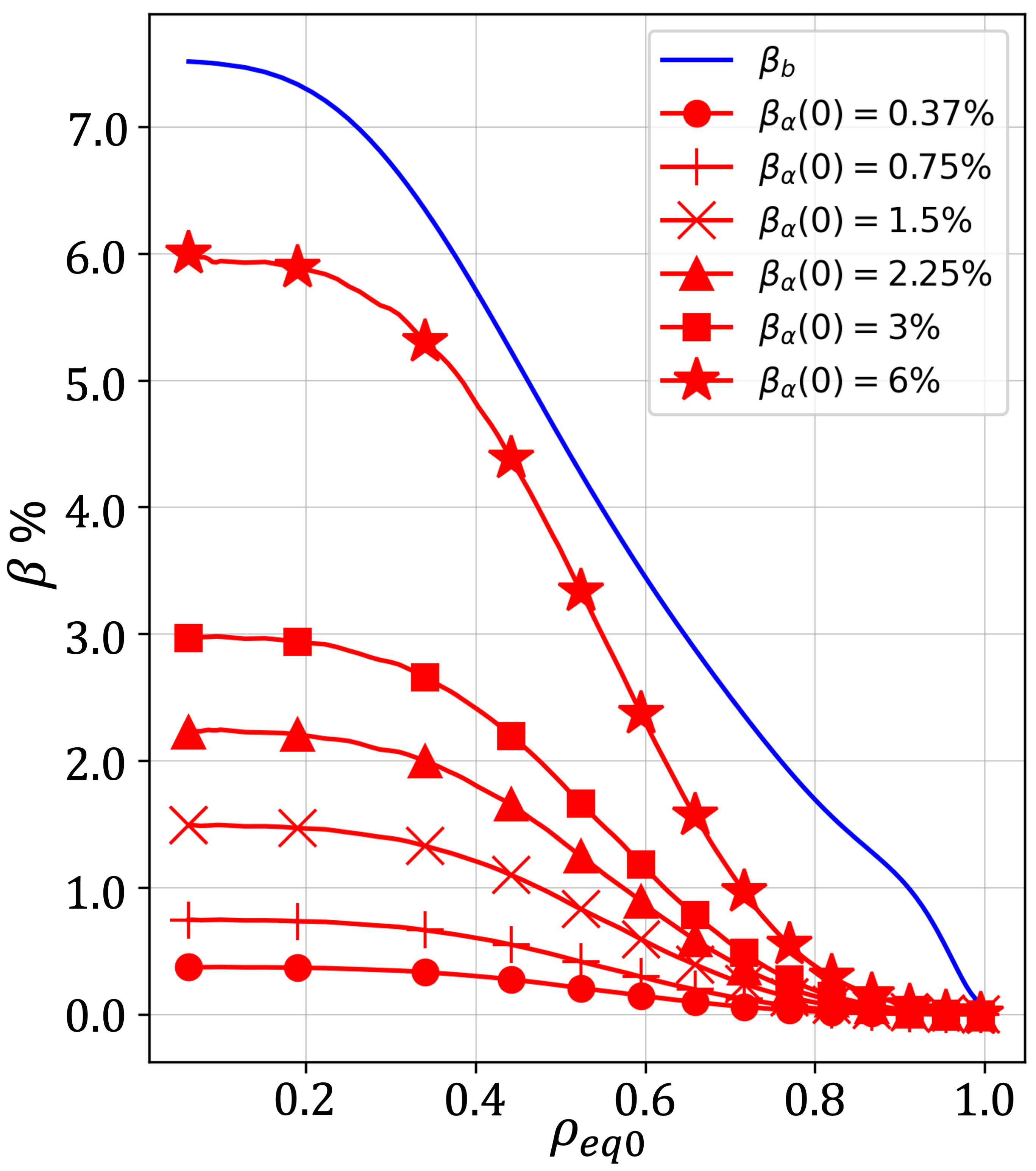}
\end{center} 
\caption{Bulk plasma $\beta_\mathrm{b}$ and alpha particle $\beta_\mathrm{\alpha}$ beta profiles. The blue solid line represents $\beta_\mathrm{b}$, which will remain unchanged. The red solid lines with markers represent the $\beta_\mathrm{\alpha}$ profiles that are used in our parameter scans.}
\label{fig:3betabh}
\end{figure}

\section{\label{sec:level3}Bulk Plasma and Alpha Particles Parameters}

\subsection{\label{sec:level3a}MHD Equilibria}
\quad
The magnetic field, plasma shape, bulk plasma pressure, density, and temperature are based on ITER hybrid scenarios described in Ref.\cite{kim2016corsica} with a flat-top plasma current of approximately $13$ MA.  The on-axis values for the magnetic field strength, electron density, and ion-electron temperature are approximately $5$ T, $8.5\times10^{19}$ $m^{-3}$, and $28$ keV, respectively.  Since this study focuses on physical understanding rather than prediction, the safety factor $q$ profile is treated as a free parameter, and the $q$ profile scans are considered without the constraint of strict experimental achievability.  These scans are constrained to the typical range for tokamak hybrid scenarios: (i) an off-axis minimum $q_\mathrm{min} \gtrsim 1$, (ii) low magnetic shear core plasma with $q\gtrsim1$, and (iii) $q_{95} \approx 4$ near the boundary. Two sets of MHD equilibrium scans are considered.  The first scan varies the safety factor at the magnetic axis, referred to as the ``$q_\mathrm{0}$ scan," while the second scan varies the radial position of the minimum safety factor $q_\mathrm{min}$, referred to as the ``$\rho_\mathrm{qmin}$ scan."   In this study, ``$\rho$" denotes the normalized toroidal flux, while the ``min", ``eq0", and ``eq" subscripts represent the $q_\mathrm{min}$ position, the axisymmetric equilibrium state, and the HC state, respectively.  In both scans, $q_\mathrm{min}$ is fixed at $1.04$, while the $\beta_\mathrm{b}(\rho_\mathrm{eq0})$ profile is fixed as shown in Fig.\ref{fig:3betabh}.   The prescribed $\beta_\mathrm{b}$ profile has the nominal value of $7.53\%$ at the magnetic axis.

In this study, the MHD equilibria for each $q$ profile are calculated using VMEC\cite{hirshman1983steepest}, a 3-D ideal MHD equilibrium solver based on the minimization of the MHD energy functional.  Both the axisymmetric $n=0$ and the HC (asymmetric $n\geq0$) equilibrium solutions are calculated.  The axisymmetric $n=0$ equilibrium serves as an initial condition for MEGA, while the HC equilibrium will be used to benchmark with the HC results calculated with MEGA.  The equilibrium calculation parameters used in this study adhere to the requirements for numerical convergence in HC equilibrium calculations, such as the need to use sufficiently large numbers of poloidal and toroidal Fourier components and radial grid points needed to achieve a converged HC equilibrium\cite{cooper2011helical}.  The MHD equilibria calculated using VMEC are shown in Fig.\ref{fig:3mhdeq}. Panels (a–d) and (e–h) show the equilibrium data for the $q_0$ and $\rho_\mathrm{qmin}$ scans, respectively. For the $q_\mathrm{0}$ scan, $q_\mathrm{0}=1.06$ (blue), $1.1$ (violet), and $1.2$ (red) are considered, while $\rho_\mathrm{qmin}$ is fixed at $0.575$, as shown in Panel (a).  For the $\rho_\mathrm{qmin}$ scan, $q_\mathrm{0}$ is fixed at $1.1$ while $\rho_\mathrm{qmin}=$ $0.456$ (green), $0.575$ (violet), and $0.675$ (cyan) are considered as shown in Panel (e).  When we vary the $q$ profile, the net plasma current will deviate from the reference value of $13$ MA and vary in the range $13$ MA $\leq I_p \leq$ $15$ MA.  It is important to note that both the $q$ profile and the pressure profile in flux coordinates are the VMEC input variables, meaning that these profiles are identical for both the axisymmetric and HC equilibria.  Panels (b, f) and (c, g) show the magnetic Poincar{\'e} plots of the HC equilibria at toroidal angles $\phi = 180^\circ$ and $0^\circ$, respectively.  Throughout this paper, the $\phi = 180^\circ$ and $0^\circ$ toroidal angles always represent the angle where the compressed flux region of HC is located on the high-field and low-field sides, respectively.  To facilitate intuitive visualization and understanding, the major radial coordinate, horizontal axis, in all figures of this paper showing a profile at the toroidal angles $\phi = 180^\circ$ will be inverted.  This inversion is intended to provide a more intuitive representation of the 3-D structure of HC by mimicking the side-view of the tokamak.  In case the HC is rotated by the non-ideal effect, the HC phase will be re-adjusted to match this convention.  In addition, Fig.\ref{fig:3mhdeq}(d,h) shows the toroidally averaged radial displacement of the magnetic axis is defined by and evaluated with the formula,

\begin{eqnarray} \label{eq:deltah}
\delta_\mathrm{HC} = \int{\frac{\sqrt{(R_{a}-R_{a0})^2 + (Z_{a}-Z_{a0})^2}d\phi}{2\pi}}.
\end{eqnarray}

\noindent
In this equation, $(R_\mathrm{a}, Z_\mathrm{a})$ and $(R_\mathrm{a0}, Z_\mathrm{a0})$ denote the positions of the magnetic axis for the HC equilibrium and the axisymmetric equilibrium, respectively.  According to the VMEC results, we can see that $\delta_\mathrm{HC}$ increases when either the $q_\mathrm{0}$ is reduced toward unity or $\rho_{qmin}$ shifts outward radially.  From our equilibrium setups and assumptions, the reduction of the $q_\mathrm{0}$ toward unity decreases the magnetic shear in the $\rho \leq \rho_\mathrm{qmin}$ region.  A larger value
of $\rho_\mathrm{qmin}$ yields a larger $q\gtrsim1$ domain size. It is important to note that the $\beta_\mathrm{b}$ profile is fixed for all equilibria.  This implies that changes in $\rho_{qmin}$ also alter the $\beta_\mathrm{b}$ gradient at $\rho_\mathrm{qmin}$.

\subsection{\label{sec:level3b}Alpha Particle Distribution Function}
\quad
In this study, the distribution function of D-T fusion-born alpha particles was estimated using the D-T fusion reaction rate profile. We assume that the alpha particles remain confined long enough to develop a slowing-down velocity distribution with a given radial profile. Based on this assumption, we calculated the alpha particle distribution from the total number of alpha particles generated by the D-T fusion reaction over the slowing-down timescale $\tau_{sd}$.  The pitch-angle distribution is assumed to be isotropic. The initial alpha particle distribution $f_{\alpha 0}$ is defined by

\begin{eqnarray} \label{eq:falpha1}
f_{\alpha 0} = \frac{1}{v^3 + v^3_{c}}\langle \sigma v\rangle_{DT}n_Dn_T\tau_{sd}\mathscr{C}.
\end{eqnarray}
\noindent

The variables in Eq.\ref{eq:falpha1} are defined as follows: D-T fusion thermal reactivity $\langle \sigma v\rangle_{DT}$, deuterium number density $n_D$, tritium number density $n_T$, slowing-down time $\tau_{sd}$, alpha particle velocity $v$, critical velocity $v_c$, and scaling factor $\mathscr{C}$.   $\langle \sigma v\rangle_{DT}$ is estimated from the empirical equation derived by H.S.Bosch and G.M. Hale\cite{bosch1992improved}.  The scaling factor $\mathscr{C}$ is an arbitrary positive real number used for the $\beta_\mathrm{\alpha}$ scan.  $\mathscr{C}$ is scanned within the range of $0.5 \leq \mathscr{C} \leq 8$, which corresponds to a range of $0.37\% \leq \beta_\mathrm{\alpha}(0) \leq 6\%$. The initial $\beta_\mathrm{\alpha}$ profiles for each $\mathscr{C}$ are plotted in Fig.\ref{fig:3betabh} as red solid lines with markers. The $\beta_\mathrm{\alpha}(0)\leq1\%$ cases are within the expected range of ITER operating values\cite{todo2014large,varela2019analysis}, while the $\beta_\mathrm{\alpha}(0)>1\%$ cases are exaggerated cases.

\section{\label{sec:level4}Helical Core Formation in the Absence of Alpha Particles}
\quad
This section aims to compare the HC quasi-steady states calculated with MEGA against those from VMEC. Since VMEC has already been validated against experiments, this comparison can ensure that MEGA can provide reliable HC-related results before proceeding with a computationally expensive kinetic simulation involving alpha particles. Since VMEC cannot account for the kinetic effects of alpha particles, we let $\beta_\mathrm{\alpha}=0\%$.   This section is divided into 3 subsections. Sections \ref{sec:level4a1}-\ref{sec:level4a2} focus the $q_\mathrm{0}$ and $\rho_\mathrm{qmin}$ equilibrium scans, respectively. These $q$ profile scans aim to demonstrate that both MEGA and VMEC produce consistent results over a wide range of the hybrid scenario's operating parameters. The results in these subsections also reveal the emergence of secondary instabilities after HC formation. We find that these secondary instabilities are a kind of resistive pressure-driven MHD modes destabilized by the steepening pressure gradient in the HC compressed flux region. The properties of these secondary modes are discussed in Section \ref{sec:level4c}. Unless otherwise stated, the electric resistivity is fixed at the moderate value of $\hat{\eta}=10^{-6}$.  The only exception is the resistivity scan in the range $10^{-7} \leq \hat{\eta} \leq 10^{-5}$ performed in Section \ref{sec:level4c3}.

\begin{figure*}[t]
\begin{center}
\includegraphics[width=1.00\linewidth]{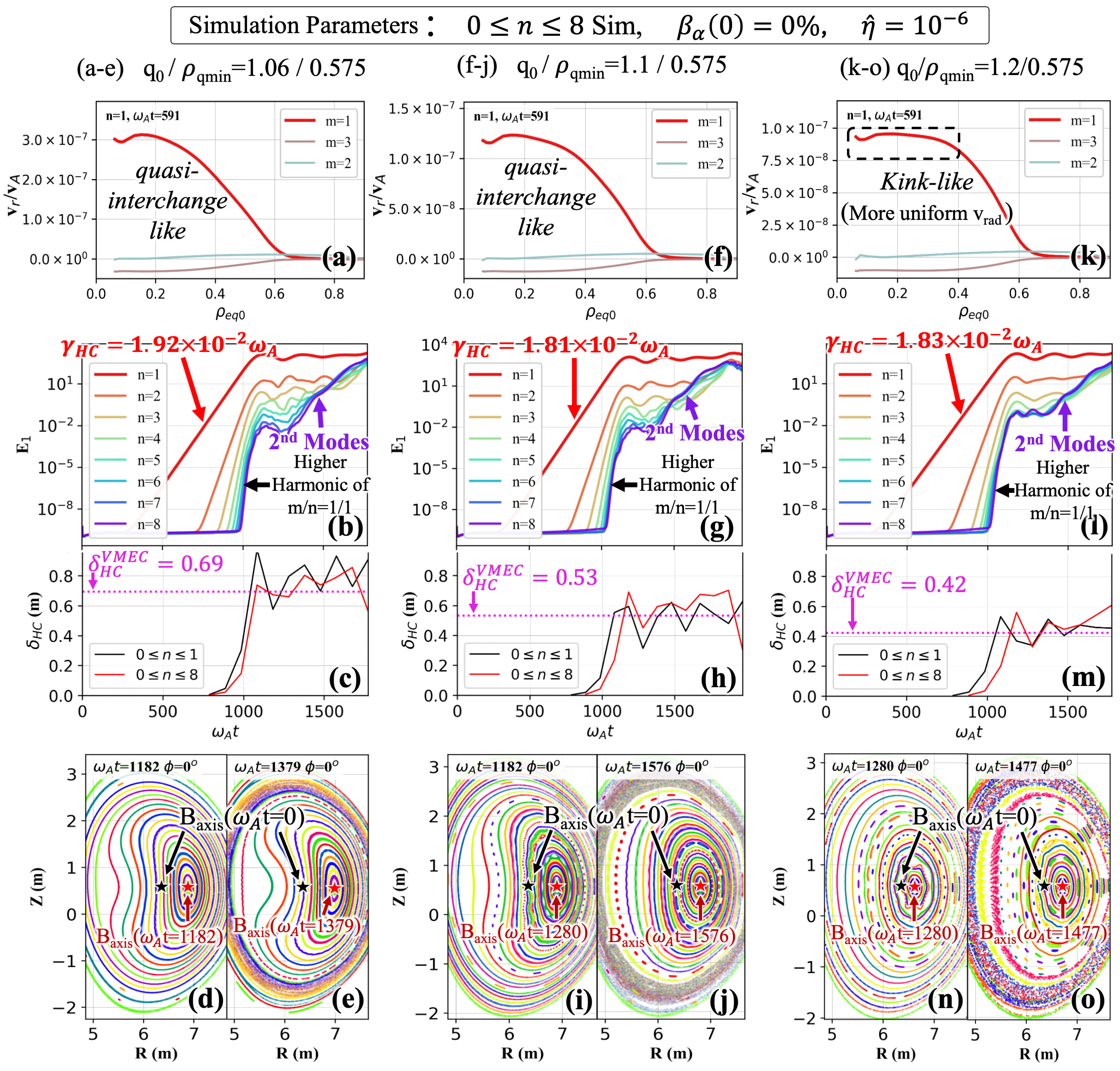}
\end{center} 
\caption{Helical core formation in the equilibria with (a-e) $q_\mathrm{0}/\rho_\mathrm{qmin}=1.06/0.575$, (f-j) $1.1/0.575$, and (k-o) $1.2/0.575$. The MEGA simulation was performed with $(N_R,N_Z)$=$(200,200)$, $\hat{\eta}=10^{-6}$, and $\beta_\mathrm{\alpha}=0\%$.  Panels (a,f,k) show the cosine components of the $n=1$ radial MHD velocity harmonics.  Panels (b,g,l) show the time evolution of $1\leq n\leq8$ mode energies $E_n$.  Panels (c,h,m) show the time evolution of the radial displacement of the magnetic axis $\delta_\mathrm{HC}$.  Panels (d,i,n) and (e,j,o) show the magnetic Poincar{\'e} plots before and after the excitation of secondary modes, respectively.}
\label{fig:4megaq0scan}
\end{figure*}

\subsection{\label{sec:level4a}$q_\mathrm{0}$ Scan}
\quad

For the HC formation in the $q_\mathrm{0}$ scan, the analyses are divided into the linear (Section \ref{sec:level4a1}) and nonlinear phases (Sections \ref{sec:level4a2}-\ref{sec:level4a3}). 

\subsubsection{\label{sec:level4a1}Linear Growth Rate and Eigenfunction} 
\quad

We begin our analysis with the linear growth phase, where we aim to ensure that HC can form spontaneously in our plasmas.  To confirm this point, we check the linear stability of the $n\leq8$ modes in our scanned equilibrium.  The simulation results for the $q_\mathrm{0}$ scan are summarized in Fig.\ref{fig:4megaq0scan}, where panels (a-e), (f-j), and (k-o) correspond to the $q_\mathrm{0}=1.06$, $1.1$, and $1.2$ equilibria, respectively.  The time evolution of the $1\leq n\leq8$ mode energies $E_n$ is shown in panels (b,g,l), where we find that the $n=1$ mode shown in red is the only linearly unstable mode. (Regarding the representation of toroidal mode number $n$ shown in this paper, the transition from red to violet, according to the rainbow spectrum, always represents the increment of $n$.)  The linear growth rate $\gamma_\mathrm{HC}$ of the $n=1$ mode for the $q_\mathrm{0}=1.06$, $1.1$, and $1.2$ equilibria are $1.83\times10^{-2}\omega_A$, $1.81\times10^{-2}\omega_A$, and $1.92\times10^{-2}\omega_A$, respectively. (For simplicity, the ``HC" subscript is used to represent both the HC and the linear $m/n=1/1$ kink/quasi-interchange mode.)

The linear eigenfunction of the $n=1$ modes represented by the radial MHD velocity profile $v_\mathrm{rad}(\rho_\mathrm{eq0})$ are shown for each equilibrium in panels (a,f,k).  The legends in these panels show the first three dominant poloidal harmonics, sorted by the peak amplitude.  The phase of each mode has been adjusted to maximize the cosine component at the peak position of the dominant poloidal Fourier component.  Only the cosine component is shown in this figure because the sine component is zero in the poloidal plane shown here.  We can confirm that the dominant harmonic is the $m/n=1/1$ mode, and two types of eigenfunction are observed.  The first is the quasi-interchange type, which is observed when the core magnetic shear becomes very weak near the $q\gtrsim1$ surface.  For the quasi-interchange type HC, the radial displacement or velocity has a centrally peaked bell-shaped structure, implying that the magnetic axis has a much higher radial displacement than the other regions.  This quasi-interchange type HC can be observed for the $q_\mathrm{0}=1.06$ and $1.10$ equilibria as shown in Fig.\ref{fig:4megaq0scan}(a,f).  The second is the kink type, which is observed at a higher value of the core magnetic shear.  The kink-type HC exhibits a more box-shaped profile within $\rho_\mathrm{qmin}$, indicating that the core region is displaced more rigidly, with the magnetic axis and surrounding flux surfaces moving at nearly the same speed.  This kink-type HC can be seen in the $q_\mathrm{0}=1.2$ equilibrium as shown in Fig.\ref{fig:4megaq0scan}(k).   These two types of HC have also been previously reported by S. Kawagoe et al.\cite{kawagoe2022two} using VMEC. 

\subsubsection{\label{sec:level4a2}Nonlinear Phase: HC Formation}
\quad

To investigate the HC formation in these equilibria, we continued our simulation until the $n=1$ mode is saturated.  In this study, we will consider that the particular nonlinear saturation state is a HC if and only if the following two conditions are satisfied simultaneously:
\begin{enumerate}
  \item The $m/n=1/1$ mode is dominant and settles in a (quasi-)steady state.
  \item Ideal MHD ``frozen-in" condition remains intact.
\end{enumerate}
Once these criteria are confirmed, the radial displacement of the magnetic axis $\delta_\mathrm{HC}$ will be calculated from field line tracing and compared with the VMEC results.  

\begin{figure}[h]
\begin{center}
\includegraphics[width=1.0\linewidth]{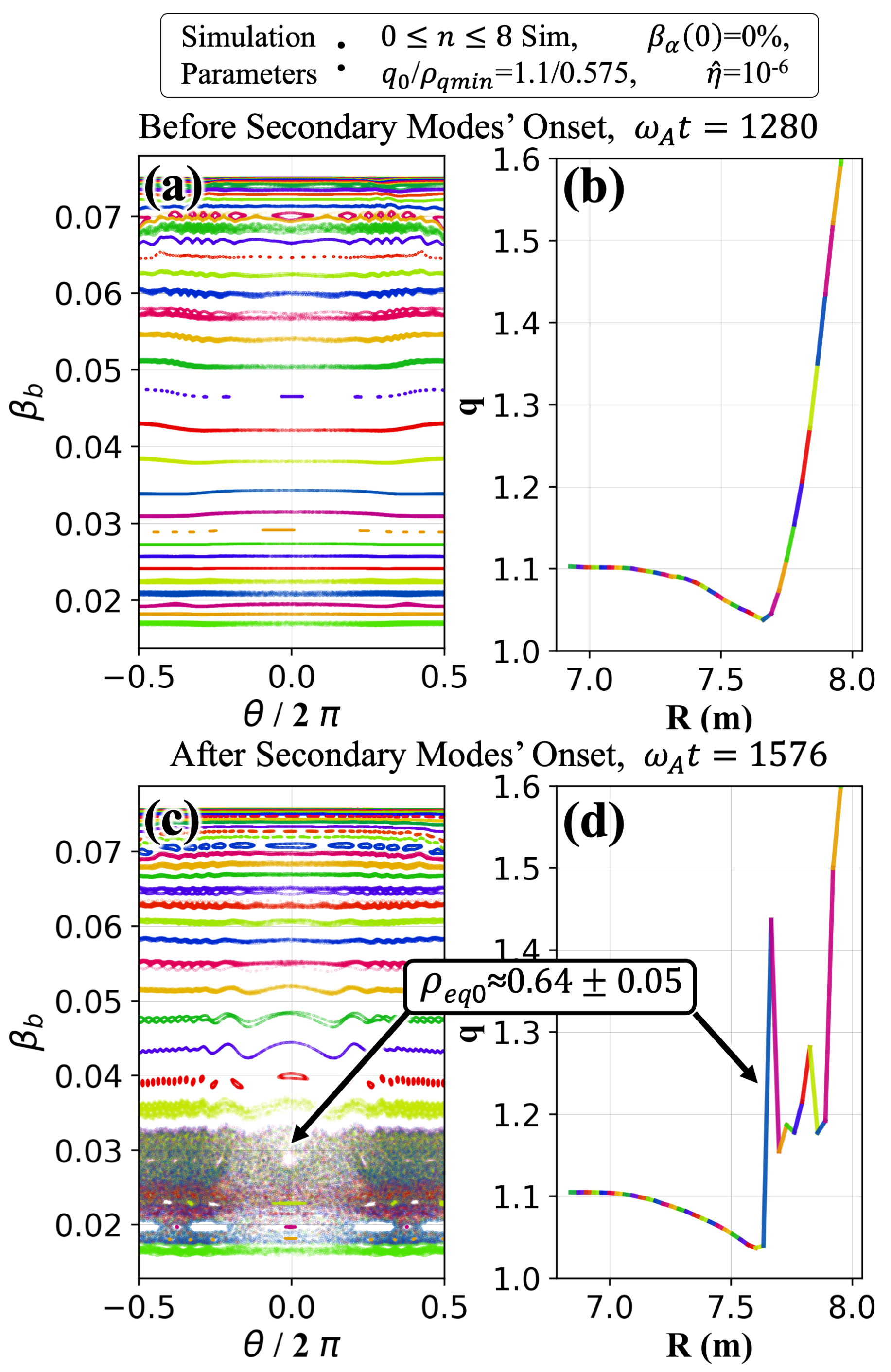}
\end{center} 
\caption{(a, c) Poloidal variation of $\beta_\mathrm{b}$ along the traced magnetic field lines, and (b, d) $q$ profile of the $q_\mathrm{0}/\rho_\mathrm{qmin} = 1.1/0.575$ equilibrium simulated with $(N_R,N_Z)$=$(200,200)$, $\hat{\eta}=10^{-6}$, and $\beta_\mathrm{\alpha} = 0\%$. The colors used in this figure represent individual magnetic field lines.  Panels (a-b) and (c-d) show the simulation results before and after the excitation of secondary modes, respectively.}
\label{fig:4megaq0scan_betabq}
\end{figure}

During the nonlinear phase (e.g., $1100 \leq \omega_At \leq 1500$ for the $q_\mathrm{0}/\rho_\mathrm{qmin} = 1.1/0.575$ equilibrium), the $n=1$ mode remains dominant for all cases as shown in Figs.\ref{fig:4megaq0scan}(b,g,l). The time variation of the $n=1$ mode energy is negligible, satisfying criterion (i). To verify criterion (ii), we examine the magnetic Poincar{\'e} plot, the pressure profile, and the safety factor profile after the HC formation. The magnetic Poincar{\'e} plots at the beginning of the nonlinear phase are shown in Figs.\ref{fig:4megaq0scan}(d,i,n). These magnetic Poincar{\'e} plots correspond to the magnetic flux surfaces at the $\phi=0^\circ$ toroidal angle.  At this timeslice, the nested magnetic flux surfaces are well-preserved in all equilibria.   Next, we need to check the bulk plasma pressure $\beta_\mathrm{b}$ and $q$ profiles during the nonlinear phase.  For the $\beta_\mathrm{b}$ profile, we showed in terms of its poloidal variation along the magnetic field line at $\phi=0^\circ$ in Fig.\ref{fig:4megaq0scan_betabq}(a).  In this figure, the vertical and horizontal axes represent the local $\beta_\mathrm{b}$ value and geometric poloidal angle, respectively.  Each line color shown in this figure corresponds to an individual magnetic field line traced at different radial positions.  Only the $q_\mathrm{0}/\rho_\mathrm{qmin} = 1.1/0.575$ results are shown because other equilibria show qualitatively similar outcomes.  It can be seen that $\beta_\mathrm{b}$ has a weak poloidal variation along the field line, mainly because $\chi_\parallel$ used in our simulation is insufficient to relax this poloidal variation within the MHD time scale.  Radial transport of the bulk plasma is not observed, indicating that the $\beta_\mathrm{b}$ profile remains mostly unchanged in the distorted flux coordinate after saturation. Similarly, the $q$ profile at the beginning of the nonlinear phase, shown in Fig.\ref{fig:4megaq0scan_betabq}(b), does not exhibit any notable flattening of the $q$ profile near the low-order rational surfaces.  These findings collectively satisfy criterion (ii) and indicate that the observed nonlinear saturation state can be classified as a HC.  The absence of notable $\beta_\mathrm{b}$ and $q$ profile flattening suggests that the free energy used to drive the $m/n=1/1$ kink/quasi-interchange mode is partially spent to displace the magnetic flux surface until the stabilization effects, such as the enhanced field line bending, become sufficient to prevent further growth.

As one continues the simulation further, the $n\geq5$ mode energies, represented by the green-to-violet color spectrum, grow after the initial saturation of the $m/n=1/1$ mode.  These modes will be addressed as the secondary modes in this study, and they were marked with the violet ``$2^{nd}$ Modes" label in Fig.\ref{fig:4megaq0scan}(b,g,l).  Based on the time evolution of the mode energies, we can preliminarily conclude that these secondary modes are linearly stable during the linear growth rate phase.  They are not driven by nonlinear mode-mode coupling, as evidenced by the nearly stationary $n=1$ mode; hence, they are likely driven unstable by the equilibrium profile modification caused by HC.  As these secondary modes grow to higher amplitudes, they can nonlinearly cause chaotization of the magnetic field lines, affecting HC in the $0\leq n\leq8$ simulation.  Their effect is illustrated in the magnetic Poincar{\'e} plots shown in Fig.\ref{fig:4megaq0scan}(e,j,o).   The chaotic magnetic field arises at the transition point between the HC and the axisymmetric region, near $q_\mathrm{min}$\footnote{The term ``chaotic" magnetic field is used to describe the destruction of nested flux surfaces caused by the overlapping of two magnetic islands with different helicities. Some works described this phenomenon as ``stochastic". However, the authors of this paper prefer ``chaotic" over ``stochastic" because randomness is not included in our MHD and guiding center equations, making our model deterministic.}.  The effects of these secondary modes on the $\beta_\mathrm{b}$ and $q$ profiles for the $q_\mathrm{0}/\rho_\mathrm{qmin} = 1.1/0.575$ case are shown in Fig.\ref{fig:4megaq0scan_betabq}(c-d), where radial mixing and the destruction of magnetic flux surfaces are evident. As these secondary modes continue to grow, they can lead to the collapse of the HC, which is indicated by the abrupt reduction of $\delta_\mathrm{HC}$ observed in the $0 \leq n \leq 8$ simulation shown in Figs.\ref{fig:4megaq0scan}(c,h). The collapse of HC is not observed for the $0 \leq n \leq 1$ simulation; therefore, its results represent the HC dynamics in the absence of these secondary modes. (For the $q_\mathrm{0}/\rho_\mathrm{qmin} = 1.2/0.575$ case, the collapse was not observed because the simulation was not sufficiently long.)  For the $q_0/\rho_{q_\mathrm{min}} = 1.1/0.575$ case, this HC collapse occurs at $\omega_A t \approx 1900$, as shown in Fig.\ref{fig:4megaq0scan}(h). Further discussion of these modes is provided in Section \ref{sec:level4c}.

\subsubsection{\label{sec:level4a3}Nonlinear Phase: MEGA and VMEC Benchmarking}
\quad

To benchmark the HC calculated with MEGA and VMEC, we compared (i) the toroidally averaged radial displacement of the HC magnetic axis $\delta_\mathrm{HC}$, (ii) the HC magnetic topology, and (iii) the deformed pressure profile.  

In the first part, we focus on the $\delta_\mathrm{HC}$ benchmarking.  For clarity, $\delta_\mathrm{HC}$ calculated with MEGA and VMEC will be denoted as $\delta_\mathrm{HC}^\mathrm{MEGA}$ and $\delta_\mathrm{HC}^\mathrm{VMEC}$, respectively.   The $\delta_\mathrm{HC}$ results for the $q_0=1.06$, $1.1$, and $1.2$ equilibria are shown in Figs.\ref{fig:4megaq0scan}(c,h,m), respectively.  For $\delta_\mathrm{HC}^\mathrm{MEGA}$, we considered the results from the $0\leq n\leq1$ and $0\leq n\leq8$ simulations, and they are shown as a black solid line and a red solid line, respectively, while $\delta_\mathrm{HC}^\mathrm{VMEC}$ is plotted as a horizontal dotted magenta line.  It is important to note that VMEC solves for the HC equilibrium as a boundary value problem; therefore, only the steady state solution of HC is calculated.  At the initial saturation time, $\delta_\mathrm{HC}^\mathrm{MEGA}$ exhibits an overshoot, exceeding that of $\delta_\mathrm{HC}^\mathrm{VMEC}$.  As time progresses, $\delta_\mathrm{HC}^\mathrm{MEGA}$ oscillates with a gradually decreasing oscillation range for a few hundred Alfv{\'e}n times.   The central value of the oscillation is slightly higher than $\delta_\mathrm{HC}^\mathrm{VMEC}$, but the maximum difference is not larger than $15\%$.  The $0\leq n\leq8$ simulation yields a slightly higher value compared to the $0\leq n\leq1$ case.  This may be attributed to the relaxation of the numerical constraint imposed by the toroidal low-pass filter, which allows the plasma to access a broader range of states.   In terms of the $q$ profile dependence, it can be seen that both codes show a consistent trend where $\delta_\mathrm{HC}$ becomes smaller with increasing $q_0$, equivalent to increasing magnetic shear.  There is one point worth noting.  The HC result for $q_\mathrm{0}/\rho_\mathrm{qmin} = 1.1/0.575$ has a comparable $\gamma_\mathrm{HC}$ value to that of $q_\mathrm{0}/\rho_\mathrm{qmin} = 1.2/0.575$, but with a noticeably higher $\delta_\mathrm{HC}$.  This paradox can be resolved by noting 2 points: (i) A correlation between the linear growth rate and the saturation level is expected only when the saturation is attributed to the flattening of the gradient with respect to the largely preserved background equilibrium.  In contrast, HC formation is a highly nonlinear process where the free energy is spent to displace and distort magnetic surfaces and the field lines on them in such a way that the total plasma energy is minimized. (ii)   From Section \ref{sec:level4a1}, we infer that the $q_\mathrm{0}/\rho_\mathrm{qmin} = 1.2/0.575$ case has a kink-like eigenfunction, meaning that the entire core plasma is displaced at the same rate, while the quasi-interchange HC observed in the $q_\mathrm{0}/\rho_\mathrm{qmin} = 1.1/0.575$ has a higher displacement near the magnetic axis region.

\begin{figure}[h]
\begin{center}
\includegraphics[width=1.00\linewidth]{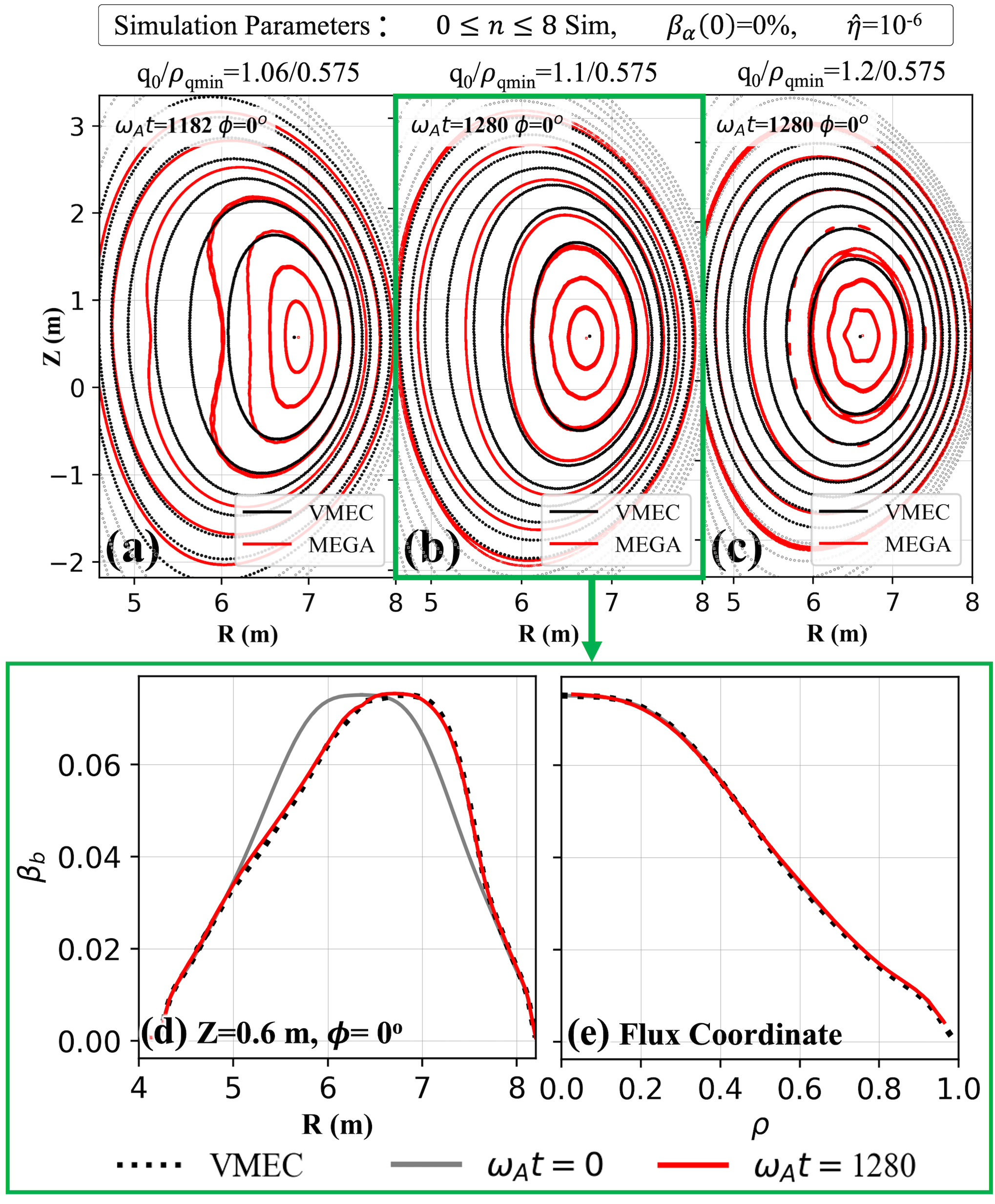}
\end{center} 
\caption{MEGA and VMEC benchmark on the (a-c) HC magnetic flux surfaces and the (d-e) pressure profile.  The MEGA simulations are performed with $(N_R,N_Z)$=$(200,200)$, $\hat{\eta}=10^{-6}$, and $\beta_\mathrm{\alpha}(0)=0\%$.  Panels (a-c) compare the magnetic Poincar{\'e} plots at toroidal angle $\phi=0^\circ$ that were obtained in the equilibria with $q_\mathrm{0}=1.06$, $1.1$, and $1.2$, respectively.  The red contours and black contours represent MEGA and VMEC results, respectively.   Panels (d-e) compare the $\beta_\mathrm{b}$ profile of the $q_\mathrm{0}/\rho_\mathrm{qmin}=1.1/0.575$ equilibrium.  Panel (d) shows the pressure profile along the major radial direction in the geometric mid-plane at $\phi = 0^\circ$, while panel (e) shows the results as functions of the normalized toroidal flux.}
\label{fig:4megavmecbench}
\end{figure}

Comparing $\delta_\mathrm{HC}$ alone is not enough to confirm the consistency between MEGA and VMEC because $\delta_\mathrm{HC}$ is a local quantity.  In the second part of this section, we compared the overall HC magnetic topology.  The magnetic Poincar{\'e} plots of the HC equilibrium calculated with MEGA and VMEC are shown in Fig.\ref{fig:4megavmecbench} as red and black solid lines, respectively.  Panels (a), (b), and (c) display results for the $q_\mathrm{0}=1.06$, $1.1$, and $1.2$ equilibria, respectively.  The selected time-slice for this comparison is selected such that the time where the oscillating $\delta_\mathrm{HC}^\mathrm{MEGA}$ closely matches $\delta_\mathrm{HC}^\mathrm{VMEC}$. For the $q_\mathrm{0}=1.1$ and $1.2$ equilibria, the magnetic flux surfaces calculated by both codes are well-aligned. However, a notable deviation is observed for the $q_\mathrm{0} = 1.06$ equilibrium in the uncompressed flux region of HC.  The magnetic flux surface calculated with MEGA takes on a bean-like shape, while VMEC predicts an oval shape. The bean-like flux surface observed in MEGA results from the convective cell of the quasi-interchange mode.  This bean-type structure can be observed in both the $0\leq n\leq 1$ and $0\leq n\leq 8$ simulations, suggesting that it is not caused by the $n>1$ modes.  These processes are absent in the steady-state solution to which VMEC is constrained. We would like to note that bean-like flux surfaces are also observed for the $q_\mathrm{0}=1.1$ equilibrium during the $\delta_\mathrm{HC}^\mathrm{MEGA}\gg\delta_\mathrm{HC}^\mathrm{VMEC}$, e.g., during the overshoot phase of HC.

For the comparison of the redistributed bulk plasma pressure $\beta_\mathrm{b}$ profiles, the MEGA and VMEC results are shown in Figs.\ref{fig:4megavmecbench}(d-e).  Panel (d) presents the $\beta_\mathrm{b}$ profile along the major radial direction, while Panel (e) shows it as a function of the flux surface.  Here, only the results for the $q_\mathrm{0}/\rho_\mathrm{qmin} = 1.1/0.575$ case are shown.  The MEGA results for the axisymmetric and HC states are shown as gray and red solid lines, respectively, while a black dotted line represents the VMEC results.  For the real space shown in Panel (d), the $\beta_\mathrm{b}$ profile along the major radial direction is plotted along the geometric mid-plane ($Z = Z_0 \approx 2.6$ m) at $\phi = 0^\circ$.  MEGA and VMEC yield a quantitative agreement.  For the comparison in the flux coordinate shown in Panel (e), it is important to note that VMEC treats $\beta_\mathrm{b}$ as a prescribed flux function, which remains fixed in the flux coordinates throughout the calculation.  We find that both codes also show a quantitative agreement in the flux coordinate.

As discussed here, VMEC and MEGA yield consistent HC solutions in terms of both $\delta_\mathrm{HC}$,  eigenfunction, and the redistributed pressure profile.  If the VMEC and MEGA results were truly consistent, one would also expect the HC equilibrium computed by VMEC to be linearly unstable to the secondary mode. Preliminary results on the linear stability of the HC equilibrium obtained from VMEC are briefly discussed in \ref{sec:apen1}.

\begin{figure}[h]
\begin{center}
\includegraphics[width=0.90\linewidth]{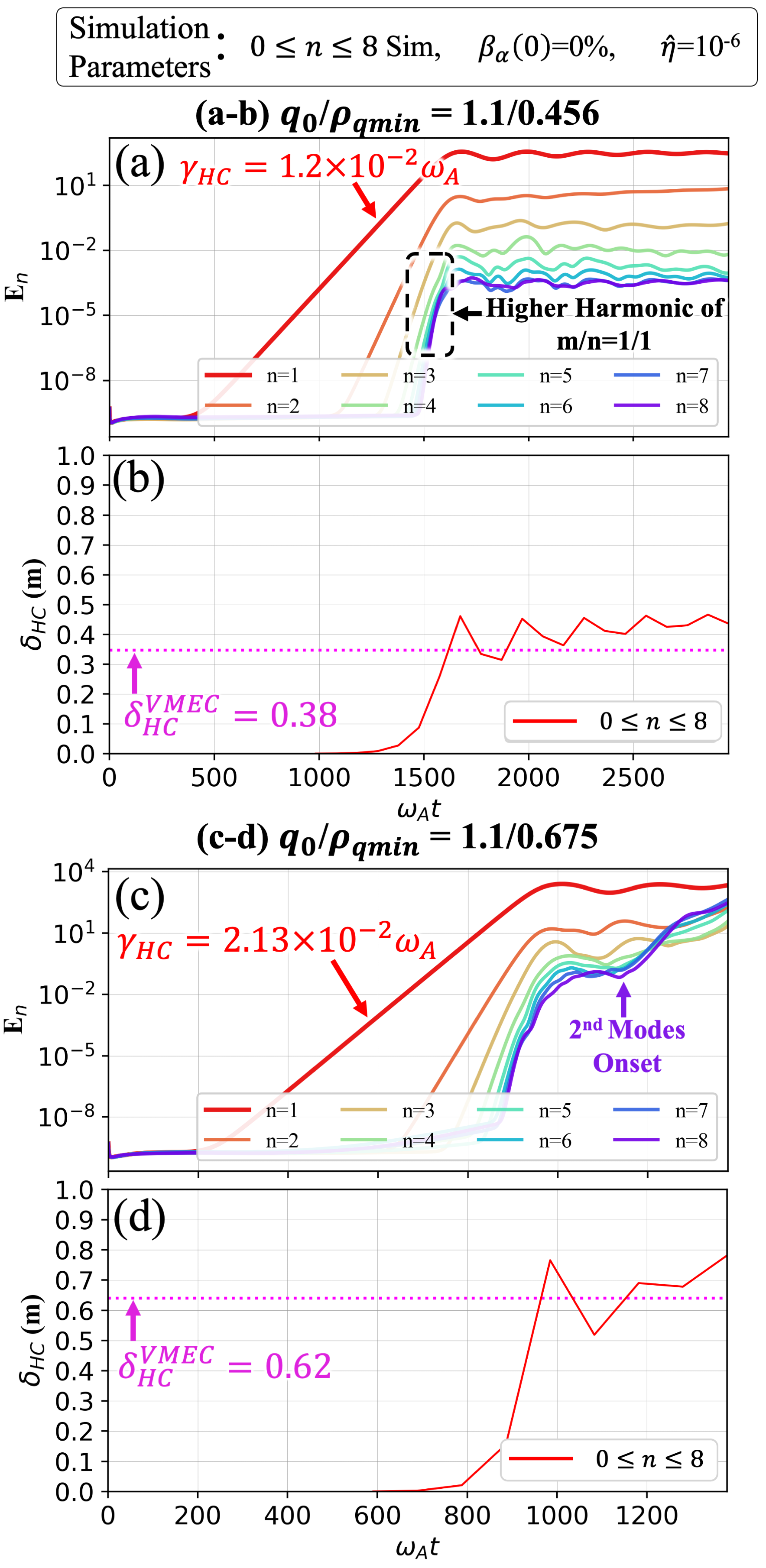}
\end{center} 
\caption{Helical core formation of the equilibria with (a-b) $q_\mathrm{0}/\rho_\mathrm{qmin}=1.1/0.456$ and (c-d) $1.1/0.675$. The MEGA simulation was performed with $(N_R,N_Z)$=$(200,200)$, $\hat{\eta}=10^{-6}$ and $\beta_\mathrm{\alpha}=0\%$.  Panels (a,c) show the time evolution of the $1\leq n\leq8$ mode energies $E_n$.  Panels (b,d) show the time evolution of the radial displacement of the magnetic axis $\delta_\mathrm{HC}$.}  
\label{fig:4megaqwscan}
\end{figure}

\begin{figure*}[t]
\begin{center}
\includegraphics[width=0.85\linewidth]{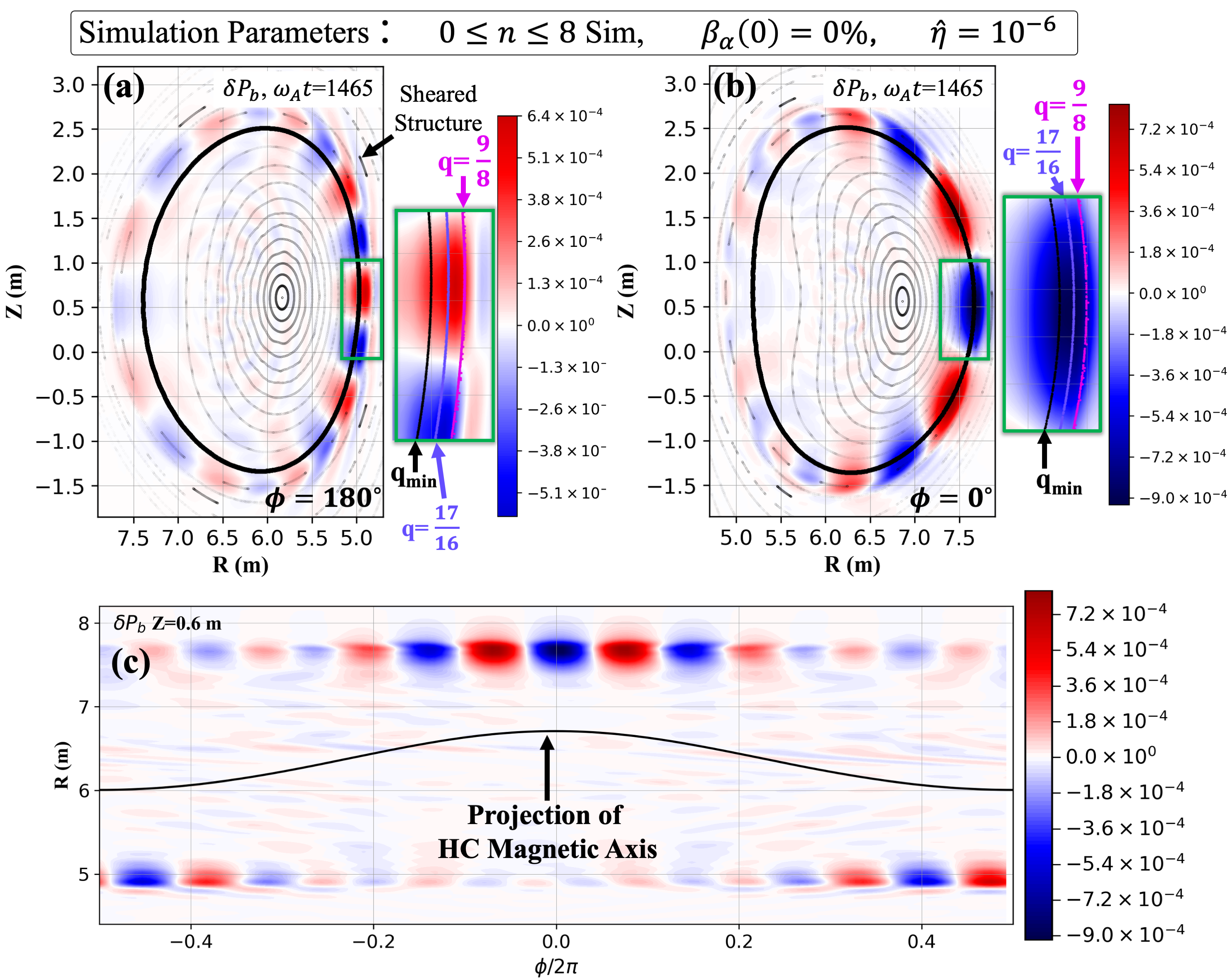}
\end{center} 
\caption{Poloidal cross section of the perturbed MHD pressure $\delta P_b$ caused by the $5 \le n \le 8$ Fourier components of the secondary mode at (a) $\phi = 180^\circ$ and (b) $0^\circ$ toroidal angles.  The MEGA simulation was performed with $(N_R, N_Z) = (200,200)$, $\hat{\eta} = 10^{-6}$, and $\beta_\mathrm{\alpha} = 0\%$ in the $q_\mathrm{0}/\rho_\mathrm{qmin} = 1.1/0.575$ equilibrium. Panel (c) presents $\delta P_b$ in the $R$-$\phi$ plane at the geometric mid-plane ($Z = Z_0 \approx 2.6$ m), where the black line indicates the projection of the HC magnetic axis onto this plane.}
\label{fig:4eta6vrad2nd_realspace}
\end{figure*}

\subsection{\label{sec:level4b}$\rho_\mathrm{qmin}$ Scan}
\quad

\begin{figure}[h]
\begin{center}
\includegraphics[width=1.00\linewidth]{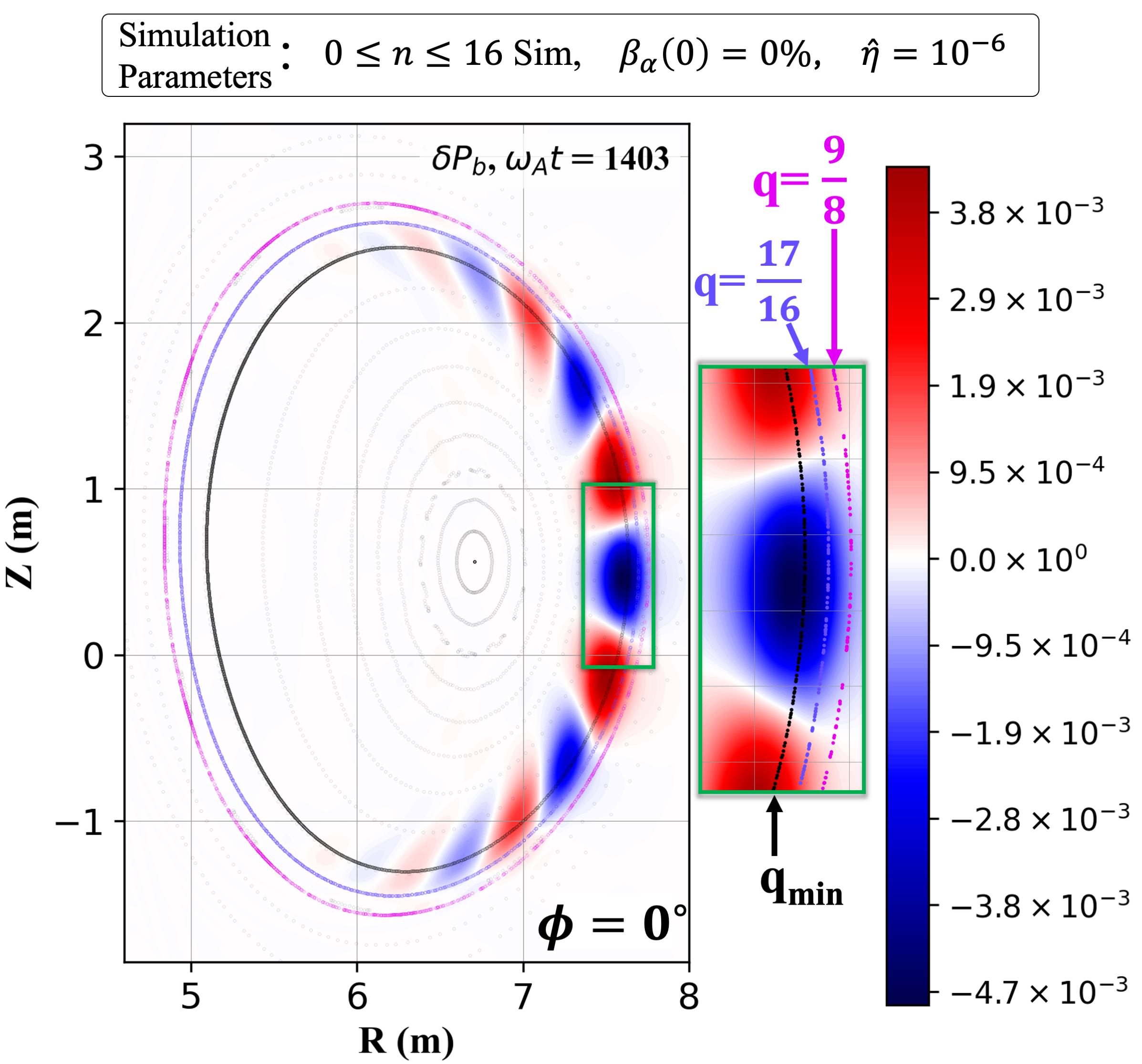}
\end{center} 
\caption{Poloidal cross section of the perturbed MHD pressure $\delta P_b$ caused by the $5 \le n \le 16$ Fourier components of the secondary mode at the $\phi = 0^\circ$ toroidal plane. The MEGA simulation was performed with $(N_R, N_Z) = (200,200)$, $\hat{\eta} = 10^{-6}$, and $\beta_\mathrm{\alpha} = 0\%$ in the $q_\mathrm{0}/\rho_\mathrm{qmin} = 1.1/0.575$ equilibrium.}
\label{fig:4eta6vrad2nd_realspace_n16}
\end{figure}

The simulation results for the $q_\mathrm{0}/\rho_\mathrm{qmin}=1.1/0.456$ and $1.1/0.675$ equilibria are shown in Fig.\ref{fig:4megaqwscan}.  The time evolutions of $n\le8$ mode energies $E_n$ for the $\rho_\mathrm{qmin} = 0.456$ and $0.675$ cases are shown in Figs.\ref{fig:4megaqwscan}(a,c), respectively.  The linear growth rate $\gamma_\mathrm{HC}$ of the $m/n = 1/1$ mode is reduced for $\rho_\mathrm{qmin} = 0.456$ but enhanced for $\rho_\mathrm{qmin} = 0.675$ compared to the $\rho_\mathrm{qmin} = 0.575$ case previously shown in Fig.\ref{fig:4megaq0scan}(g).  Similar to the $q_\mathrm{0}$ scan equilibria, the energy of the $n=1$ mode exhibits a quasi-steady state and preserves the frozen-in flux condition so that we can regard this as the HC. In terms of the radial displacement of the HC magnetic axis $\delta_\mathrm{HC}^\mathrm{MEGA}$, $\delta_\mathrm{HC}^\mathrm{MEGA}$ increases with the size of the low magnetic shear $q\gtrsim1$ region, as shown in Figs.\ref{fig:4megaqwscan}(b,d).  When compare with VMEC, $\delta_\mathrm{HC}^\mathrm{MEGA}$ has a slightly higher value by approximately $18\%$ and $10\%$ for the $\rho_\mathrm{qmin}=0.456$ and $0.675$ cases, respectively.  We also found that the secondary modes are absent in the simulation starting from the $\rho_\mathrm{qmin}=0.456$ equilibrium.  The dependence on the secondary mode stabilities will be elaborately discussed in the next section.

\subsection{\label{sec:level4c}Properties of Secondary Mode}
\quad

As reported in Section \ref{sec:level4a2}, the HC state calculated with MEGA can sometimes be merely transient and collapse within an MHD time scale. We find that this collapse—marked by an abrupt reduction in $\delta_\mathrm{HC}$, as shown in Figs.\ref{fig:4megaq0scan}(c,h), is triggered by the secondary mode that becomes unstable after the HC formation.  To understand why HC states calculated with MEGA are not long-lived in some equilibria, we investigate the physical aspects of these secondary modes.  In Section \ref{sec:level4c1}, the spatial structure of the secondary mode will be discussed.  In Sections \ref{sec:level4c2}-\ref{sec:level4c3}, the dependence of the secondary mode on the electric resistivity $\hat{\eta}$ and the $q$ profile will be discussed.

\subsubsection{\label{sec:level4c1}Mode Structure}
\quad

Representing and Fourier-decomposing a mode in magnetic flux coordinates is a common and useful approach for analyzing dynamics in magnetically confined plasmas. In MEGA, such an analysis can currently be performed only by projecting the perturbed MHD fields from cylindrical coordinates $(R,\phi,Z)$ onto the magnetic flux coordinates $(\Psi_{\rm P,eq0},\phi,\vartheta_{\rm eq0})$ of the initial equilibrium state at $\omega_A t = 0$; namely, in the unperturbed axisymmetric equilibrium magnetic field ${\bm B}_{\rm eq0} = {\bm \nabla}\Psi_{\rm P,eq0} \times{\bm \nabla}\phi + B_{\phi,{\rm eq0}}(\Psi_{\rm P,eq0}){\bm \nabla}\phi$. Here, this is appropriate during the exponential growth phase where the HC displacement $\delta_\mathrm{HC}$ is still small, as during the first 1000 Alfv\'{e}n times in Fig.~\ref{fig:4megaq0scan}(a,f,k) of Section~\ref{sec:level4a2} above. However, during the nonlinear phase, where the plasma is evolving toward the HC equilibrium state and $\delta_\mathrm{HC}$ becomes significant, $(\Psi_{\rm P,eq0},\phi,\vartheta_{\rm eq0})$ do no longer constitute magnetic coordinates and results like those for $t\omega_{\rm A0} > 1000$ in Fig.~\ref{fig:4megaq0scan}(a,f,k) become difficult to interpret. Since we currently do not have a suitable method to analyze MEGA simulation data in the proper magnetic coordinates of the HC state, the following analysis of the secondary mode structure will be performed in real space only. Even then, one should keep in mind that the mode structures we obtained are distorted by a low-pass Fourier filter that has been applied along the geometric toroidal angle $\phi$, which implies that the results are convoluted across a range of the HC's flux surfaces (recall Section~\ref{sec:level2b}).

The poloidal cross sections of the $5\leq n \leq8$ perturbed MHD pressure $\delta P_b$ at $\phi = 180^\circ$ and $0^\circ$ toroidal angles are shown in Figs.\ref{fig:4eta6vrad2nd_realspace}(a,b), respectively. The $q = 9/8$ and $17/16$ magnetic flux surfaces are overlaid in magenta and violet, while other arbitrary flux surfaces are plotted using gray color. These magnetic Poincar{\'e} plots are calculated from the magnetic field data after HC formation but just before the onset of magnetic field chaotization. Regarding the radial position, the peaks of $\delta P_b$ are located slightly closer to the $q = 17/16$ surface than the $q=9/8$ surface, which suggests that some higher-$n$ components that were eliminated by our $n\leq 8$ filter may have been a significant part of the mode (as will be confirmed shortly below and in \ref{sec:apen1}). Concerning the poloidal and toroidal structures, the secondary mode has the strongest fluctuation along the compressed flux region of the HC, in contrast to usual ballooning modes, which are typically localized in the bad curvature region (low-field side) for all toroidal angles.  This secondary mode shares some similarity with the high-$n$ ballooning modes observed during the fast sawtooth relaxation in Refs.\cite{park1995high, nishimura1999onset}, where they are destabilized by the local MHD pressure steepening caused by the $m/n=1/1$ kink mode.  We also plotted the secondary mode structure in the $(R,\phi)$ plane at the geometric mid-plane in Fig.\ref{fig:4eta6vrad2nd_realspace}(c). This structure resembles that shown in Fig.3(b) of Ref.\cite{park1995high}.

The toroidally and poloidally localized structure of the secondary mode persists during its growth phase (prior to its saturation), suggesting that it is a single coherent mode.   To produce such a localized structure, this mode should comprise a broad spectrum of Fourier components with different toroidal and poloidal mode numbers that constructively and destructively interfere in the compressed and uncompressed flux regions, respectively.   This expectation is consistent with the synchronous growth of the $5\leq n \leq8$ Fourier component energies in Figs.\ref{fig:4megaq0scan}(b,g,l) and Fig.\ref{fig:4megaqwscan}(c).

As we have already noted at the beginning of this section, our toroidal low-pass filter alters the secondary mode's structure by truncating high-n components. To demonstrate this, we extend the maximum toroidal mode number of the low-pass filter to $n = 16$.  To prevent the rapid growth of modes with the $9 \leq n \leq 16$ Fourier components during the linear growth phase, the $0 \leq n \leq 16$ simulation is initialized using the $0 \leq n \leq 8$ results obtained after HC formation but before the onset of the secondary mode.  The poloidal cross-section of $\delta P_b$ composed from $5\leq n \leq 16$ at the $\phi=0^\circ$ toroidal angle is shown in Fig.\ref{fig:4eta6vrad2nd_realspace_n16}.  The (magenta) $q = 9/8$, (violet) $17/16$, and (black) $q_\mathrm{min}$ magnetic flux surfaces are overlaid. We find that the fluctuation becomes more localized, and the peak of $\delta P_b$ shifts closer to the $q_\mathrm{min}$ surface than in the $0\leq n\leq8$ simulation.  It is evident that our present simulation yields only an approximate form of the secondary mode.  To study the exact form, shorter wavelength components ($n>16$) should be considered; however, modeling such a short-wavelength structure using the MHD model may not be physically accurate.

In terms of mode parity, the secondary mode structure presented in real space shows that the secondary mode appears to largely have an interchange parity; however, a small inversion of $\delta P_b$ can be locally observed in some positions.  For instance, in the expanded section of Fig.\ref{fig:4eta6vrad2nd_realspace}(a).  The strongly sheared poloidal structure is also evident in the top and bottom parts of this panel, indicating a mixture of interchange and tearing parity.  These features may be closely connected to the magnetic chaos that appears in the presence of the secondary mode, but the details remain to be worked out.

Lastly, the quantitative prediction of the magnetic chaos observed in the present simulation may not be accurate due to the limitations of the simulation model and the numerical approach, particularly our imperfect toroidal Fourier filter.  A detailed discussion of this issue is provided in \ref{sec:apen2}.

\begin{figure}[h]
\begin{center}
\includegraphics[width=1.00\linewidth]{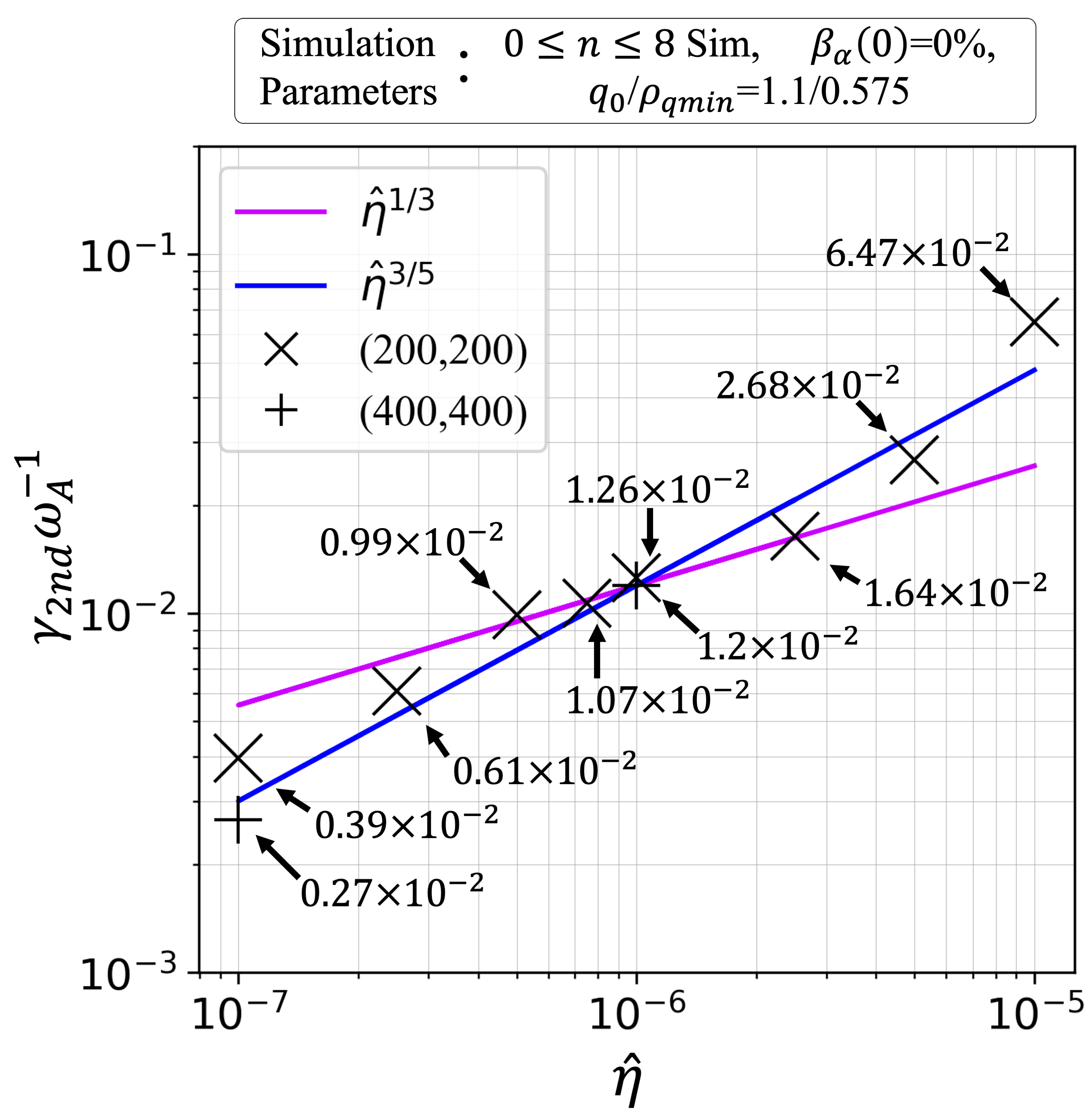}
\end{center}  
\caption{Dependence of the growth rate of the secondary mode $\gamma_{2nd}$ on $\hat{\eta}$ for the $q_\mathrm{0}/\rho_\mathrm{qmin}=1.1/0.575$ equilibrium with $\beta_\mathrm{\alpha}=0\%$.  ``$\times$" and ``+" symbols represent the results obtained with ($N_R$,$N_Z$)=($200$,$200$) and ($400$,$400$), respectively.  For comparison and orientation, the scalings $\hat{\eta}^{1/3}$ (resistive kink and ballooning modes) and $\hat{\eta}^{3/5}$ (tearing mode) are plotted as a violet and blue solid lines, respectively.}
\label{fig:4etascan2nd}
\end{figure}

\begin{figure}[h]
\begin{center}
\includegraphics[width=1.0\linewidth]{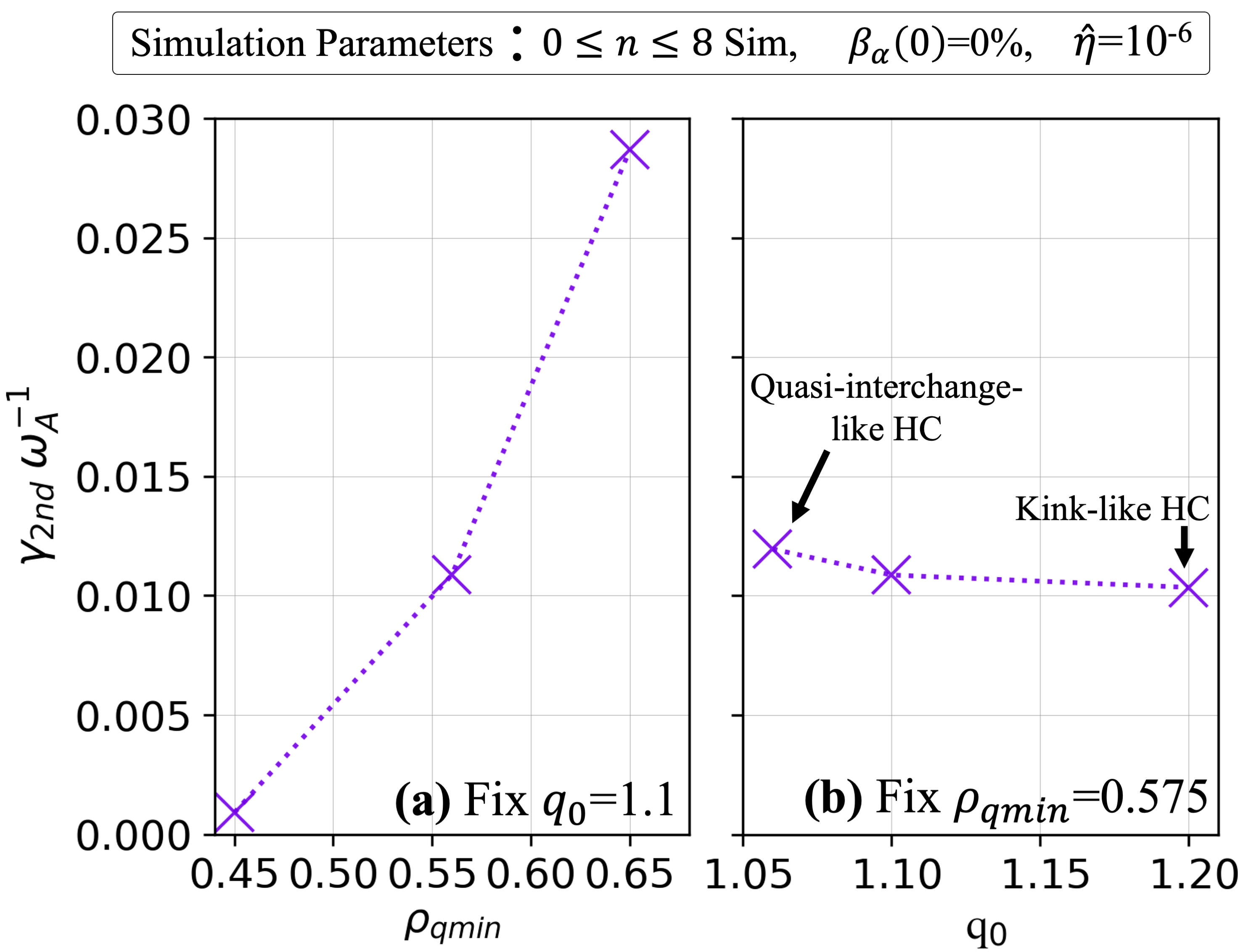}
\end{center} 
\caption{Dependence of the growth rates of the secondary mode on (a) $\rho_\mathrm{qmin}$ and (b) $q_\mathrm{0}$, calculated with $(N_R,N_Z)$=$(200,200)$, $\hat{\eta}=10^{-6}$, and $\beta_\mathrm{\alpha} = 0\%$.}
\label{fig:4megaqwscan_nl}
\end{figure}

\subsubsection{\label{sec:level4c2}Dependence of Secondary Mode Stabilities on Plasma Resistivity $\eta$}
\quad

Here, we investigate the dependence of the secondary mode on the plasma resistivity $\eta$.  To quantify the stability of the secondary mode, we measure its exponential growth rate $\gamma_{2nd}$.  Since the $5\leq n \leq8$ Fourier components seem to be a single eigenmode, $\gamma_{2nd}$ will be measured only from the $n=8$ component, a dominant Fourier component.  Unlike a linear growth rate, a growth rate measured during the nonlinear phase is influenced by mode-mode couplings and profile modifications.  In other words, $\gamma_{2nd}$ is effectively measured from different and evolving equilibria. From this reason, the stability assessment based on $\gamma_{2nd}$ should not be considered quantitatively accurate, unlike similar analyses performed during the linear growth phase. 

We perform the $\hat{\eta}$ scan within the $10^{-7} \leq \hat{\eta} \leq 10^{-5}$ range for the $q_\mathrm{0}/\rho_\mathrm{qmin}=1.1/0.575$ case.  The simulation results calculated from our normal simulation parameters are shown in Fig.\ref{fig:4etascan2nd} using the ``$\times$"  marker.  We find that $\gamma_\mathrm{2nd}$ increases with $\hat{\eta}$, suggesting that the secondary mode is a kind of resistive mode.  In the low resistivity range, the numerical dissipation caused by the discrete grid may become comparable to the effect of resistivity.  To ensure that this numerical dissipation is sufficiently low, the convergence test is carried out using a higher poloidal resolution ($N_R$,$N_Z$)=($400$,$400$), and its results are plotted using ``+"  markers.  At $\hat{\eta} = 10^{-6}$, the results from both $(200, 200)$ and $(400, 400)$ poloidal resolutions agree, confirming the convergence.  However, at $\hat{\eta} = 10^{-7}$, $\gamma_{2nd}$ calculated using $(N_R, N_Z) = (200, 200)$ is approximately $44\%$ larger than the $(N_R, N_Z) = (400, 400)$.

\begin{figure}[t]
\begin{center}
\includegraphics[width=1.0\linewidth]{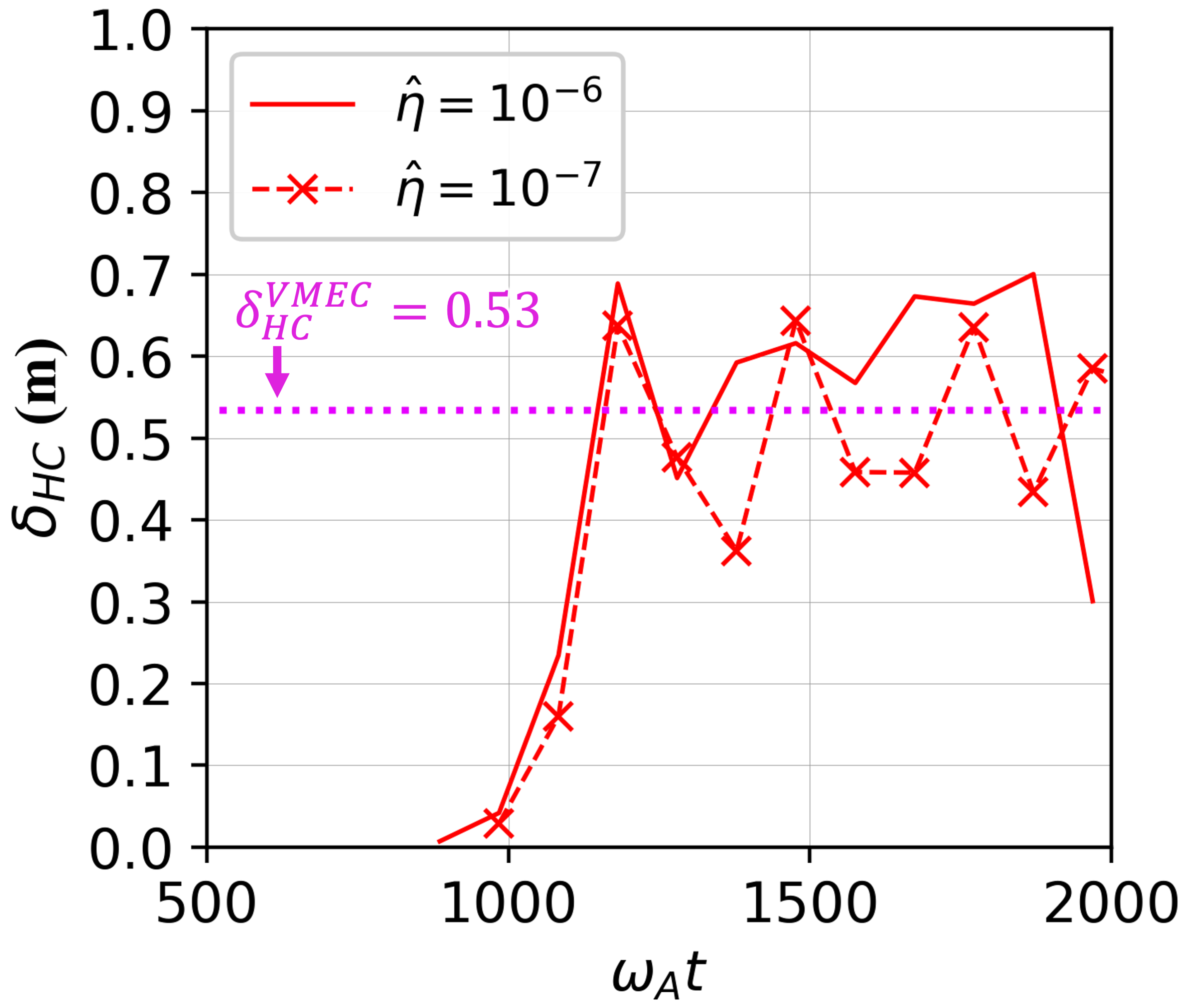}
\end{center} 
\caption{Time evolution of $\delta_\mathrm{HC}^\mathrm{MEGA}$ calculated with $\hat{\eta}=10^{-6}$ and $10^{-7}$.  The results for $\hat{\eta}=10^{-6}$ and $10^{-7}$ cases are represented with a red solid line and red dashed line with ``$\times$" markers, respectively. }
\label{fig:5deltah_eta}
\end{figure}

Lastly, we compared $\gamma_\mathrm{2nd}$ with the scaling laws derived from the linear theories of resistive ballooning mode and tearing mode.  The linear growth rate of the resistive kink and ballooning mode scales with $\gamma \propto \hat{\eta}^{1/3}$\cite{garcia1999spatiotemporal}, while the tearing mode scales with $\gamma \propto \hat{\eta}^{3/5}$\cite{furth1963finite}.  These scalings are overlaid on Fig.\ref{fig:4etascan2nd} as the violet and blue solid lines, respectively.  Overall, we observe some resemblance of $\gamma_{2nd}$ with that of the tearing mode scaling with some deviation around $\hat{\eta} \sim10^{-6}$.  One should note that the overall proximity to $\hat{\eta}^{3/5}$ is a mere curiosity.  There seems to be no reason to expect adherence to any particular such scaling here because these theoretical scalings are derived based on linearized equations with respect to the fixed equilibrium, and they did not account for the toroidal mode coupling with HC.  In addition, $\gamma_{2nd}$ is calculated from an evolving HC where there are changes in the plasma profiles and magnetic field. This issue is briefly shown in Fig.\ref{fig:5deltah_eta} where we compared the time evolution of $\delta_\mathrm{HC}^\mathrm{MEGA}$ calculated with $\hat{\eta}=10^{-6}$ and $10^{-7}$.  $\delta_\mathrm{HC}^\mathrm{MEGA}$ calculated with $\hat{\eta}=10^{-6}$ is higher than that of the $\hat{\eta}=10^{-7}$.  These results confirm that the secondary mode is sufficiently strong to modify the equilibrium profile at the instance when $\gamma_\mathrm{2nd}$ is measured, which complicates the interpretation of Fig.\ref{fig:4etascan2nd} and similar analyses.  However, qualitatively, the resistive dependence discussed here is a robust result; therefore, we can consider the secondary mode as a kind of resistive mode.  (There is also an ideal branch that seems to be suppressed by the toroidal low-pass filter.  The existence of the ideal secondary mode can be seen in Fig.\ref{fig:APPENetascan} shown in \ref{sec:apen1}.)

\subsubsection{\label{sec:level4c3}Dependence of Secondary Mode Stability on $q$}
\quad

The results from Sections \ref{sec:level4a} and \ref{sec:level4b} show that the secondary mode is suppressed in the $q_\mathrm{0}/\rho_\mathrm{qmin}=1.1/0.45$ equilibrium.  The dependencies of $\gamma_{2nd}$ on $\rho_\mathrm{qmin}$ and $q_\mathrm{0}$ are shown in Figs.\ref{fig:4megaqwscan_nl}(a) and (b), respectively, and can be summarized as follows: 

\begin{enumerate} 
\item $\rho_\mathrm{qmin}$ Scan: The growth rate of the secondary mode becomes higher as the width of the low magnetic shear $q\gtrsim1$ region increases (higher $\rho_\mathrm{qmin}$). In contrast, shifting $\rho_\mathrm{qmin}$ inward leads to complete suppression of the secondary mode, as briefly discussed in Section \ref{sec:level4b}.  From the results shown in Fig.\ref{fig:3mhdeq}(h) and Fig.\ref{fig:4megaqwscan}(d), $\delta_\mathrm{HC}$ increases with increasing radial width of the $q\gtrsim1$ region.  The wider $q\gtrsim1$ region means that more flux surfaces will be compressed, while a higher $\delta_\mathrm{HC}$ leads to a stronger compression. These two effects collectively lead to a stronger steepening of the bulk plasma pressure gradient within the HC-compressed flux region. It should be noted that this dependence may be biased by our use of a fixed initial $P_b$ profile because changing $\rho_{qmin}$ position also changes the bulk plasma pressure gradient in the region that will be compressed by the HC.

\item $q_{0}$ Scan: $\gamma_{2nd}$ exhibits only a weak dependence on $q_\mathrm{0}$. $\gamma_{2nd}$ of the secondary mode increases by a mere $20\%$ relative to the $q_\mathrm{0} = 1.06$ equilibrium with that of the $q_\mathrm{0} = 1.2$ equilibrium.  This modest increase stands in stark contrast to the larger change in $\delta_\mathrm{HC}$.  In the $q_\mathrm{0} = 1.06$ equilibrium, $\delta_\mathrm{HC}$ is approximately $1.5$ to $2$ times larger than that of the $q_\mathrm{0} = 1.2$ equilibrium.  While a larger $\delta_\mathrm{HC}$ may imply stronger pressure gradient steepening, this only holds if the HC eigenfunction remains similar. The linear eigenfunction of the $m/n=1/1$ mode shown in Figs.\ref{fig:4megaq0scan}(a,k) show that the $q_\mathrm{0} = 1.06$ equilibrium has a quasi-interchange-like HC, whereas the $q_\mathrm{0} = 1.2$ equilibrium has a kink-like HC. Consequently, the larger $\delta_\mathrm{HC}$ in the $q_\mathrm{0} = 1.06$ (quasi-interchange-like HC) case causes pressure gradient steepening primarily in the core region, while having a weaker influence on the pressure gradient near the location of the secondary mode (around $q_\mathrm{min}$).  This is one likely reason for the lack of correlation between $\delta_\mathrm{HC}$ and $\gamma_\mathrm{2nd}$.
\end{enumerate}

\begin{figure}[h]
\begin{center}
\includegraphics[width=1.0\linewidth]{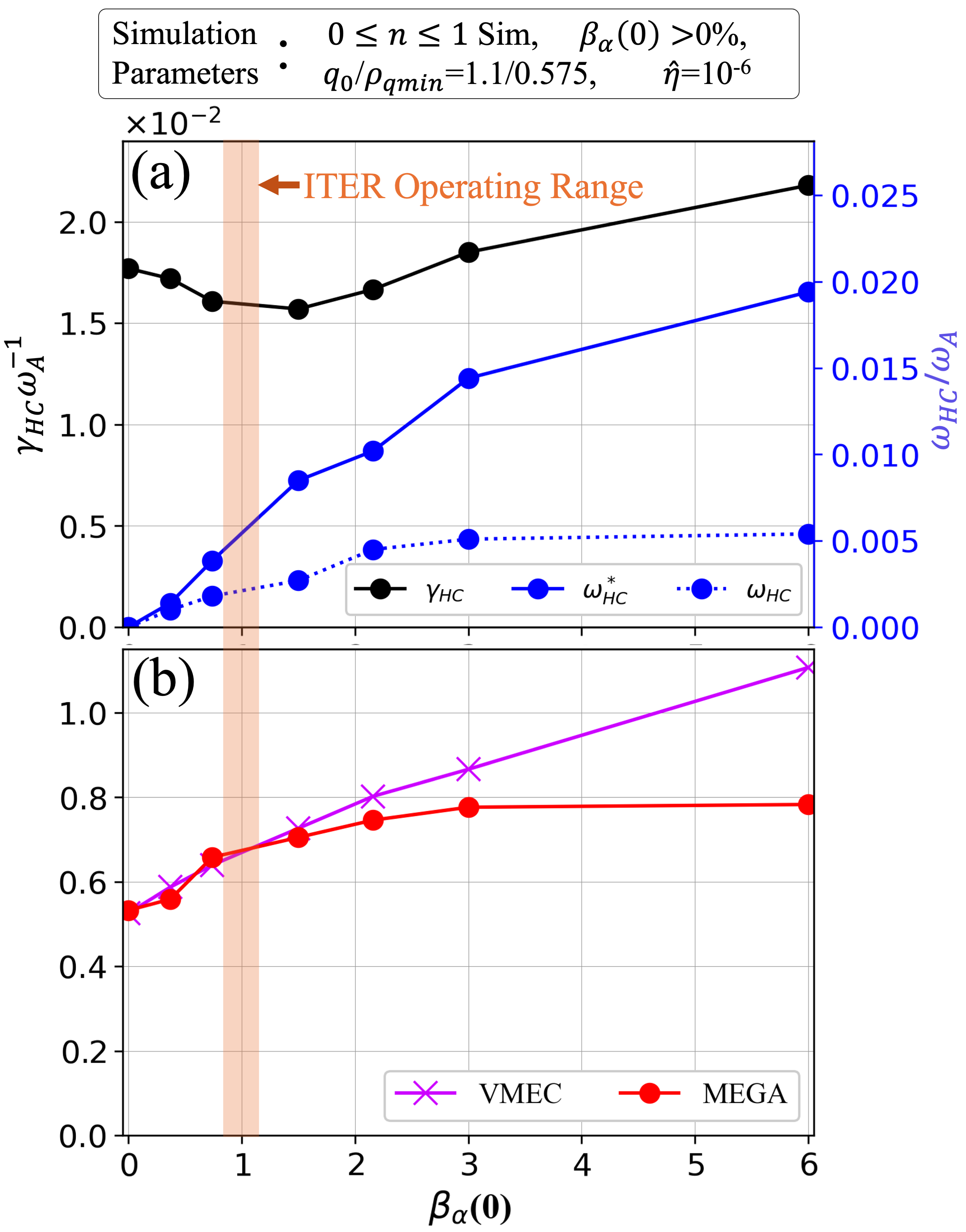}
\end{center} 
\caption{Dependence of the linear growth rate (a) $\gamma_\mathrm{HC}$, real frequency $\omega_\mathrm{HC}$, and (b) radial displacement of the magnetic axis $\delta_\mathrm{HC}$ of the $m/n=1/1$ mode on $\beta_\mathrm{\alpha}$ in the $q_\mathrm{0}/\rho_\mathrm{qmin}=1.1/0.575$ equilibrium.}
\label{fig:5gammafreqhdis_beta}
\end{figure}

\section{\label{sec:level5}Helical Core Formation in the Presence of Alpha Particle for the $q_\mathrm{0}/\rho_\mathrm{qmin} = 1.1/0.575$ Equilibrium}
\quad
This section focuses on the interactions between fusion-born alpha particles and the HC formation in ITER-relevant plasma parameters (and somewhat beyond) for the $q_\mathrm{0}/\rho_\mathrm{qmin} = 1.1/0.575$ equilibrium.  This equilibrium is selected because the HC equilibrium state calculated with VMEC and the HC quasi-steady state calculated with MEGA are best in terms of quantitative agreement, as previously discussed in Section \ref{sec:level4a3}. This section is divided into four subsections.  Section \ref{sec:level5a} discusses the dependence of the linear growth rate, linear real frequency, and quasi-steady state $\delta_{HC}$ on $\beta_\alpha$ from the macroscopic perspective. Section \ref{sec:level5b} provides additional physical detail at the level of individual alpha test particle orbits in the HC magnetic field.  Section \ref{sec:level5c} focuses on the nonlinear dynamics of the HC, in the presence of alpha particles, and the alpha particle confinement after HC formation.  Finally, Section \ref{sec:level5d} deals with the effect of secondary instabilities on alpha particles.

\subsection{\label{sec:level5a}Dependence of $\gamma_\mathrm{HC}$, $\omega_\mathrm{HC}$, and $\delta_\mathrm{HC}$ on $\beta_\mathrm{\alpha}$}
\quad

The $\beta_\mathrm{\alpha}$-dependence of the $m/n=1/1$ mode's linear growth rate and real frequency in the $q_\mathrm{0}/\rho_\mathrm{qmin}=1.1/0.575$ equilibrium is shown in Fig.\ref{fig:5gammafreqhdis_beta}(a).  In the $0\% \le \beta_\mathrm{\alpha}(0) \le 1.0\%$ range, the linear growth rate $\gamma_\mathrm{HC}$ is reduced by roughly $10\%$. We suspect that this is a manifestation of the known stabilizing influence of trapped alpha particles on kink/quasi-interchange MHD modes\cite{porcelli1991fast,fu2006global}. When $\beta_\mathrm{\alpha}$ is increased beyond $\beta_\mathrm{\alpha}(0) \geq 1.5\%$, the growth rate $\gamma_{HC}$ is found to increase, which suggests that this is a different branch of the mode, similar to the EP-driven fishbone \cite{mcguire1983study,chen1984excitation,coppi1986theoretical}. Interestingly, the linear real frequency $\omega_\mathrm{HC}^*$ gradually increases with $\beta_\mathrm{\alpha}$.  The lack of a clear frequency jump near the minimum of $\gamma_\mathrm{HC}$ around $\beta_\mathrm{\alpha}(0)\sim1.5\%$ suggests that the MHD branch smoothly transitions into the EP branch. This smooth transition is likely related to the fact that the initial plasma state in our setup is situated far above the HC stability threshold.  It then remains to explain why the frequency $\omega_\mathrm{HC}^*$ continues to increase as we raise $\beta_\mathrm{\alpha}$ beyond $1.5\%$, in spite of the fact that our $\beta_\alpha$-scan is effectively a density scan (because we fix the form of the alpha velocity distribution). For a fishbone or, more generally, EPM-type mode, one may expect the real frequency to depend primarily on the orbit frequency of the resonant particles rather than their density. A possible explanation for the continued increase of $\omega_\mathrm{HC}^*$ in Fig.\ref{fig:5gammafreqhdis_beta}(a) is the increasing diamagnetic drift frequency, which is captured here by the alpha particle magnetization current; namely, the last term in Eq.\ref{eq:jalpha}. Performing the simulation without this term, yields the frequency $\omega_\mathrm{HC}$ that is plotted in Fig.\ref{fig:5gammafreqhdis_beta}(a) as a blue dotted line with filled circles. One can see $\omega_\mathrm{HC}$ is nearly constant for $\beta_\mathrm{\alpha} \gtrsim 2\%$, which confirms our above expectations.

The $\beta_\mathrm{\alpha}$-dependence of the HC's magnetic axis displacement $\delta_\mathrm{HC}$ is shown in Fig.\ref{fig:5gammafreqhdis_beta}(b). One can see that $\delta_\mathrm{HC}$ increases monotonically with $\beta_\mathrm{\alpha}$, even in the range $\beta_\mathrm{\alpha} \lesssim 1.5\%$ where the alpha particles have a stabilizing influence on the mode's linear growth $\gamma_\mathrm{HC}$.

Such a lack of correlation between $\gamma_\mathrm{HC}$ and $\delta_\mathrm{HC}$ is not unexpected in this study. A correlation between the linear growth rate and the saturation level is expected only when a mode saturates via the depletion of the initially destabilizing local gradient (where $\gamma_\mathrm{HC}$ was measured), while the background equilibrium is largely preserved. In contrast, a HC is not a mere ``mode'' in the sense of a coherent fluctuation pattern that exists within the initial equilibrium. Instead, we view a HC as constituting both the process and the result of a global reorganization of the plasma into a new non-axisymmetric equilibrium. The rate and path of such a transition to another minimum energy state can, in principle, be fairly independent of the initial instability that triggered the transition. It should also be noted that $\delta_\mathrm{HC}$ is merely a local and, thus, incomplete measure of the global HC displacement, which can be kink- or quasi-interchange-like, with possible further modifications due to fast ions. Recall also the related discussion in Section~\ref{sec:level4a3} for the MHD limit. In the case of Fig.\ref{fig:5gammafreqhdis_beta}, the new force balance in the quasi-steady HC is also influenced by alpha particles.

To assess the influence of alpha particles, in particular the role of kinetic effects, it is instructive to compare the radial displacement of the HC magnetic axis $\delta_\mathrm{HC}$ calculated by MEGA with that of VMEC.  Since VMEC uses an ideal MHD model, the effects of alpha particles are incorporated only as an additional scalar pressure. This scalar pressure is flux-surface-averaged, which is expressed as
\begin{eqnarray} \label{eq:dhvmec}
\beta_{MHD}(\rho_{eq0})=\beta_{b0}(\rho_{eq0})+\frac{1}{2\pi}\int_0^{2\pi}\beta_{\alpha0}(\rho_{eq0},\theta)d\theta.
\end{eqnarray}
\noindent
The radial displacement of the HC magnetic axis obtained by this method is denoted as $\delta_\mathrm{HC}^\mathrm{VMEC,\alpha}$.  This VMEC result represents the HC solution in the limit where the kinetic effects of alpha particles, their finite orbit width (along with the associated pressure anisotropy and mean flows), and alpha particle transport are neglected.  In Fig.\ref{fig:5gammafreqhdis_beta}(b), $\delta_\mathrm{HC}^\mathrm{VMEC,\alpha}$ is shown as a magenta solid line with crosses.  In the $0\%<\beta_\mathrm{\alpha}(0)<1.5\%$ range, $\delta_\mathrm{HC}^\mathrm{MEGA}$ and $\delta_\mathrm{HC}^\mathrm{VMEC,\alpha}$ quantitatively agree.  This agreement between MEGA (nonlinear resistive MHD equations coupled with EP drift kinetic equation) and VMEC (ideal MHD equilibrium) implies that the non-ideal MHD and EP kinetic effects are weak or reversible during the quasi-steady state.  The situation changes at higher values of the alpha pressure, here $\beta_\mathrm{\alpha}(0)\geq1.5\%$, where the rise of $\delta_\mathrm{HC}^\mathrm{MEGA}$ with $\beta_\mathrm{\alpha}$ becomes much weaker than that of $\delta_\mathrm{HC}^\mathrm{VMEC,\alpha}$.  At $\beta_\mathrm{\alpha}(0)\approx3\%$, $\delta_\mathrm{HC}^\mathrm{MEGA}$ reaches an upper bound of about $0.8$ m, while $\delta_\mathrm{HC}^\mathrm{VMEC,\alpha}$ continues to increase.  These differences indicate that non-ideal, kinetic alpha, or radial transport effects become increasingly significant and irreversible.

To obtain a better understanding of the dependence of $\delta_\mathrm{HC}$ on $\beta_\mathrm{\alpha}$, the mechanisms via which alpha particles and the HC interact with each other must be examined.  In the following Section \ref{sec:level5b}, we illustrate how a HC with intact magnetic surfaces modifies the guiding center drift orbits of confined alpha test particles. The insights won will be useful in the subsequent discussion of alpha particle confinement and HC dynamics in Section \ref{sec:level5c}.

\begin{figure*}[t]
\begin{center}
\includegraphics[width=0.95\linewidth]{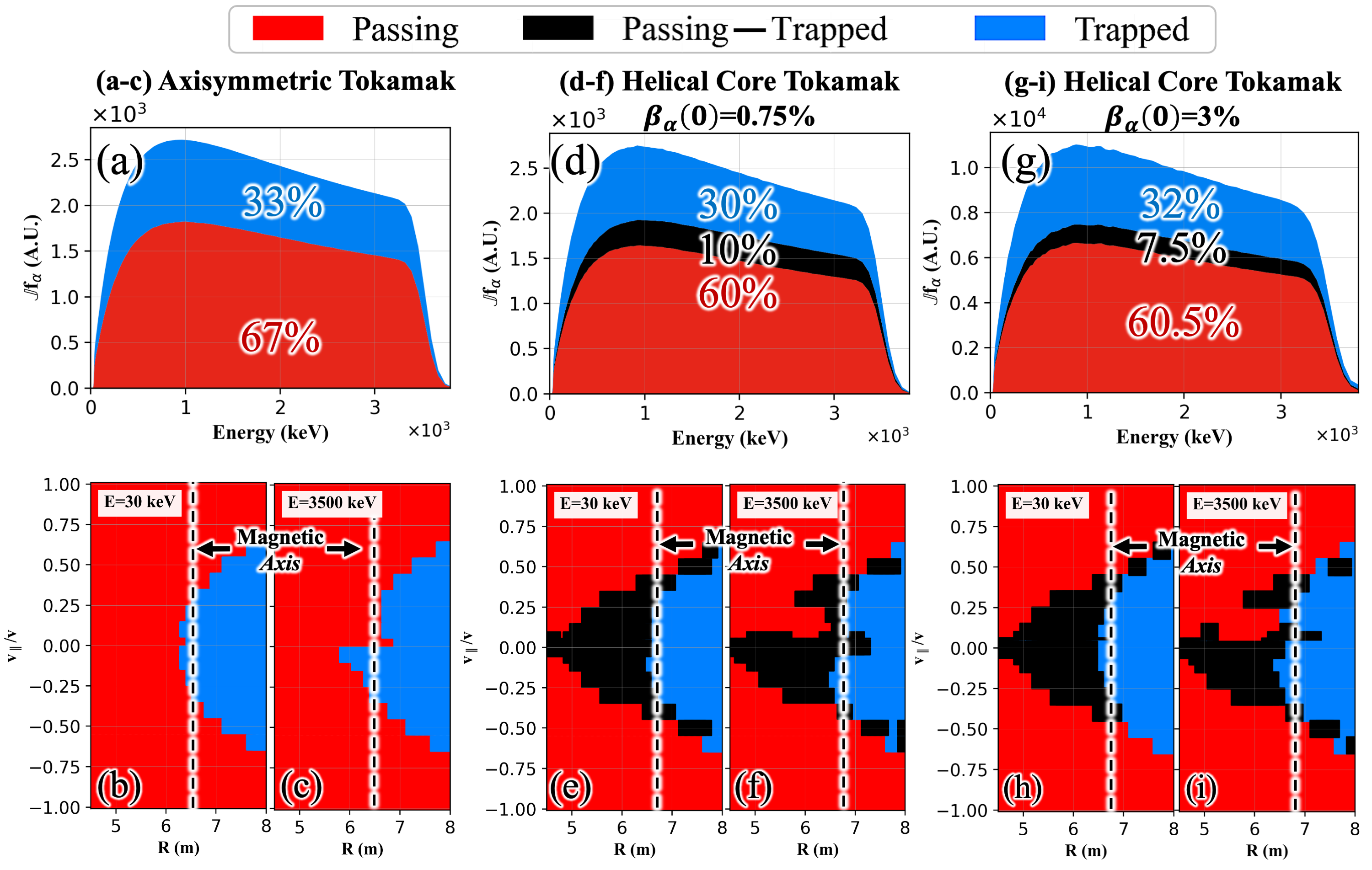}
\end{center} 
\caption{Orbit classification of the alpha particles in the magnetic fields of an (a-c) axisymmetric equilibrium, (d-f) moderate $\delta_\mathrm{HC}$, (g-i) large $\delta_\mathrm{HC}$ configuration.  The red, blue, and black colors represent the passing, trapped, and passing-trapped transitional orbits, respectively.  The axisymmetric and moderate $\delta_\mathrm{HC}$ equilibria represented by the $\beta_\mathrm{\alpha}(0)=0.75\%$ case at the initial and nonlinear phases, respectively.  In contrast, the large $\delta_\mathrm{HC}$ equilibrium refers to the nonlinear phase of the $\beta_\mathrm{\alpha}(0)=3\%$ case.
The upper row shows the histogram of alpha particle energy $h_\mathrm{\alpha}(E)=\mathbb{J}(E)f(E)$, where $\mathbb{J}(E)$ and $f(E)$ are the Jacobian of transformation and distribution function, respectively.  The percentages shown in panels (a, d, g) represent the population percentages of each orbit type.
The lower row shows the localization for each orbit type in the reduced orbit coordinate space ($R,v_\parallel/v,E$), where each point represents the initial coordinates of an alpha test particle when it is launched from the magnetic midplane at $\phi = 0$. In the axisymmetric case (a-c), these mid-plane-based orbit coordinates ($R,v_\parallel/v,E$) are constants of motion equivalent to $(P_\phi,\mu,E)$ \cite{bierwage2025construction}.}
\label{fig:5testorbit_fedis}
\end{figure*}

\begin{figure}[t]
\begin{center}
\includegraphics[width=0.95\linewidth]{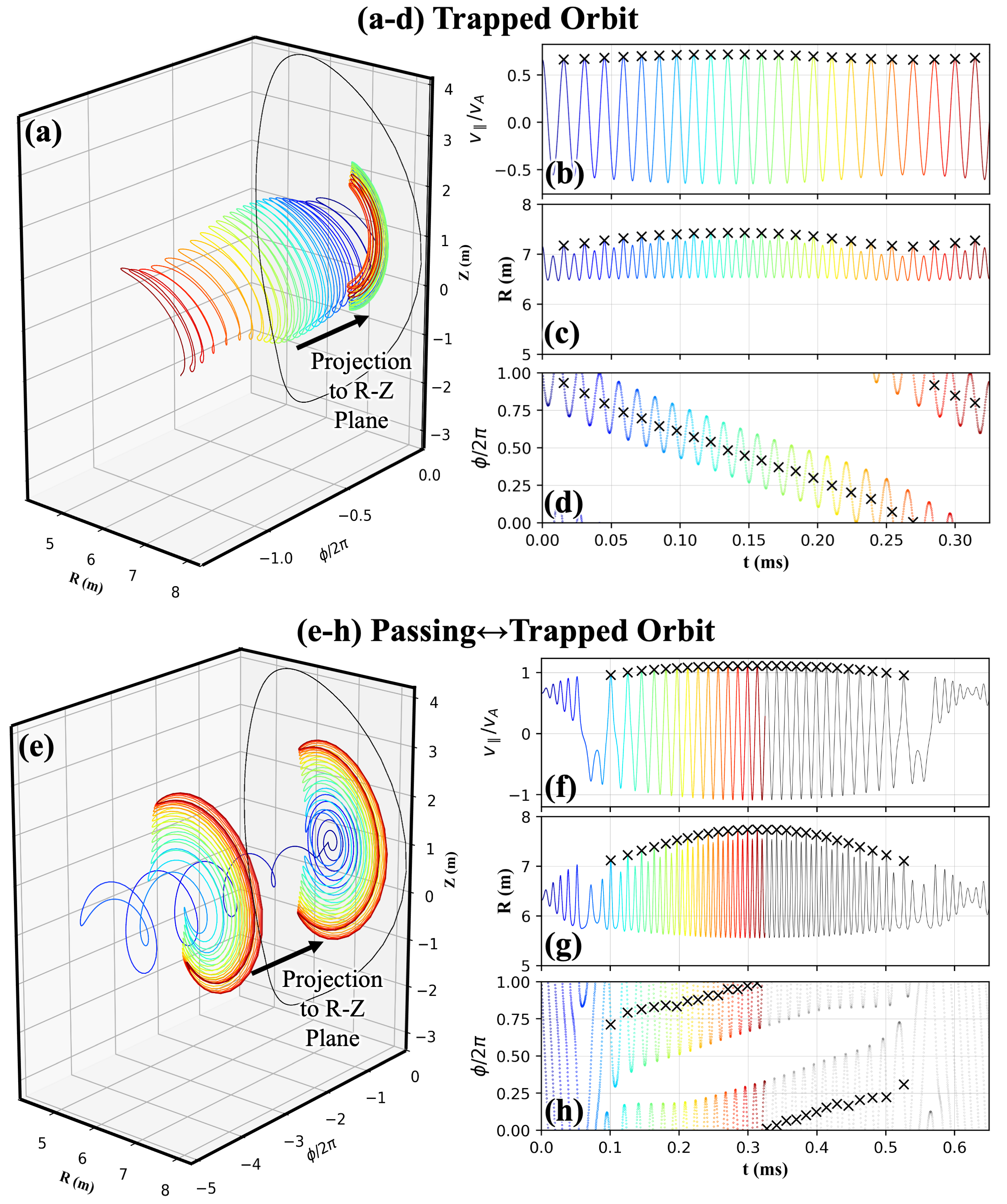}
\end{center} 
\caption{Representative trajectories of test particles residing on (a-d) trapped orbits and (e-h) ``passing-trapped transitional'' orbits in the HC equilibrium with $\delta_\mathrm{HC}=0.64$ m ($\beta_\alpha = 0.75\%$).  These test particles have an energy of $E_\alpha=3500$ keV.  Panels (a,e) on the left show the orbits of these particles in 3-D. The panels on the right-hand side show the time traces of (b,f) the parallel velocity $v_\parallel(t)/v_{\rm A}$, (c,g) the major radial position $R(t)$, and (d,h) the toroidal angular position $\phi(t)$.  The transition from blue to red colors according to the rainbow spectrum represents the forward direction in time. The black ``$\times$" symbols mark the times and places at which the banana orbit reaches its major radial peak $R_{\rm b,max}$ during each bounce period.}
\label{fig:5testtraptran}
\end{figure}

\subsection{\label{sec:level5b}Effects of HC Toroidal Asymmetric Magnetic Field on Alpha Particle Orbits}
\quad

In axisymmetric tokamak equilibria, one distinguishes between passing and toroidally mirror-trapped orbits.\footnote{Passing orbit populations can be further divided into sub-classes known as circulating and stagnation orbits, and trapped orbits can be divided into potato and banana orbits, depending on whether or not the orbit's poloidal contour encloses the magnetic axis. Here, we will not make such a distinction.}  Toroidal asymmetries in the magnetic field give rise to new classes of particle orbits, such as the helically ripple-trapped orbit observed in stellarators and heliotrons.  In this section, we demonstrate that as much as $10\%$ of the alpha particle population may populate non-standard orbits in an ITER plasma with HC. The properties of these orbits and their possible influence on the HC dynamics are discussed.

\subsubsection{\label{sec:level5b1}Alpha Test Particle Orbits in Axisymmetric Equilibrium}
\quad

We begin our analysis with the axisymmetric equilibrium state, using the particle and magnetic field data from the initial time step of the MEGA simulation for the $\beta_\mathrm{\alpha} = 0.75\%$ case of Fig.\ref{fig:5gammafreqhdis_beta}.  The results are shown in Fig.\ref{fig:5testorbit_fedis}(a–c). Panel (a) presents the histogram of alpha particle energy, $h(E)=\mathbb{J}(E)f(E)$, where $\mathbb{J}(E)$ and $f(E)$ are the Jacobian of transformation and distribution function, respectively.  Panels (b) and (c) show how these two orbit classes are distributed in the reduced space of mid-plane-based orbit coordinates ($R,v_\parallel/v,E$). Each point represents the initial kinetic energy $E$, pitch $v_\parallel/v$, and major radial position $R$ of an alpha test particle that has been launched from the magnetic mid-plane at $\phi=0^\circ$. The magnetic midplane is defined by the condition ${\vec B}\cdot{\nabla}B = 0$. We note that, in the axisymmetric case of Fig.\ref{fig:5testorbit_fedis}(b,c), ($R,v_\parallel/v,E$) are constants of motion equivalent to $(P_\phi,\mu,E)$ \cite{bierwage2025construction}, where $P_\phi$ is the canonical toroidal angular momentum and $\mu$ the magnetic moment of an alpha particle's guiding center. (For details, see Sections 2.2 and 3.4 of Ref.\cite{bierwage2025construction}).
The alpha particle orbit space slices in panels (b) and (c) were taken at kinetic energies $E = 30$ keV and $3500$ keV, respectively. They contain all the familiar features; most notably, the localization of trapped orbits on the low-field side and in regions with small pitch $v_\parallel / v$. At the higher energy in panel (c), the increased asymmetry relative to the horizontal ($v_\parallel = 0$) axis is due to larger magnetic drifts and larger numbers of stagnation and potato orbits within the passing and trapped populations.

\begin{figure}[t]
\begin{center}
\includegraphics[width=1.0\linewidth]{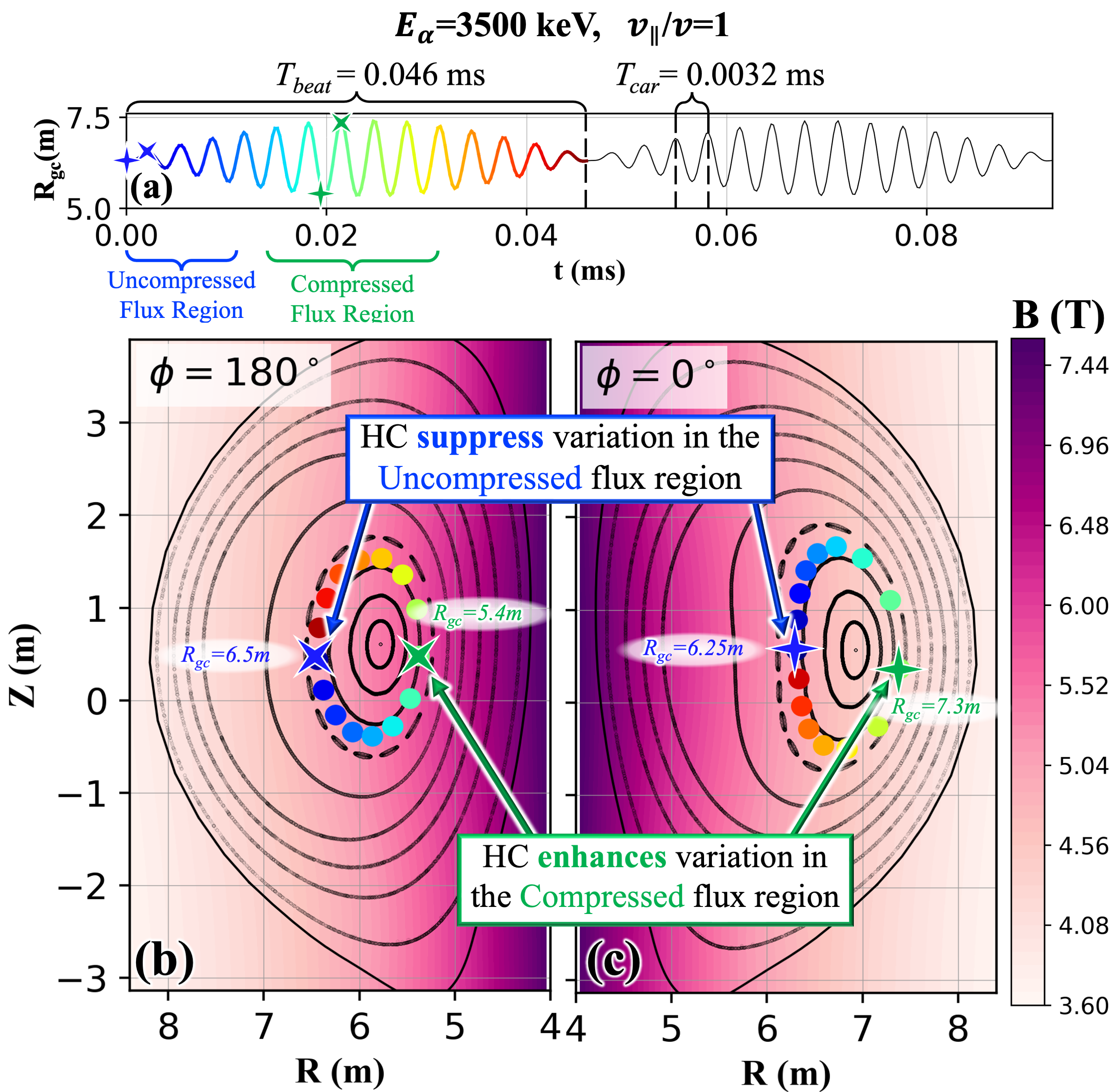}
\end{center} 
\caption{Passing alpha test particle orbit with $E_\alpha=3500$ keV and initial $v_\parallel/v=1.0$ in the HC equilibrium with $\delta_\mathrm{HC}=0.64$ m.  Panel (a) shows the time evolution of the major radial position of the test particle, which can also be translated into the magnetic field strength perceived by the particle according to $B\propto\frac{1}{R}$.  Panels (b) and (c) show the Poincar{\'e} plot of the test particle overlaid with the HC magnetic flux surface at the $\phi=180^\circ$ and $0^\circ$ toroidal angles, respectively.  The transition from blue to red colors according to the rainbow spectrum represents the forward direction in time.}
\label{fig:5hcripple}
\end{figure}

\subsubsection{\label{sec:level5b2}Alpha Test Particle Orbits in HC Equilibrium}
\quad

Let us now inspect the situation in the quasi-steady HC state obtained with MEGA for $\beta_\mathrm{\alpha}=0.75\%$ ($\delta_{HC}=0.64$ m). 
The histogram of alpha particle energy of this case is shown in Fig.\ref{fig:5testorbit_fedis}(d), and the pitch-radius distribution of orbit classes at $E = 30$ keV and $3500$ keV can be seen in panels (e) and (f), respectively. We note that due to the lack of toroidal symmetry, the mid-plane-based orbit coordinates ($R,v_\parallel/v,E$) no longer constitute a set of constants of motion. Only the kinetic energy $E$ is still conserved in panels (e) and (f), because these plots were generated for a time-independent snapshot of the magnetic field. Meanwhile, the data in panel (d) were taken from the actual MEGA simulation, where all fields are time-dependent, so that the kinetic energy $E$ of an alpha particle is also a dynamic variable.

Fig.\ref{fig:5testorbit_fedis}(d) shows that, during the quasi-steady state of the HC, the passing and trapped particles account for $60\%$ and $30\%$, respectively.  The remaining $10\%$ corresponds to a new orbit type that periodically transitions between passing and trapped states. In Fig.\ref{fig:5testorbit_fedis}, the areas populated by such transitional orbits are drawn in black and labeled ``Passing-Trapped Transitional''.

Concrete examples of trapped and transitional orbits in the HC equilibrium are shown in Figs.\ref{fig:5testtraptran}(a-d) and (e-h), respectively (Passing particles are not shown, as they simply follow magnetic flux surfaces with only small deviations.)  Panels (a,e) show the respective 3-D guiding center trajectories in cylinder coordinates: $(R(t),\phi(t),Z(t))$. Panels (b,f), (c,g), and (d,h) show the time traces of the guiding center's parallel velocity $v_\parallel(t)$, major radial position $R(t)$, and toroidal angle $\phi(t)$, respectively.  The transition from blue to red, according to the rainbow spectrum, represents the forward direction in time. For the trapped orbit, one can see that the banana orbit in Fig.\ref{fig:5testtraptran}(a-d) is well-confined and preserves its identity while the spatial form of its orbit contour varies under the influence of the HC's toroidal asymmetry.  The relative complexity of the transitional orbit can be observed in panels (e–h). Initially, this test particle follows the trajectory of a passing orbit for several toroidal transits ($t\leq 0.1$ ms, shown in blue). At $t=0.1$ ms, it transitions into a trapped orbit, accompanied by a large radial excursion. The trapped orbit phase can be discerned by the black ``$\times$" symbols in Figs.\ref{fig:5testtraptran}(f-h), which mark the times at which the banana orbit reaches its maximal major radial position $R_{\rm b,max}$ during each bounce period. As the particle enters deeper into the mirror-trapped domain, the variation in $R(t)$ increases continuously as shown in Fig.\ref{fig:5testtraptran}(g). Around $t=0.3$ ms, the magnitude of oscillation in $R(t)$ reaches a maximum and subsequently decreases as the test particle approaches the trapped-passing boundary and eventually transits back to a passing orbit at $t=0.55$ ms. Recalling that the orientation of the HC has been set up such that the magnetic axis is shifted maximally outward in $R$ at $\phi=0$, one can infer from the time trace of the toroidal angle $\phi(t)$ and from the clustering of the black ``$\times$" symbols for $R_{\rm b,max}$ around $\phi \approx 0$ or $2\pi$ in Fig.\ref{fig:5testtraptran}(h) that the trapped phase coincides with the time interval when the test particle is located in the compressed flux region. (Orbits similar to our passing-trapped transitional orbits were also mentioned in analyses of LHD plasmas \cite{spong20153d}, but there may be differences in the underlying mechanisms that require further investigation.)

Returning to Figs.\ref{fig:5testorbit_fedis}(e-f), where the approximate domains populated by transition orbits are drawn black, one can see that they occupy a significant portion of what used to be the passing and trapped domains surrounding the trapped-passing boundary in the axisymmetric case of Figs.\ref{fig:5testorbit_fedis}(b-c). Figs.\ref{fig:5testorbit_fedis}(e-f) also show an increasing asymmetry with respect to the horizontal ($v_\parallel = 0$) line when the particle energy increases. In part, this asymmetry may be due to inaccuracies in determining the magnetic midplane in the HC state or in identifying transition orbits. However, physical reasons are also conceivable.  It should be noted that different orbit-type distributions would be obtained with test particles launched from different toroidal angles.

To understand the origin of transitional orbits, we discuss in the following subsection the structure of the asymmetric magnetic field and its effect on the motion of charged particles.

\begin{figure*}[h]
\begin{center}
\includegraphics[width=1.00\linewidth]{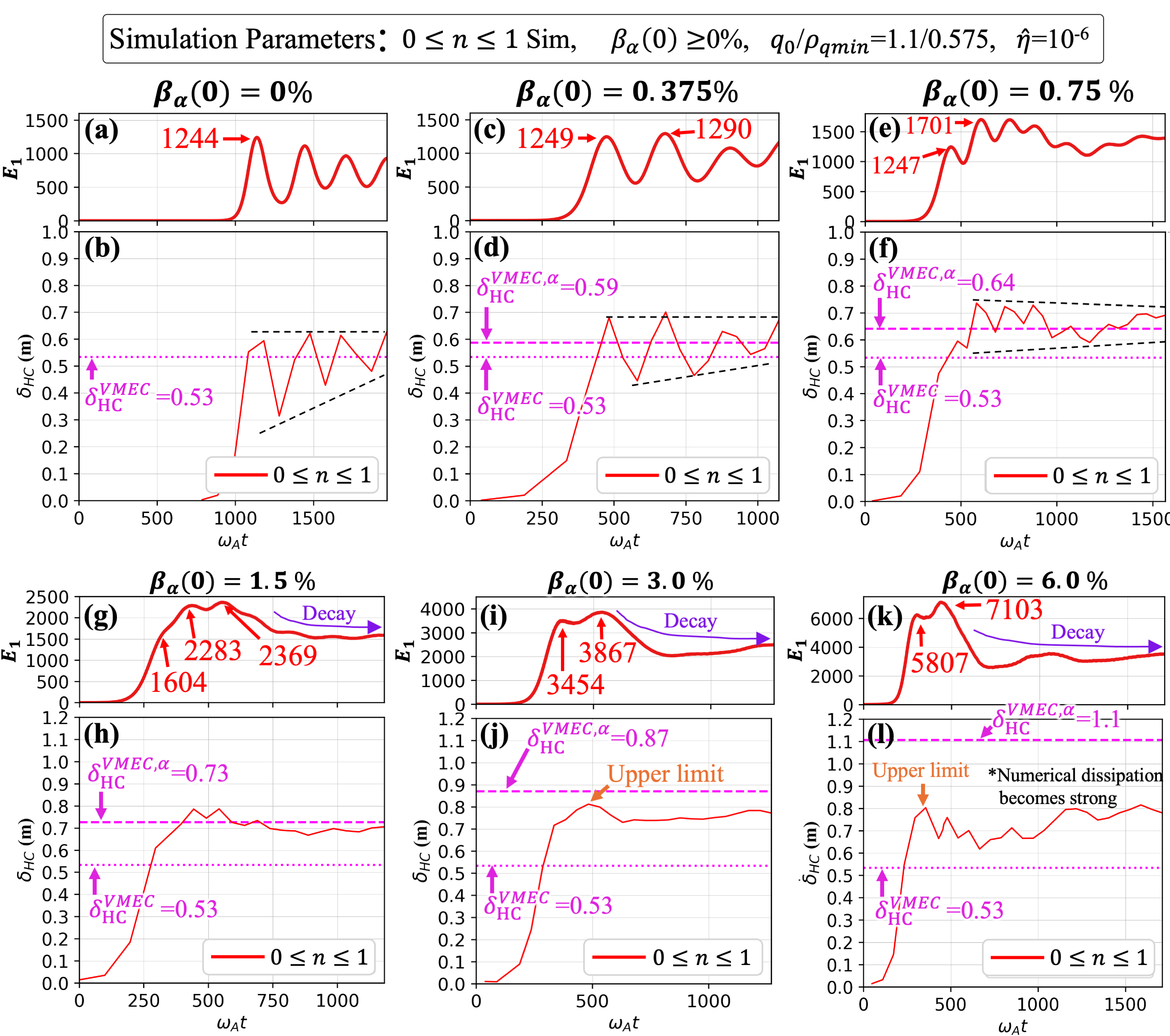}
\end{center} 
\caption{Time evolution of (a,c,e,g,i,k) the $n=1$ mode energy $E_1$ and (b,d,f,h,j,l) $\delta_\mathrm{HC}$ in the $q_\mathrm{0}/\rho_\mathrm{qmin}=1.1/0.575$ equilibrium with $\beta_\mathrm{\alpha}>0\%$.  The $0\leq n\leq1$ MEGA simulation was performed with $\hat{\eta}=10^{-6}$.  Panels (a-b) show the results for $\beta_\mathrm{\alpha}(0)=0\%$, (c-d) for $0.375\%$, (e-f) for $0.75\%$, (g-h) for $1.5\%$, (i-j) for $3\%$, and (k-l) for $6\%$.  The labeled red arrows indicate the value of $E_1$ at the primary and secondary saturation points in arbitrary units.}
\label{fig:5betascan_energyhdis_n1}
\end{figure*}

\subsubsection{\label{sec:level5b3}HC-Induced Magnetic Field Variation}
\quad

The basic mechanism responsible for the modification of trapped orbits and the emergence of transitional orbits as seen in Fig.\ref{fig:5testtraptran} is the modulation of the magnetic field strength $B$ along a particle orbit. In contrast to the magnetic ripples associated with discrete toroidal field coils in a tokamak or the twisted coils in stellarators, the $B$ field strength in our simulations remains nearly unaltered by the HC in the lab frame, where its gradient is still predominantly in the major radial direction. For illustration, the contours of the magnetic field strength in our HC equilibrium are shown in Fig.\ref{fig:5hcripple}(b,c) at $180^\circ$ and $0^\circ$ toroidal angles, respectively.  Nonetheless, the $m/n=1/1$ displaced magnetic flux surfaces and associated distortion of the particle orbits cause an additional modulation of the magnetic field strength in the particle frame.  To illustrate this effect,  Fig.\ref{fig:5hcripple} shows the trajectory of a test passing particle confined within the core region.  Panel (a) shows the time evolution of the particle’s major radial position, which is inversely proportional to the magnetic field strength along its orbit ($B \propto 1/R$).   The dynamic of the particle’s major radial position exhibits fast and slow variations, resembling a beat wave.  

To understand this beat motion, we need to note that the particle tends to remain in the uncompressed, compressed, and intermediate regions of the HC configuration for several toroidal transits because the safety factor profile is close to unity ($q \gtrsim 1$).  This is illustrated in Fig.\ref{fig:5hcripple}(b,c), where colored Poincar{\'e} plots of our passing test particle trajectory are plotted together with black Poincar\'{e} plots of the magnetic flux surfaces at toroidal angles $\phi = 180^\circ$ and $0^\circ$, respectively.   The rainbow color gradient represents the temporal position of this test particle, according to the time evolution of the major radial position shown in panel (a).  This test particle is launched from the HC uncompressed flux region at $\phi = 0^\circ$, indicated by the blue arrow as shown in panel (c). After a half toroidal transition ($\Delta\phi = 180^\circ$), the particle moved to the position marked by the blue arrow in panel (b).  Clearly, this particle remains in the uncompressed region of HC, and the change in its major radial position is relatively small ($|\Delta R_\mathrm{gc}|=|6.25-6.5|=0.25$ m).  As time progresses, this particle slowly precesses poloidally from the uncompressed flux region to the compressed flux region, while experiencing an increasing variation in the major radial position and, thus, magnetic field strength.  When this particle reaches the compressed flux region, as indicated by the green arrow, the variation of the major radial position and magnetic field strength reaches a maximum ($|\Delta R_\mathrm{gc}|=|7.3-5.4|=1.9$ m).  Based on this physical picture, the observed beat can be attributed to the small difference in the toroidal and poloidal orbit frequencies, $f_\phi$ and $f_\theta$, of a particle, while the amplitude of the beat corresponds to the HC displacement.  To confirm this point, $f_\phi$ and $f_\theta$ are approximated using $\frac{v_\alpha}{2\pi R_0}$ and $\frac{v_\alpha}{2\pi R_0q}$, respectively.  Given that $v_\alpha=1.29\times10^{7}$ m/s ($3500$ keV alpha particle), $R_0=6.5$ m, and $q=\frac{q_0+q_{min}}{2}=1.07$, we obtain $f_\phi=317$ kHz and $f_\theta=296$ kHz.  From these frequencies, the periods of beat and carrier are $T_\mathrm{beat}=0.048$ ms and $T_\mathrm{car}=0.003$ ms, respectively.  These periods are in good quantitative agreement with the time evolution of the major radial position in Fig.\ref{fig:5testorbit_fedis}(a).  In summary, the magnetic field variation perceived by the particle is reduced (destructive interference) when the particle resides within the uncompressed flux region and enhanced (constructive interference) in the compressed flux region.

The HC-induced magnetic field variation discussed above can explain the emergence of transitional orbits like the one shown in Figs.\ref{fig:5testtraptran}(e-h). This test particle was launched from the uncompressed flux region of the HC, where the magnetic field variation is weak.  According to the time evolution of the major radial position shown in Fig.\ref{fig:5testtraptran}(g), the passing orbit phase ($t\leq0.1$, ms shown in blue) experiences a small variation, reflecting the relatively uniform magnetic field strength experienced by the particle in that region of the plasma. As the particle precesses poloidally to the compressed flux region, it experiences an increasing variation in the magnetic field strength.  When the magnetic field variation becomes sufficiently large and the particle's parallel kinetic energy is sufficiently small, the particle is trapped by the magnetic mirror.  During the trapped phase, the particle is poloidally localized at the low-field side of the plasma.  As time progresses, the precessional drift slowly carries the particle back to the uncompressed region, gradually reducing the magnetic field variation, which allows the particle to transit back to a passing orbit.  One consequence of this orbit type is that the particle tends to remain spatially and temporally localized at specific toroidal and poloidal positions.  For instance, during the trapped orbit phase, the particle spends more time at toroidal angles where the HC compressed flux region lies on the low-field side, exhibiting a large radial excursion.

Lastly, we find that the HC-induced magnetic field variation, particularly the field suppression in the uncompressed flux region, is not strictly proportional to the HC displacement $\delta_\mathrm{HC}$. Minimum magnetic field variation occurs when the uncompressed flux region is, on average, aligned with the geometric center of the plasma. If one increases $\delta_\mathrm{HC}$ further from this point, the magnetic field variation in the uncompressed flux region becomes stronger. We think that the consequence of this behavior can be observed in Figs.\ref{fig:5testorbit_fedis}(g-i), which show the energy distribution and orbit classification for the case with $\beta_\mathrm{\alpha}(0) = 3\%$. Despite a larger displacement ($\delta_\mathrm{HC}^\mathrm{MEGA} \approx 0.8$ m), the population of passing-trapped transitional orbits does not increase, but instead shows a slight reduction when compared to the $\beta_\mathrm{\alpha}(0)=0.75\%$ case  ($\delta_\mathrm{HC}^\mathrm{MEGA} \approx 0.65$ m) in Figs.\ref{fig:5testorbit_fedis},(d-f).

\subsection{\label{sec:level5c}HC Stability and Alpha Confinement}
\quad

In this section, we investigate the time evolution of the $n=1$ mode energy, $\delta_\mathrm{HC}^\mathrm{MEGA}$, and the pressure profiles, aiming to clarify the dependence of $\delta_\mathrm{HC}$ on $\beta_\mathrm{\alpha}$.  We begin with the examination of the $\beta_\mathrm{\alpha}(0) \leq 0.75\%$ case, where alpha particles slightly reduce the linear growth rate $\gamma_\mathrm{HC}$ of the $m/n=1/1$ MHD mode.  After that, we discuss cases with $\beta_\mathrm{\alpha}(0)\geq1.5\%$, where alpha particles drive an $m/n=1/1$ EPM.

\subsubsection{\label{sec:level5c1}$\beta_\mathrm{\alpha}(0)\leq0.75\%$ Cases}
\quad

The time evolution of $E_1$ and $\delta_\mathrm{HC}^\mathrm{MEGA}$ during the nonlinear phase is shown in Figs.\ref{fig:5betascan_energyhdis_n1}.  Panels (a-b) show the results for $\beta_\mathrm{\alpha}(0)=0\%$, (c-d) for $0.375\%$, and (e-f) for $0.75\%$.  The $\beta_\mathrm{\alpha}(0)=0\%$ results shown in panels (a-b) are the same as those shown in Figs.\ref{fig:4megaq0scan}(g-h). The results are plotted again for ease of comparison.  The two black dashed lines shown in Figs.\ref{fig:5betascan_energyhdis_n1}(b,d,f) indicate the direction of convergence of $\delta_\mathrm{HC}^\mathrm{MEGA}$ toward its quasi-steady state value.

For the $\beta_\mathrm{\alpha}(0)=0.375\%$ and $0.75\%$ cases, the energy of the $n=1$ mode $E_1$ does not simply oscillate with a decreasing amplitude.  Instead, the mode continues to grow after a primary saturation level.  For the $\beta_\mathrm{\alpha}(0)=0.375\%$ case, the $n=1$ mode energy at the secondary saturation increases slightly from $1249$ to $1290$, indicated by the red labels in Fig.\ref{fig:5betascan_energyhdis_n1}(c).  The enhancement of the secondary saturation level becomes more prominent for the $\beta_\mathrm{\alpha}(0)=0.75\%$ case, where the saturation level increases from $1247$ to $1701$.  After the secondary saturation, the $n=1$ mode energy slightly decays and, on average, remains at a value higher than the primary saturation level.  $\delta_\mathrm{HC}^\mathrm{MEGA}$ also follows similar behaviors with the $n=1$ mode energy in these two cases.  These multi-stage saturation processes may be a manifestation of differences in the ways the bulk plasma and alpha particle respond to the change in equilibrium.  The passing alpha particles are likely to have a similar response to the bulk plasmas, while the particles on trapped and transitional particles can respond differently and require a longer time.  The period of the pulsation of $\delta_\mathrm{HC}^\mathrm{MEGA}$ that can be observed in Fig.\ref{fig:5betascan_energyhdis_n1}(e) is approximately $500$ Alfv{\'e}n times, equivalent to $0.43$ ms, which is remarkably close to the transition period between the passing and trapped particle shown in Figs.\ref{fig:5testtraptran}(e-h).  However, at present we can only speculate about such a connection between passing-trapped transitions and the pulsation dynamics of $\delta_\mathrm{HC}$, which remains to be verified through an examination of the collective effects of all particles.  One should also remember that the orbit analysis in Section \ref{sec:level5b} was performed in a static snapshot of the HC magnetic field.  During the course of the actual MEGA simulation, the HC grows and rotates with a non-zero frequency, which may alter the particle orbits.

\begin{figure}[t]
\begin{center}
\includegraphics[width=1.00\linewidth]{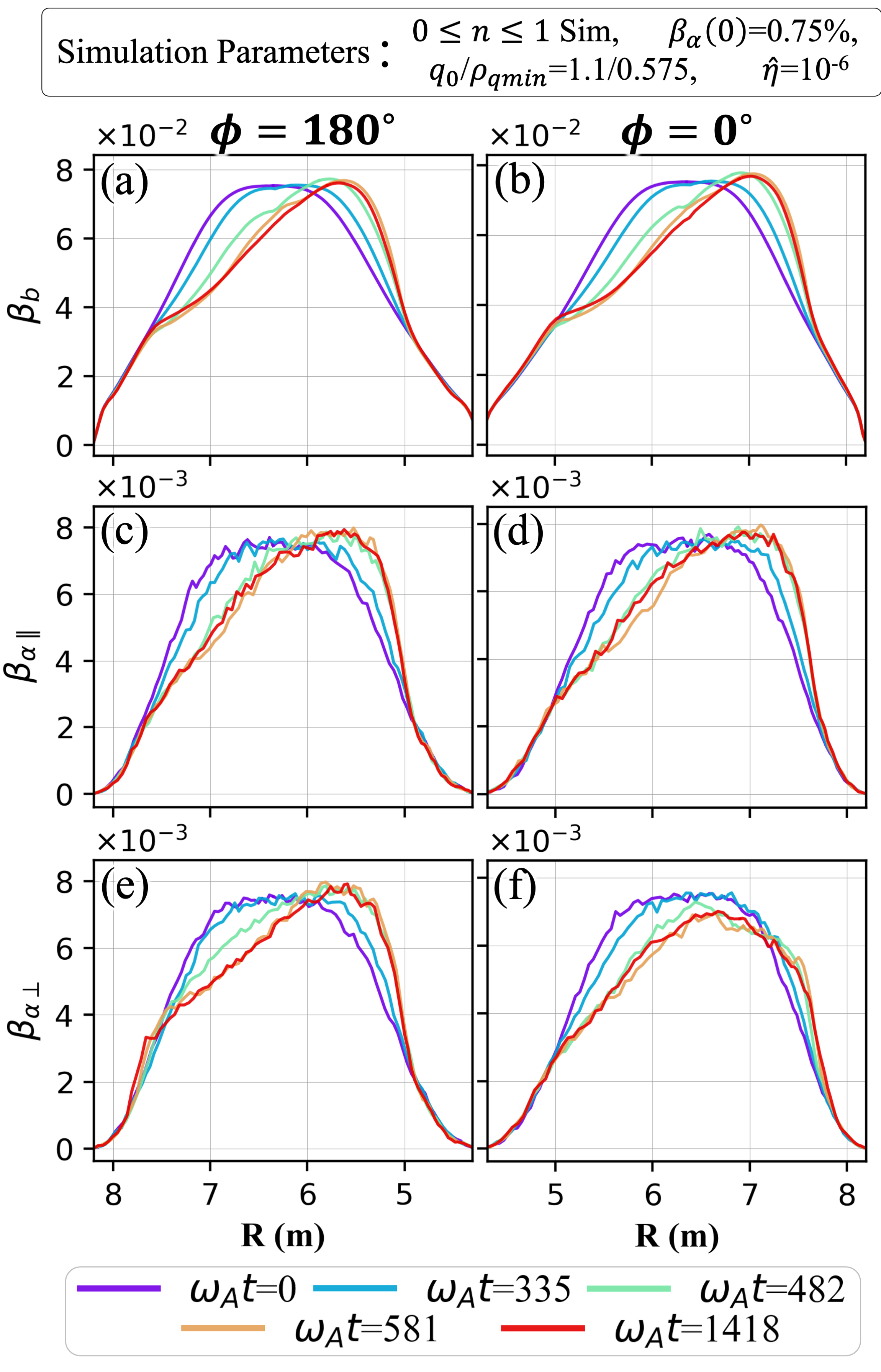}
\end{center} 
\caption{Time evolution of (a-b) $\beta_\mathrm{b}$, (c-d) $\beta_\mathrm{\alpha\parallel}$, and (e-f) $\beta_\mathrm{\alpha\perp}$ along the geometric mid-plane during the HC onset for the $\beta_\mathrm{\alpha}(0) = 0.75\%$ case as shown in Figs.\ref{fig:5betascan_energyhdis_n1}(e-f).  Panels (a,c,e) and (b,d,f) correspond to the $\phi=180^\circ$ and $0^\circ$ toroidal angles, respectively.  The radial profile at $\phi = 180^\circ$ is plotted with the inverted horizontal axis.}
\label{fig:5quasilinear_75}
\end{figure}

During the quasi-steady state, $\delta_\mathrm{HC}^\mathrm{MEGA}$ relaxes and oscillates around $\delta_\mathrm{HC}^\mathrm{VMEC}$, showing the quantitative agreement as depicted previously in Fig.\ref{fig:5betascan_energyhdis_n1}(b).   This agreement implies that the kinetic effects of the alpha particles and their radial transport are not significant.  To confirm this point, the bulk plasma and alpha pressure profiles during the HC formation for the $\beta_\mathrm{\alpha}=0.75\%$ are investigated.  The time evolution of these profiles along the geometric mid-plane at the $\phi=180^\circ$ and $0^\circ$ toroidal angles for the $\beta_\mathrm{\alpha}(0)=0.75\%$ case is shown in Fig.\ref{fig:5quasilinear_75}.  The pressure profiles at the initial state, the HC transition state, and the quasi-steady state are plotted with rainbow solid lines, where the transition from violet to red, following the rainbow spectrum, represents the forward direction in time.  At the toroidal angles (a,c,e) $\phi= 180^\circ$ and (b,d,f) $0^\circ$, the core plasma pressure shifts toward the high-field and low-field sides, respectively. (Please note that the radial profile at $\phi = 180^\circ$ is plotted with the inverted horizontal axis.) The profiles of (a-b) $\beta_\mathrm{b}$, (c-d) $\beta_\mathrm{\alpha\parallel}$, and (e-f) $\beta_\mathrm{\alpha\perp}$ gradually deform with the displaced magnetic flux surface.  Notable radial mixing is not observed in $\beta_\mathrm{b}$ and $\beta_\mathrm{\alpha\parallel}$. Only a minor reduction is observed in $\beta_\mathrm{\alpha\perp}$ at $\phi=0^\circ$ but not at $\phi=180^\circ$.   This toroidally localized flattening of $\beta_\mathrm{\alpha\perp}$ may be related to the passing-trapped transitional orbits that we discussed in Section \ref{sec:level5b} above, which can undergo large radial excursions when they are localized in the low-field side of the HC-compressed flux region.  Nonetheless, this flattening of $\beta_\mathrm{\alpha\perp}$ is toroidally localized and small, and the overall peaked $\beta_\mathrm{b}$ and $\beta_\mathrm{\alpha}$ profiles are largely preserved.  The deviation from the scalar MHD pressure profile used in VMEC remains marginal, which is consistent with the good quantitative agreement between the two codes in this parameter regime.  These results also indicate that the alpha particles affect the quasi-steady state of HC mainly via their fluid response, here in the form of the deformed $\beta_\alpha$ profile.

\subsubsection{\label{sec:level5c2}$\beta_\mathrm{\alpha}(0)\geq1.5\%$ Cases}
\quad

At higher values of $\beta_\mathrm{\alpha}$, MEGA simulations yield lower values of $\delta_\mathrm{HC}$ than VMEC, which suggests that non-ideal and kinetic effects become prominent and irreversible. For $\beta_\mathrm{\alpha}\geq3\%$, $\delta_\mathrm{HC}^\mathrm{MEGA}$ remains near an upper limit of approximately $0.8$ m.  To understand the underlying reason, we analyzed the $1.5\% \leq \beta_\mathrm{\alpha} \leq 6\%$ cases in a similar way as the $\beta_\alpha \leq 0.75\%$ cases in Section~\ref{sec:level5c1} above and summarized the results in the lower part of Fig.\ref{fig:5betascan_energyhdis_n1}. The time evolution of the $n=1$ mode energy $E_1$ and $\delta_\mathrm{HC}^\mathrm{MEGA}$ for the $\beta_\mathrm{\alpha}(0)=1.5\%$, $3\%$, and $6\%$ cases are shown in Figs.\ref{fig:5betascan_energyhdis_n1}(g-h), (i-j), and (k-l), respectively.  With increasing $\beta_\mathrm{\alpha}$, the alpha particles resonantly drive the $n=1$ mode energy to a higher value.  After the secondary or tertiary saturation, the $n=1$ mode energy decays significantly towards a level lower than that of the primary saturation level, indicating that a stable equilibrium state or peaked pressure profile cannot be maintained.  For $\delta_\mathrm{HC}^\mathrm{MEGA}$, it does not increase above what appears to be an upper bound of approximately $0.8$ m as shown in Figs.\ref{fig:5betascan_energyhdis_n1}(h,i,l).  This upper bound does not correspond to the averaged radial distance between the axisymmetric magnetic axis and the $q_\mathrm{min}$ position (the physical limit of the $m/n=1/1$ eigenmode).

\begin{figure}[t]
\begin{center}
\includegraphics[width=1.00\linewidth]{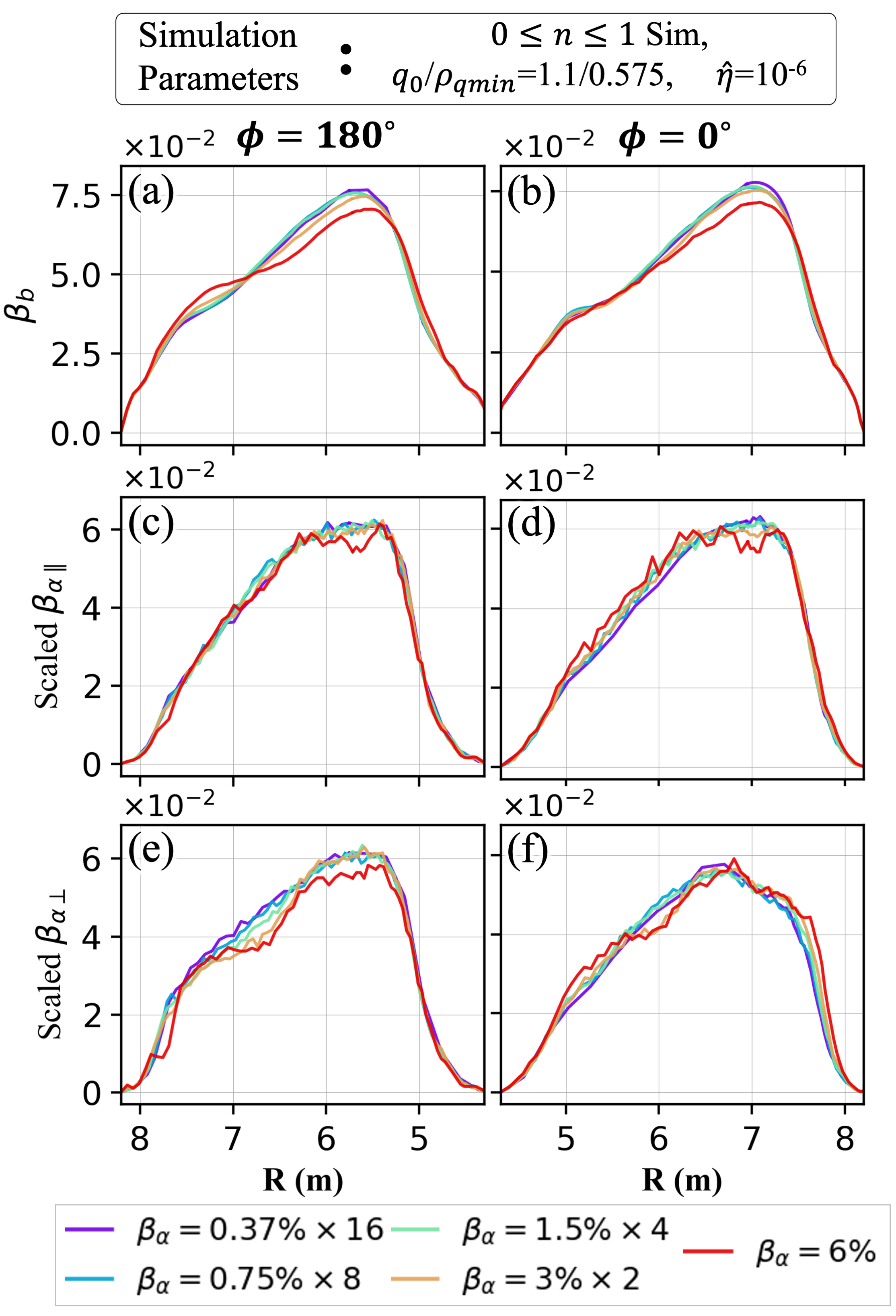}
\end{center} 
\caption{Comparison of the scaled (a-b) $\beta_\mathrm{b}$, (c-d) $\beta_\mathrm{\alpha\parallel}$, (e-f) $\beta_\mathrm{\alpha\perp}$ profiles along the geometric mid-plane during the quasi-steady state for different $\beta_\mathrm{\alpha}(0)$ values. Panels (a,c,e) and (b,d,f) correspond to the $\phi=180^\circ$ and $0^\circ$ toroidal angles, respectively.  The radial profile at $\phi = 180^\circ$ is plotted with the inverted horizontal axis.}
\label{fig:5quasilinear_compare}
\end{figure}

To understand the upper bound of $\delta_\mathrm{HC}^\mathrm{MEGA}$, we compared the geometric mid-plane profiles of $\beta_\mathrm{b}$, $\beta_\mathrm{\alpha\parallel}$, and $\beta_\mathrm{\alpha\perp}$ during the quasi-steady state for all values of $\beta_\mathrm{\alpha}(0)$ in our scan, as shown in Fig.\ref{fig:5quasilinear_compare}. For ease of comparison, the $\beta_{\alpha,\parallel}$ and $\beta_{\alpha,\perp}$ profiles for each case are scaled up to match that of the $\beta_\mathrm{\alpha}(0)=6\%$ case.  For examples, the $\beta_\mathrm{\alpha}$ profiles of the $\beta_\mathrm{\alpha}(0)=0.375\%$ is scaled by $16$ times.  For $\beta_\mathrm{\alpha}(0)=0.375\%$ (violet) and $0.75\%$ (blue), the differences are negligible.  When $\beta_\mathrm{\alpha}$ increases to a higher value, the difference becomes more apparent.  We find that both bulk plasma and parallel alpha pressure profiles for the $\beta_\mathrm{\alpha}(0)\geq 3\%$ cases are flattened and have a lower peak value.  For the $\beta_\mathrm{\alpha\perp}$ profile, a clear profile redistribution can be seen in the uncompressed flux region of HC, which differs from that of $\beta_\mathrm{\alpha\parallel}$.  The associated pressure anisotropy is not captured by VMEC.  The flattening of $\beta_\mathrm{b}$ and $\beta_\mathrm{\alpha}$ explain why $\delta_\mathrm{HC}^\mathrm{MEGA}$ does not increase beyond a certain upper limit in Figs.\ref{fig:5betascan_energyhdis_n1}(j,l): the profile flattening in MEGA provides a path to a new energy minimum that is inaccessible by VMEC, so that $\delta_\mathrm{HC}^\mathrm{MEGA}$ and $\delta_\mathrm{HC}^\mathrm{VMEC,\alpha}$ diverge for $\beta_\alpha > 1.5\%$ in Fig.\ref{fig:5gammafreqhdis_beta}(b).   

\begin{figure}[t]
\begin{center}
\includegraphics[width=0.90\linewidth]{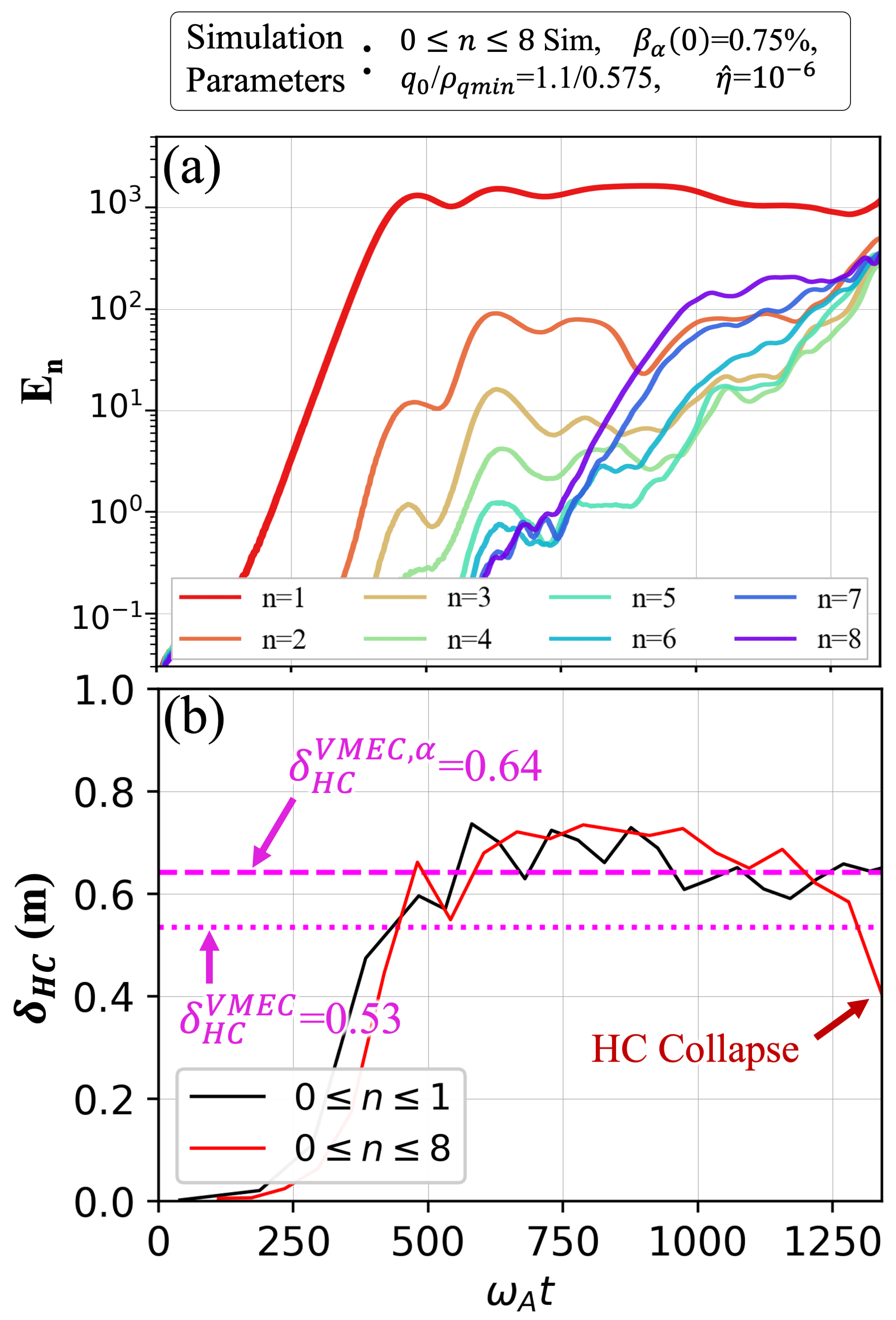}
\end{center} 
\caption{Time evolution of (a) the mode energies $E_n$ and (b) $\delta_\mathrm{HC}$ in the $q_\mathrm{0}/\rho_\mathrm{qmin} = 1.1/0.575$ equilibrium with $\beta_\mathrm{\alpha}(0) = 0.75\%$ simulated by MEGA with $0\leq n \leq 8$. }
\label{fig:5betascan_energyhdis_n8}
\end{figure}

\subsection{\label{sec:level5d}Effects of the Secondary Mode on Alpha Particle Confinement}
\quad

\begin{figure*}[h]
\begin{center}
\includegraphics[width=1.00\linewidth]{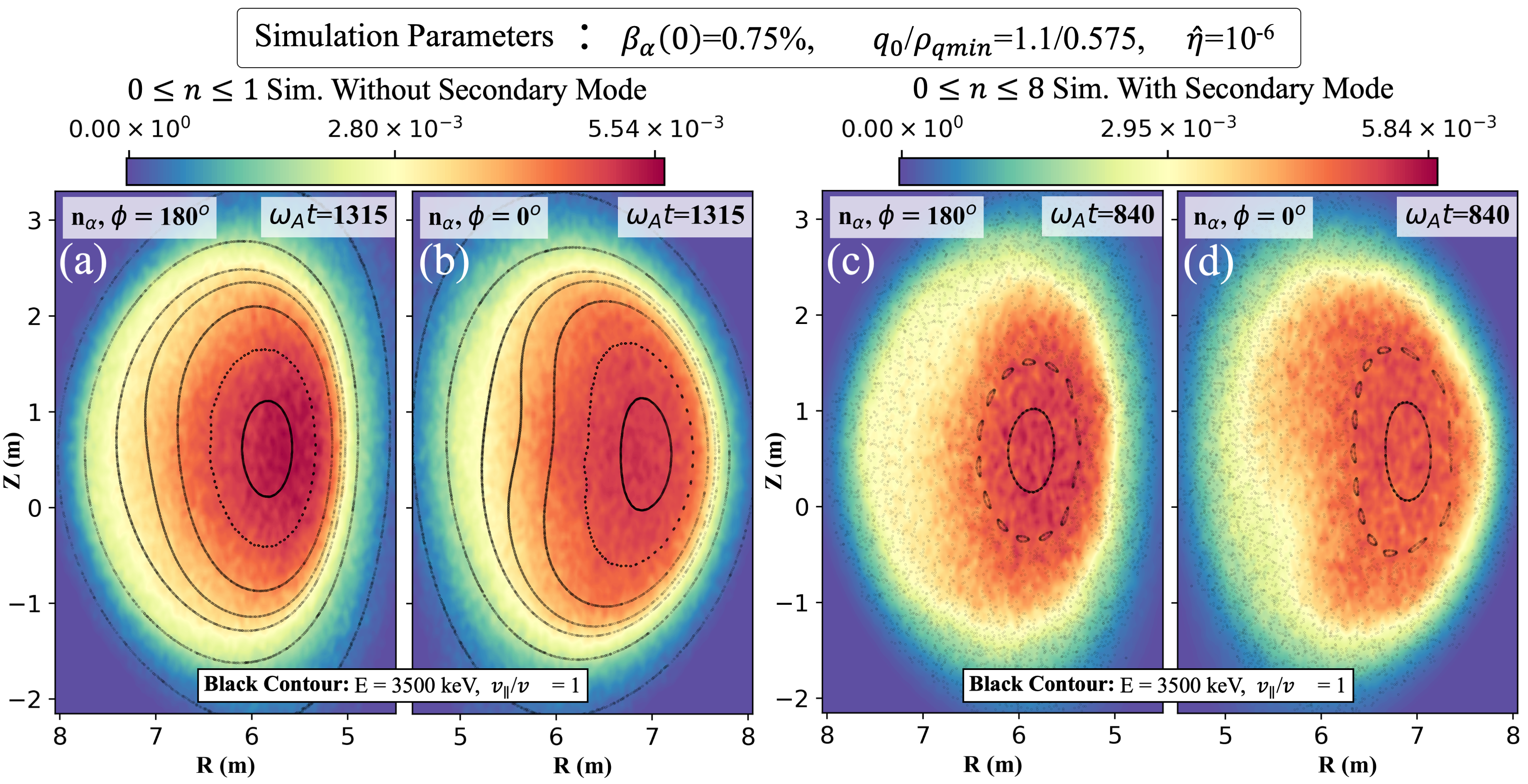}
\end{center} 
\caption{Poloidal cross-section of the alpha particle density field $n_\alpha(R, Z)$ during the HC quasi-steady state obtained with MEGA.  The MEGA simulation was performed in the $q_\mathrm{0}/\rho_\mathrm{qmin}=1.1/0.575$ equilibrium with $\beta_\mathrm{\alpha}(0)=0.75\%$ and $\beta_\mathrm{b}(0)=7.53\%$.  Panels (a–b) and (c–d) show the results for the $0 \leq n \leq 1$ and $0 \leq n \leq 8$ simulations, respectively, while Panels (a, c) and (b, d) correspond to the toroidal angles $\phi = 180^\circ$ and $\phi = 0^\circ$, respectively. The radial coordinate at $\phi = 180^\circ$ is plotted with an inverted horizontal axis.  The Poincar{\'e} plot represents the orbit of co-passing alpha particles with $E=3500$ keV and $v_\parallel/v=1$.}
\label{fig:5mega_nalpha_n1n8_betaA075}
\end{figure*}

In Sections \ref{sec:level5a}-\ref{sec:level5c}, we applied a low-pass filter that removes $n>1$ components and thereby suppressed the secondary MHD modes discussed in Section \ref{sec:level4}.  Recall that the secondary mode consists of several Fourier components with different helicities capable of causing the chaotization of the magnetic field line.  Unlike bulk ions, alpha particles can have a non-negligible orbit width, which means that the drift orbit chaos may differ from the magnetic chaos (it may be enhanced or reduced as the location and width of resonances vary).  Although the magnetic chaos predicted by our model may not be quantitatively accurate, it is still instructive to qualitatively examine the extent to which the secondary mode influences alpha particles.  

Fig.\ref{fig:5betascan_energyhdis_n8} shows the time evolution of the $1\leq n\leq8$ mode energies and $\delta_\mathrm{HC}^\mathrm{MEGA}$ for the $\beta_\mathrm{\alpha}(0)=0.75\%$ case.  It can be seen that the $5\leq n \leq 8$ Fourier components become increasingly unstable during the nonlinear phase, indicating the destabilization of the secondary mode.  As the secondary mode saturates, the $\delta_\mathrm{HC}^\mathrm{MEGA}$ experiences an abrupt reduction, indicating the collapse of HC.  To evaluate the impact of the secondary mode on the alpha particle confinement, Fig.\ref{fig:5mega_nalpha_n1n8_betaA075} compares the differences in the alpha particle density field $n_\alpha(R, Z)$ after HC formation between the (a-b) $0\leq n\leq1$ and (c-d) $0\leq n\leq8$ simulations.  In both simulations, the alpha density profiles at $\phi = 180^\circ$ and $0^\circ$ are shown in panels (a, c) and (b, d), respectively. The Poincar{\'e} plot of co-passing alpha particles with $E=3500$ keV and $v_\parallel/v=1$ is shown as overlaid black contours.  In the $0\leq n\leq1$ simulation (This belongs to the same simulation as the $\beta_\mathrm{\alpha}(0)=0.75\%$ case discussed in Section \ref{sec:level5a}.), panels (a-b), the alpha contours are distorted by the HC but remain mostly sharp and nested. Alpha particles with different energies and pitch angles show similar behavior (not shown here); however, trapped and passing-trapped transitional alpha particle orbits are more complicated than the passing particles and cannot be represented with a simple Poincar{\'e} plot in the R-Z plane, as discussed in Section \ref{sec:level5b}.  These nested alpha contours are consistent with $n_\alpha(R, Z)$, where no notable signs of radial mixing are observed.  In contrast, the $0\leq n\leq8$ simulation in panels (c-d) shows a clear signature of radial mixing in $n_\alpha(R, Z)$ and the chaotization of the alpha particles' trajectories. 

\begin{figure*}[t]
\begin{center}
\includegraphics[width=1.00\linewidth]{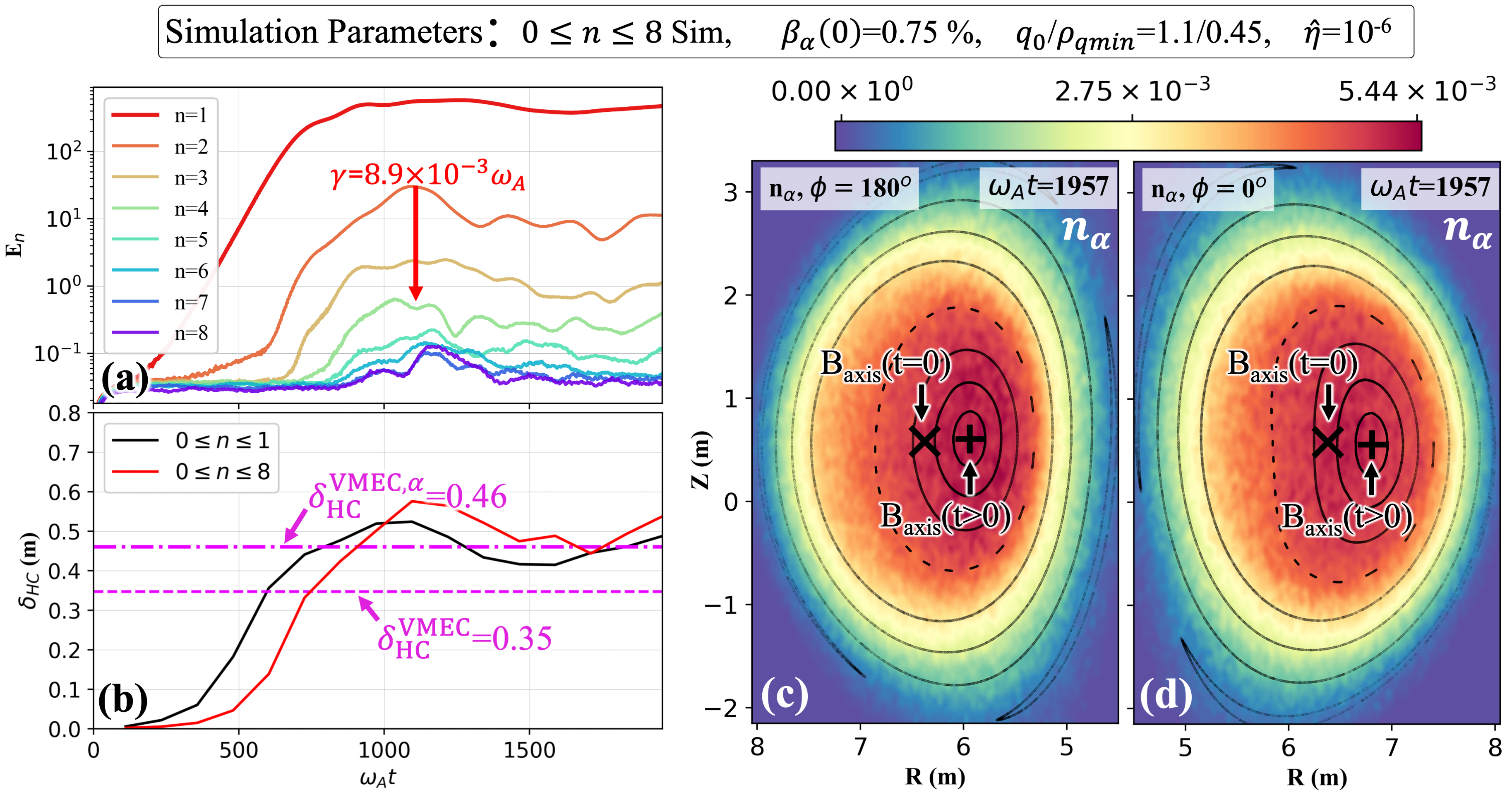}
\end{center} 
\caption{Helical core formation of the $q_\mathrm{0}/\rho_\mathrm{qmin}=1.1/0.456$ equilibrium with $\beta_\mathrm{\alpha}(0)=0.75\%$.  The $0\leq n\leq8$ MEGA simulation was performed with $\hat{\eta}=10^{-6}$: (a) Time evolution of $1\leq n\leq8$ mode energies.; (b) Time evolution of radial displacement of the magnetic axis $\delta_\mathrm{HC}$.; Alpha particle density field $n_\alpha$ during the quasi-steady state at (c) $\phi=180^\circ$ and (d) $0^\circ$ toroidal angles, respectively.  The radial coordinate at $\phi = 180^\circ$ is plotted with an inverted horizontal axis.}
\label{fig:5rho45_energyhdisPoincare}
\end{figure*}

\section{\label{sec:level6}Effects of Alpha Particles on HC Formation in the $q_\mathrm{0}/\rho_\mathrm{qmin} = 1.1/0.456$ Equilibrium}
\quad

The results shown in Figs.\ref{fig:5gammafreqhdis_beta} and \ref{fig:5betascan_energyhdis_n8} for the $q_\mathrm{0}/\rho_\mathrm{qmin}=1.1/0.575$ equilibrium suggest that the alpha particles can enhance $\delta_\mathrm{HC}$, and its quasi-steady values calculated with MEGA quantitatively agree with those calculated with VMEC for $\beta_\mathrm{\alpha}(0) \leq 1\%$.  However, in our resistive MHD model, this equilibrium is unstable to the secondary mode, which can lead to the radial mixing of alpha particles. In Sections \ref{sec:level4b} and \ref{sec:level4c}, we found that the secondary mode was completely suppressed when $\rho_\mathrm{qmin}$ was reduced, as in the $q_\mathrm{0}/\rho_\mathrm{qmin}=1.1/0.456$ equilibrium without alpha particles.  Here, we test whether the secondary mode remains stable in the case with alpha particles and also test the robustness of our findings that the alpha particles tend to enhance $\delta_\mathrm{HC}$ during the quasi-steady state.  The simulation is performed for the $\beta_\mathrm{\alpha}(0)=0.75\%$ case with $0\leq n\leq8$ and moderate plasma resistivity $\hat{\eta}=10^{-6}$.   

The time evolution of the $1\leq n\leq8$ mode energies and $\delta_\mathrm{HC}$ are shown in Figs.\ref{fig:5rho45_energyhdisPoincare}(a-b), respectively.  The linear growth rate of the $m/n=1/1$ kink/quasi-interchange mode $\gamma_\mathrm{HC}$ is $8.9\times10^{-3}\omega_A$ which is lower than in the $\beta_\mathrm{\alpha}=0\%$ case ($1.2\times10^{-2}\omega_A$) shown in Fig.\ref{fig:4megaqwscan}(a) by roughly $25\%$; however, the $\delta_\mathrm{HC}$ calculated with MEGA increases, and its quasi-steady value quantitatively agrees with that of VMEC.  After the HC formation $\omega_At>900$, the energy of the $5\leq n\leq 8$ modes are negligible as shown in Fig.\ref{fig:5rho45_energyhdisPoincare}(a). This means that the secondary mode remains stable (at least for the duration of our simulation) even in the presence of alpha particles.  The poloidal cross-section of the alpha particle density field, $n_\alpha(R, Z)$, at the final time step, is shown in Figs.\ref{fig:5rho45_energyhdisPoincare}(c-d).  Panels (c-d) show $n_\alpha(R,Z)$ at the $\phi=180^\circ$ and $0^\circ$ toroidal angles, respectively.  The overlaid black contours represent the magnetic Poincar{\'e} plot.  Due to the absence of the secondary mode, the nested magnetic flux surfaces are maintained, and the alpha particles remain well-confined.  These results demonstrate that it is feasible to have a HC state that maintains MHD stability and provides good alpha particle confinement within the ITER operating $\beta_\mathrm{\alpha}$ range.

\section{\label{sec:level7}Conclusion and Discussion}
\quad

This study numerically investigated the effects of D-T fusion-born alpha particles on the HC formation in an ITER-scale hybrid scenario, namely plasmas whose central region has a weak magnetic shear and $q \gtrsim 1$.   Our main tool is the nonlinear MHD-PIC simulation code MEGA.  We also compared the quasi-steady state solution of HC calculated by MEGA with that of VMEC, a 3-D MHD equilibrium code. The contribution of alpha particles is included in VMEC only as an additional scalar MHD pressure.  Comparisons between the HC solutions from these two codes, which employ different physics models, yielded initial insights into the effect of alpha particles on HC formation and {\it vice versa}.  Based on the results of our systematic parameter scans with these two codes, where we varied the $q$ profile and the alpha particle pressure $\beta_\mathrm{\alpha}$, we can answer the three key research questions that were posed in Section \ref{sec:level1}. In a nutshell, we found that (1) HC can form in plasmas with ITER-relevant values of $\beta_\mathrm{\alpha}$, (2) alpha particles tend to enhance the saturation level of a HC where the steady-state is determined mainly from the net bulk plasma and alpha pressure profiles, and (3) alpha particles remain well confined as long as the HC state does not trigger secondary instabilities (in our case, in the form of a resistive pressure-driven MHD mode).  Answers (2) and (3) are discussed in more detail in the following Sections \ref{sec:level7a} and \ref{sec:level7b}, respectively.

\subsection{\label{sec:level7a}Interplay between Alpha Particles and HC}
\quad

Our key findings can be summarized as follows:
\begin{enumerate} 
\item Sections \ref{sec:level5a} and \ref{sec:level5c1}: Within the ITER operating range $\beta_\mathrm{\alpha}\leq 1\%$, alpha particles weakly reduce the linear growth rate $\gamma_\mathrm{HC}$ of the $m/n = 1/1$ kink/quasi-interchange mode. However, the radial displacement of the magnetic axis in the quasi-steady state, $\delta_\mathrm{HC}$, is enhanced.  The quasi-steady value of $\delta_\mathrm{HC}^\mathrm{MEGA}$ agrees quantitatively with $\delta_\mathrm{HC}^\mathrm{VMEC,\alpha}$, suggesting that non-ideal MHD effects, kinetic effects, and alpha particle transport are negligible.  This agreement implies that the steady state solution is determined mainly from the ideal MHD energy minimization (i.e., plasma frozen-in with the field) in this regime.
\item Sections \ref{sec:level5a} and \ref{sec:level5c2}: In the cases with exaggerated $\beta_\mathrm{\alpha}>1\%$, alpha particles resonantly drive the $m/n=1/1$ mode.  $\delta_\mathrm{HC}^\mathrm{MEGA}$ continues to increase with $\beta_\mathrm{\alpha}$ until it reaches a limit where a stable HC cannot be maintained with the initial peaked $\beta_\mathrm{b}$ and $\beta_\mathrm{\alpha}$ profiles. The fact that this limit is set by stability and transport processes is confirmed by the absence of such a limit in $\delta_\mathrm{HC}^\mathrm{VMEC,\alpha}$.
\item Section \ref{sec:level5b}: While the HC does not have a significant effect on the gradient of the magnetic field strength $B \propto 1/R$, the displaced guiding center orbits ${\bm X}_{\rm gc}(t)$ in the HC are subject to an additional beat-type modulation of $B$ along ${\bm X}_{\rm gc}(t)$. Taking into account the fact that $q$ is close to unity in the domain of the HC, it is found that particles residing in the uncompressed flux region remain close to the plasma's geometric center and thus experience only a weak modulation of $B$ during each toroidal transit.\footnote{One might also sat that the helicity of the HC largely cancels the helicity of the initial equilibrium field in this region. In ``optimal'' cases where this cancellation is perfect, particles would travel on perfectly toroidal orbits with constant $R$.}
The reverse is true for particles residing in the HC's compressed flux region, further away from the plasma's geometric center. The enhanced modulation of $B(R_{\rm gc}(t))$ towards the compressed flux region causes particles near the passing–trapped boundary to transition between passing and trapped orbits periodically.
\end{enumerate}

These key findings may inform the design of HC plasma scenarios and control schemes; however, we find that there are still some points and open questions that require further investigation:
\begin{itemize} 
\item {\it (Un)importance of drift-kinetic effects:} The ITER reference case considered in this paper features a high plasma current $13$ MA $< I_p <$ $15$ MA, which leads to a small alpha particle drift orbit width $\Delta_{\rm orbit}$.  When $\Delta_{\rm orbit}$ is negligible, the alpha particle pressure profile can be well represented as a flux function. In our case, drift-kinetic effects were ignorable (so that MEGA and VMEC made similar predictions) when $\beta_\alpha \lesssim 1.5\%$, but this parameter window is expected to be smaller in plasmas with lower current (as in the proposed new ITER baseline\cite{barabaschi2025iter}).
\item {\it HC dynamics on the transport time scale:} Our study only considers the effect of alpha particles on HC formation within the MHD time scale (a few 100 Alfv\'{e}n times).  On the longer transport time scale, the effects of sources, collisions, and sinks can be significant, and they can impact the steady state of HC in the long run. When simulating the long-term HC dynamics self-consistently, the source, collision, and sink models must also be consistent with the HC configuration.
\item {\it EP phase space dynamics:} A thorough analysis of alpha particle dynamics in phase space or in reduced sets of suitable (exact or adiabatic) invariants remains to be carried out. This will be necessary to understand the HC-EP interactions (such as resonances and net energy exchange) and to confirm some of the assertions made in the present work. In cases where alpha particles remain well confined after HC formation, which implies that the configuration is omnigenious, we expect that the canonical toroidal angular momentum $P_\phi$ is conserved relatively well on average.  The extent to which a suitably averaged $P_\phi$ can be used for reliable interpretative analyses requires further study.
\end{itemize} 

\subsection{\label{sec:level7b}Alpha Particle Confinement and Control of Secondary Mode}
\quad
After the HC formation, the good alpha particle confinement regime is limited by the possible onset of secondary instabilities, here in the form of a resistive pressure-driven MHD mode destabilized by the steepening of the pressure gradient in the compressed flux region of a HC. In the Fourier representation, this mode consists of a broad spectrum of short-wavelength Fourier components that constructively interfere along the HC compressed flux region. Within the capabilities of our resistive MHD model (e.g., missing microscopic physics) and numerical constraints (e.g., imperfect filtering), our preliminary findings concerning the secondary mode stability may be summarized as follows:

\begin{enumerate} 
\item {\it Resistivity dependence (Sections \ref{sec:level4c1}-\ref{sec:level4c2}):} In our simulations, the secondary mode grows more rapidly at higher plasma resistivity $\eta$.  The $\eta$ range used in this study, which is several orders of magnitude higher than the Spitzer resistivity for ITER core plasma parameters, can be viewed as a proxy for other non-ideal processes, such as the magnetic diffusion caused by electromagnetic microturbulence.
\item {\it Profile dependence (Sections \ref{sec:level4b} and \ref{sec:level4c3}):} When we fix the centrally peaked profile of the bulk plasma pressure $P_{\rm b}$ and vary the $q_\mathrm{min}$ radius, $\rho_\mathrm{qmin}$, we find that the secondary mode tends to become more unstable with increasing $\rho_\mathrm{qmin}$.
\end{enumerate}

\noindent It must be noted that these analyses of secondary instabilities have several caveats as discussed below.

First, the profile dependence of the secondary mode stability is difficult to study systematically because the results depend on how the parameter scans are performed, and it is unclear what the most realistic procedure would be. In reality, a changing $q$ profile can be expected to alter the heating profiles from fusion-born alpha particles and external sources. Changes in the $q$ profile will also affect self-organization processes in the plasma, including the global structure of the HC (quasi-interchange or kink type) that is the subject of the present work. These and other factors will influence the degree of pressure steepening in the compressed flux region for a given displacement $\delta_{\rm HC}$ of the magnetic axis. As stated in item (ii), in the present study, we ignored the connection between $q$ and heating profiles, and chose to fix the centrally peaked $P_b$ profile while varying $\rho_\mathrm{qmin}$. The position of $q_{\rm min}$ determines the size of the region of low magnetic shear ($q \gtrsim 1$) and is thus more or less the boundary of the HC domain. This means that moving $\rho_\mathrm{qmin}$ relative to a fixed nonuniform $P_b$ profile, as we have done, necessarily has a strong influence on the resulting pressure gradient steepening associated with the HC's compression of magnetic flux surfaces and, thus, potential secondary instabilities. The observed trends will thus depend on the choice of reference profiles, and our finding that the secondary mode tends to become more unstable with increasing $\rho_\mathrm{qmin}$ is clearly case-dependent. This implies that strategies for avoiding such secondary instabilities are also likely to be sensitive to the self-consistent interdependence of $q$ and pressure profiles.

Second, our analysis of the secondary mode stability in this paper was limited to a relatively narrow range of low toroidal mode numbers $n\leq8$, which is insufficient for an accurate representation of this mode (see the Appendices) and is likely to quantitatively affect both the resistivity-dependence and profile dependence. The analysis should include shorter wavelength components, and it should preferably be done using a global electromagnetic turbulence code.

Last but not least, the toroidal low-pass Fourier filter used to eliminate higher-$n$ components ceases to function as intended when the HC displacement $\delta_{\rm HC}$ becomes substantial. This is because the filter was applied in the direction of the geometric toroidal angle and was thus misaligned with the HC flux surfaces. See Section~\ref{sec:level2b} and the Appendices for details.

In summary, the results presented in this study can only suggest the existence and qualitative properties of a secondary mode in a HC configuration and its potential impact, including the possibility of magnetic chaos and resulting local reduction of bulk and alpha particle confinement, and possibly even the collapse of the HC state.

\subsection{\label{sec:level7c}Future Works}
\quad

The present paper dealt with the HC far from marginal stability, because that was the situation we obtained for the nominal pressure profile of the ITER reference case we used.  The effect of alphas in the case where the MHD HC is marginally unstable is studied in a separate paper.

The next step of our study will focus on Alfv{\'e}n eigenmodes (AEs) and energetic-particle modes (EPMs) in a HC equilibrium, where the hybrid simulation will be initiated from the HC equilibrium calculated with VMEC.  Using the insights won in the present work, we should be able to design cases where secondary instabilities are weak or absent (e.g., choosing a relatively small $q_\mathrm{min}$ radius located in a region with a weak bulk pressure gradient).  In addition, we would like to investigate further the HC formation in the cases with an anisotropic EP pressure.  The anisotropic pressure case is important because neutral beam (NB) and ion-cyclotron resonance frequency (ICRF) heating will be used in fusion reactors to promote bulk ion heating and to tailor both the safe factor and plasma rotation profiles.  

For the secondary mode, namely the resistive pressure-driven MHD mode in some HC configurations, it should be studied using electromagnetic turbulence codes that can handle short-wavelength structures correctly, such as EUTERPE\cite{kleiber2024euterpe}.

\ack
Numerical computations were performed on the JFRS-1 supercomputer of the International Fusion Energy Research Centre (Project ID: MHDEP3D) and HPE SGI8600 in the National Institutes for Quantum Science and Technology (Project ID: PG24527).  The authors of this work would like to thank Prof.\ Dr.\ Yasushi Todo (NIFS, Japan) for providing MEGA  (Some part of the original code was modified in this study), and Dr.\ Shuhei Sumida (QST), Prof.\ Dr.\ Kouji Shinohara (Tokyo University, Japan), Dr.\ Haruki Seto (QST, Japan), Dr.\ Gakushi Kawamura (QST, Japan), and Dr.\ Hogun Jhang (KFE, Republic of Korea) for fruitful discussions.

\clearpage
\appendix

\section{\label{sec:apen1}MHD Instabilities in HC Equilibrium calculated with VMEC}
\quad

In Section \ref{sec:level4}, we observed the destabilization of the secondary mode after the HC formation. The resistivity scan shown in Fig.\ref{fig:4etascan2nd} indicated that this secondary mode could be classified as a kind of resistive mode.  If the HC state calculated with MEGA and VMEC is truly consistent, the resistive pressure-driven modes should also be linearly unstable in the HC equilibrium calculated with VMEC. In this Appendix, we performed an additional MEGA simulation using the HC equilibrium calculated by VMEC. These results also serve as a preliminary test for our future study, where we intend to study EP-driven MHD modes in HC equilibrium.

\begin{figure}[h]
\begin{center}
\includegraphics[width=1.00\linewidth]{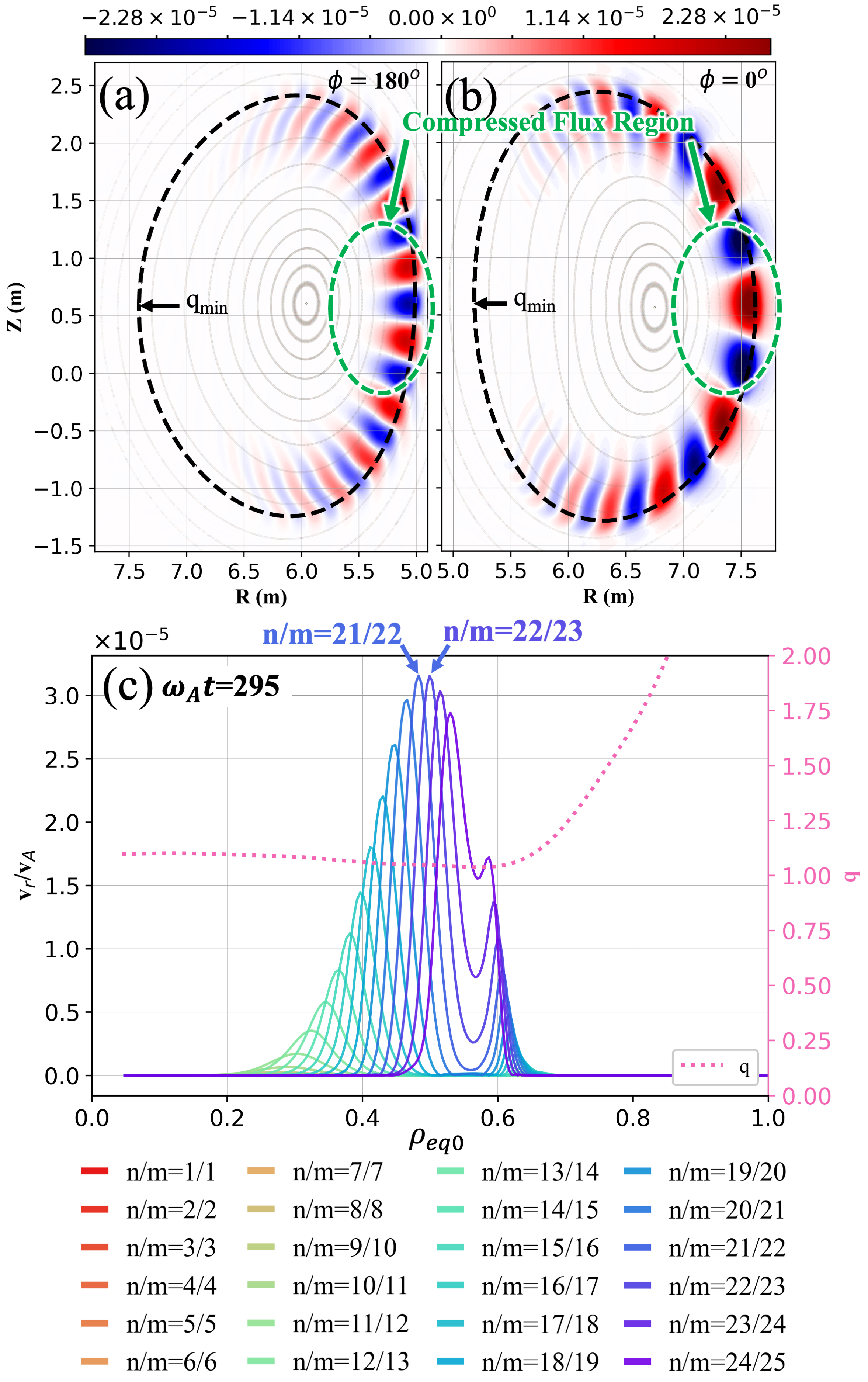}
\end{center} 
\caption{Spatial structure of the linearly unstable MHD mode in HC equilibrium: (a-b) Poloidal cross-section of the perturbed MHD plasma pressure $\delta P_b$ at $\phi=180^\circ$ and $0^\circ$ toroidal angles, respectively.  The radial coordinate at $\phi = 180^\circ$ is plotted with an inverted horizontal axis.; (c) Radial MHD velocity profiles of the dominant $n\leq24$ modes calculated in the $q_\mathrm{0}/\rho_\mathrm{qmin}=1.1/0.575$ HC equilibrium with $\hat{\eta}=10^{-7}$ and ($N_R$,$N_Z$,$N_\phi$) = ($400$,$400$,$384$). We note that our toroidal low-pass filter was not used in the simulation discussed in this Appendix.}
\label{fig:APPENhcprof}
\end{figure}

As discussed in Section \ref{sec:level2}, MEGA employed a simple low-pass filter along the geometrical toroidal direction to suppress short-wavelength modes artificially.  If we initiate the simulation from the HC equilibrium, the geometrical toroidal angle at a given $(R,Z)$ position will cross a range of magnetic surfaces, so that perturbations relative to the HC will appear convoluted with the HC itself.  Our simple low-pass filter along $\phi$ then no longer works as intended and can produce artifacts.  One method to resolve this issue is to perform a low-pass filter along the flux coordinate (e.g., Boozer or VMEC straight field line coordinate), where the MHD field data will be converted back and forth between the cylindrical and flux coordinates, respectively.  However, we find that such a method has a non-negligible grid conversion error, leading to artificial smoothening and residual magnetic monopole.  We have not yet found a suitable replacement for the MEGA low-pass filter; therefore, the low-pass filter will not be used in this Appendix.  

\begin{figure}[h]
\begin{center}
\includegraphics[width=1.00\linewidth]{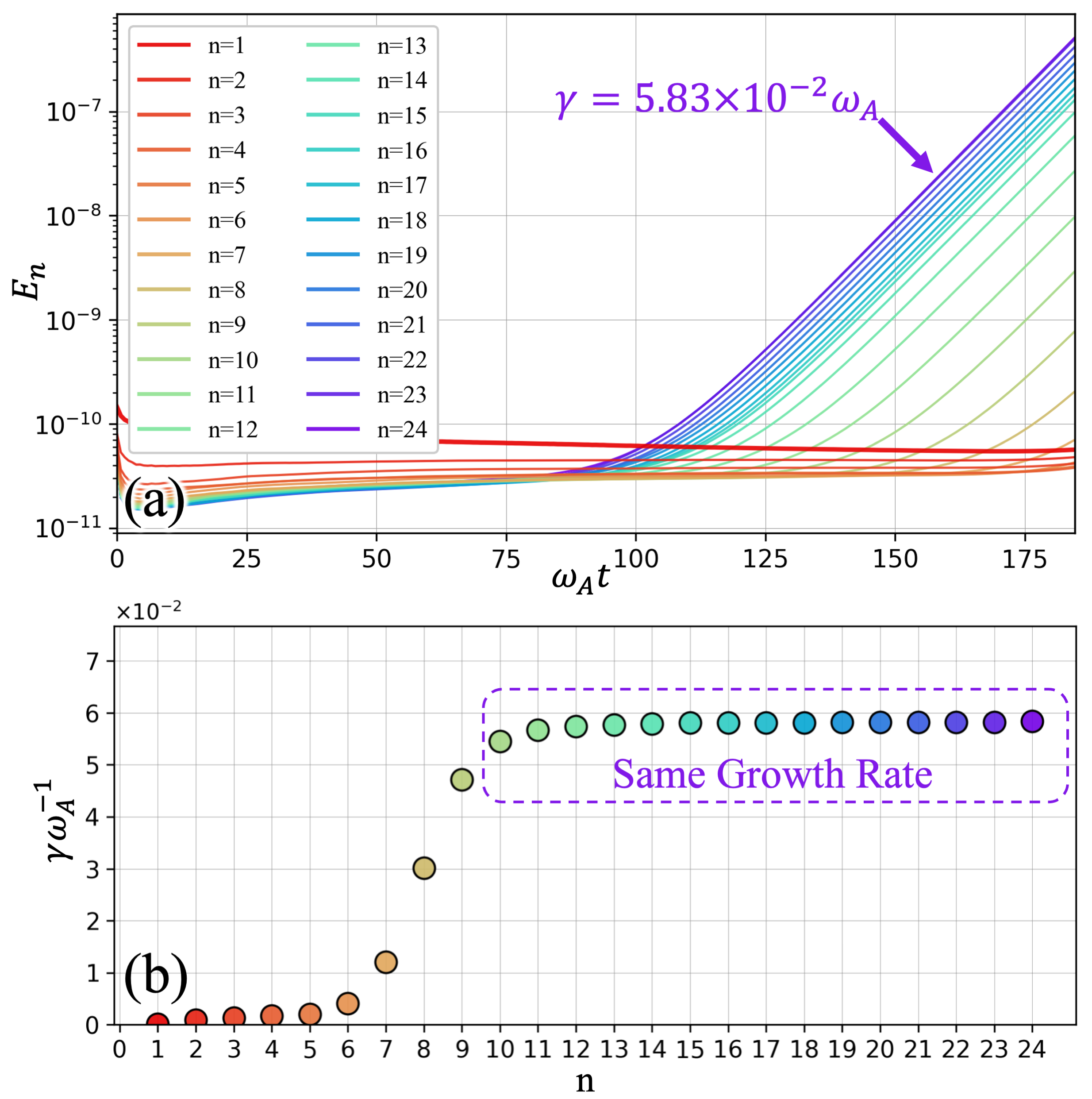}
\end{center} 
\caption{(a) Time evolution of the $n\leq24$ mode energies calculated in the $q_0/\rho_{qmin}=1.1/0.575$ HC equilibrium with $\hat{\eta}=10^{-7}$ and ($N_R$,$N_Z$,$N_\phi$) = ($400$,$400$,$384$). (b) Linear growth rates of $n\leq24$ modes measured during $160 \leq \omega_At \leq 180$.}
\label{fig:APPENhctevol}
\end{figure}

We focus on the MHD simulation of the $q_\mathrm{0}/\rho_\mathrm{qmin} = 1.1/0.575$ HC equilibrium with $\beta_\mathrm{\alpha}=0\%$.  The cylindrical grid resolution and plasma resistivity are ($N_R$, $N_Z$, $N_\phi$)=($400$, $400$, $384$) and $\hat{\eta}=10^{-7}$, respectively.  We used a higher resolution here to account for the weaker dissipation at short wavelengths.  Here, we only consider the plasma within the $\rho_\mathrm{eq0}\leq0.7$ flux surface to further increase the core resolution and reduce computational resources.  We will refer to this type of trimmed simulation domain as the ``core-only" simulation.  The flux coordinate used to analyze the mode energy and structure in this Appendix is the VMEC straight-field-line coordinate.  

\begin{figure}[h]
\begin{center}
\includegraphics[width=0.95\linewidth]{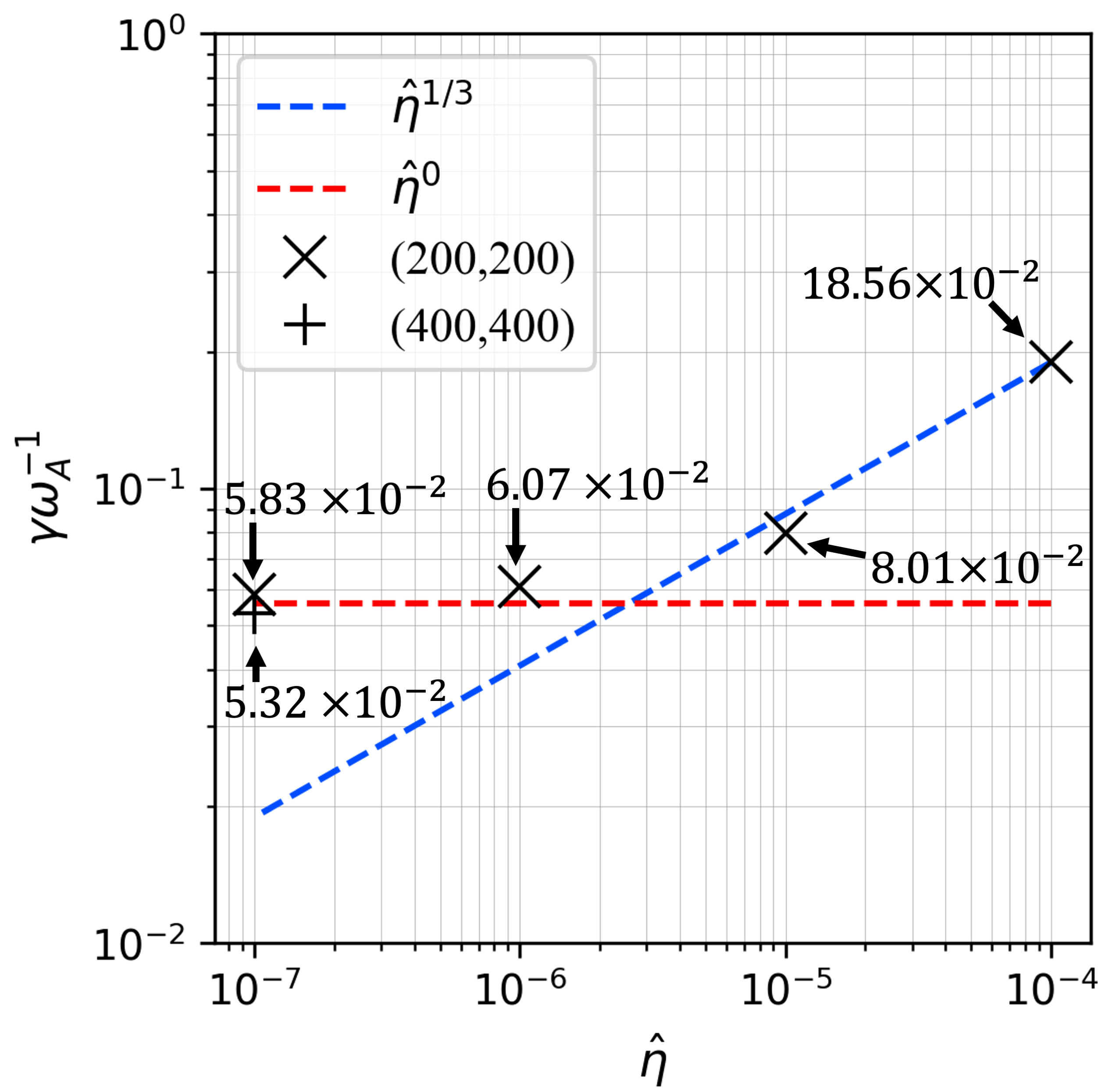}
\end{center} 
\caption{Dependence of the linear growth rate of the most unstable mode on $\hat{\eta}$ for the $q_\mathrm{0}/\rho_\mathrm{qmin}=1.1/0.575$ HC equilibrium.   ``$\times$" and ``+" symbols represent the results calculated with ($N_R$,$N_Z$)=($200$,$200$) and ($400$,$400$), respectively.  Blue dotted line represents the $\hat{\eta}^{1/3}$ scaling, while the red dotted line shows the ideal scaling.}
\label{fig:APPENetascan}
\end{figure}

The poloidal cross-section of the perturbed $\delta P_b$ at $\phi=180^\circ$ and $0^\circ$ toroidal angles are shown in Figs.\ref{fig:APPENhcprof}(a-b), respectively, while the overlaid black contours represent the magnetic Poincar{\'e} plot.  (Please note that the radial profile at $\phi = 180^\circ$ is plotted with the inverted horizontal axis.) The fluctuation is strongest along the compressed flux region of HC, similar to the $0\leq n \leq8$ and $0\leq n\leq 16$ secondary mode shown in Figs.\ref{fig:4eta6vrad2nd_realspace} and \ref{fig:4eta6vrad2nd_realspace_n16}, respectively.  In terms of toroidal and poloidal localization, the mode presented in this Appendix is more localized than the $0 \leq n \leq 8$ secondary mode, but slightly broader than the $0 \leq n \leq 16$ secondary mode. One possible explanation is the effect of the imperfect toroidal low-pass filter used in the main paper, which may decorrelate the coupling among individual Fourier harmonics near the core region. However, this remains speculative.  Its radial position is also shifted inward, as it is no longer constrained by the imperfect toroidal low-pass filter. This shift reveals the actual radial location with the maximum bulk pressure gradient.

Unlike the results discussed in the main paper, the flux coordinate is prepared from the HC equilibrium; therefore, the Fourier representation of the perturbation along the flux surface can be done correctly.  The Fourier analysis presented in this Appendix is limited to toroidal modes in the $0 \leq n \leq 24$ range, and the radial MHD velocity profiles during the linear growth phase are shown in Fig.\ref{fig:APPENhcprof}(c), revealing a broad spectrum of unstable Fourier components. We find that each toroidal mode number has a single dominant poloidal harmonic with $m=n+1$.   The radial location of the linearly unstable $10\leq n \leq 24$ modes corresponds to the $m/n=(n+1)/n$ rational surface localized near $q_\mathrm{min}$.  From Fig.\ref{fig:APPENhctevol}, the $10\leq n\leq24$ Fourier components are growing at the same rate at $\gamma=5.83\times10^{-2}\omega_A$, which is much higher than that of the $0\leq n\leq8$ secondary mode reported in Section \ref{sec:level4c2} ($\gamma_\mathrm{2nd}=1.1\times10^{-2}\omega_A$ for $\hat{\eta}=10^{-6}$). It should be noted that the linear growth of the $n \leq 9$ modes has not yet fully developed at the time of this measurement. If the simulation were continued for a longer time, the range of Fourier components with a similar growth rate would likely expand until nonlinear effects become significant.  The difference between the present growth rate and the one found in Section \ref{sec:level4c2} is not unexpected because it is not constrained to the $0\leq n \leq8$ dynamics.  (We suspect that the low-pass filter gives the mode an unnatural shape that prevents it from growing self-consistently and optimally in the given configuration.) Since all of these Fourier components grow at the same rate rather than obeying summation rules of nonlinear mode-mode coupling, it is clear that they represent a single coherent eigenmode of the non-axisymmetric HC configuration. The broad spectrum of Fourier components merely implies that the basis functions (sines and cosines) yield an inefficient representation of such a helically localized mode. 

Finally, we perform a plasma resistivity $\hat{\eta}$ scan for this HC equilibrium, with values of $\hat{\eta} = 10^{-4}$, $10^{-5}$, $10^{-6}$, and $10^{-7}$. The dependence of the linear growth rate of the $n=21$ Fourier component on $\hat{\eta}$ for the $q_\mathrm{0}/\rho_\mathrm{qmin} = 1.1/0.575$ HC equilibrium is shown in Fig.\ref{fig:APPENetascan}. The resistive kink/ballooning mode ($\hat{\eta}^\frac{1}{3}$) and the ideal scaling lines are represented as blue and red dashed lines, respectively.   In the $3\times 10^{-6} \leq \hat{\eta} \leq 10^{-4}$ range, the linear growth rate of the mode scales with $\hat{\eta}^{\frac{1}{3}}$, showing a resistive nature.  As one lowers the resistivity further, particularly $\hat{\eta} < 3\times 10^{-6}$, the linear growth rate becomes almost independent of $\hat{\eta}$, suggesting the transition to ideal mode.  This transition was not observed for the secondary mode reported in the main paper, Fig.\ref{fig:4etascan2nd}.  One possible explanation is that the secondary mode is ideally unstable in the case without the low-pass filter, whereas the $n \leq 8$ secondary mode reported in Section \ref{sec:level4c2} is ideally stable. Again, the difference in MHD stability between the two cases can be understood from the difference in their mode structures.  The structure of the secondary mode reported in this Appendix is computed more accurately with a larger number of Fourier components, so we expect it to have a higher degree of compatibility with the magnetic geometry and pressure field of the plasma it lives in than the low-pass-filtered mode in Section \ref{sec:level4c1}.

Most data points (``+'' symbols) in Fig.\ref{fig:APPENetascan} were obtained with ($N_R$, $N_Z$) = ($200$, $200$) grid points in the reduced poloidal plane, which is approximately equivalent to the resolution of the $(400,400)$ mesh in the full domain used in Fig.\ref{fig:4etascan2nd} of the main paper. As a convergence test, the case with $\hat{\eta}=10^{-7}$ was also computed with a doubled resolution ($400$, $400$), which would correspond to a $(800,800)$ mesh in the full domain. The result (``$\times$'' symbol) agrees well with the one obtained with the original resolution, so we consider them to be numerically converged. Note that this agreement is better than what we saw in Fig.\ref{fig:4etascan2nd} of the main paper (full plasma domain, low-pass filtered, equivalent resolution).

The results presented in this Appendix show that high-$n$ resistive or ideal pressure-driven MHD mode can also be excited in the HC equilibrium of VMEC.  Apart from the quantitative differences attributed to low-pass filtering, as discussed above, the characteristics of these modes closely resemble those of the secondary mode discussed in Section \ref{sec:level4}, further confirming the agreement between the HC equilibria calculated with MEGA and VMEC.

\section{\label{sec:apen2}Performance of Toroidal Low-pass Filter during HC Formation}
\quad

\begin{figure*}[h]
\begin{center}
\includegraphics[width=1.00\linewidth]{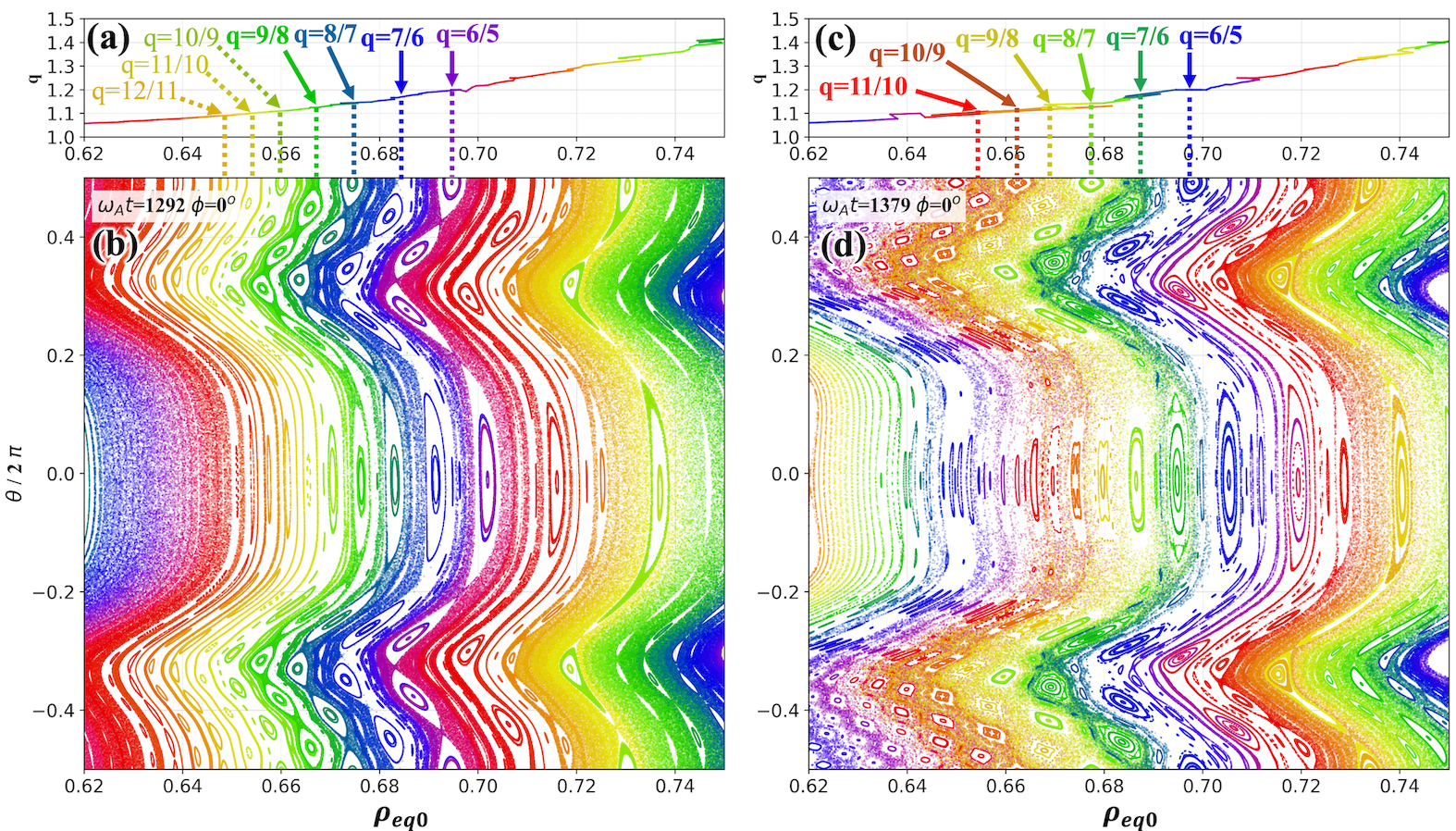}
\end{center} 
\caption{Magnetic Poincar{'e} plots in ($\rho_{eq0}$, $\theta$) space calculated after HC formation and the onset of secondary mode with the $0\leq n\leq8$ low-pass filter. Panels (a–b) show the magnetic structure before the onset of magnetic chaos, while panels (c–d) show it shortly after chaotization occurs. Panels (a) and (c) display the safety factor profile along $\theta/2\pi=1/2$, and panels (b) and (d) present the corresponding magnetic Poincar{'e} plots.}
\label{fig:APPENchaos1}
\end{figure*}

As discussed in Section \ref{sec:level2}, the low-pass filter implemented in MEGA is used to maintain numerical stability by artificially removing short-wavelength modes and to isolate the effects of the secondary MHD modes on HC.   This low-pass filter applies a Fourier decomposition along the geometric toroidal angle, and it is valid when the magnetic flux surface remains close to a toroidally axisymmetric equilibrium state, as in studies of Alfv{\'e}n eigenmodes.  However, this is not the case in our HC formation studies, where the plasma and magnetic field evolve toward a toroidally asymmetric state.  Within the HC region, any arbitrarily fixed point in the $R$-$Z$ plane will intersect different magnetic flux surfaces along the geometrical toroidal direction.  This issue and its consequences on our secondary mode are discussed in this Appendix based on the results presented in Section \ref{sec:level4c}.

In Section \ref{sec:level4c}, where we studied the properties and the consequences of the secondary mode, our low-pass filter intends to limit the growth of $n>8$ modes.  If our low-pass filter functions as intended, we should not observe any $n>8$ modes in real space after HC formation.  For simplicity, we only consider the MHD simulation of $0\leq n\leq8$ in the $q_\mathrm{0}/\rho_\mathrm{qmin}=1.1/0.575$ equilibrium shown in Fig.\ref{fig:4megaq0scan}(f-j).  Here, we re-cast the magnetic Poincar{\'e} plot at $\phi=0^\circ$ during the nonlinear phase of this case into the ($\rho_{eq0},\theta$) space, where $\rho_{eq0}$ and $\theta$ are the square root of the normalized toroidal flux at $\omega_At=0$ and geometric poloidal angle, respectively\footnote{The ($\rho_{eq0},\theta$) space is used because it allows to present the poloidal mode structure near $q_\mathrm{min}$ in a visual-friendly manner.}.  These results are shown in Fig.\ref{fig:APPENchaos1}, where panels (a-b) and (c-d) show the results slightly before and after the chaotization. Panels (a,c) show the local safety factor profile along $\theta/2\pi=1/2$.  In both time slices, the magnetic flux surface is not a vertical line ($\rho_{eq0}= {\rm const}.$) because of the $m/n=1/1$ HC and its sideband. (Even in the $\rho>\rho_\mathrm{qmin}$ region, the HC can cause a flux surface displacement of around $\delta\rho_{eq0}\approx0.02$.) The situation before chaotization is shown in panels (a-b), where we observe the growth of the $m/n=9/8$, $8/7$, $7/6$, and $6/5$ magnetic islands within $0.66 \leq \rho_{eq0} \leq 0.7$. The origin of these islands is currently unclear; it could be physical (pre-existing tearing components or parity mixing) or numerical (e.g., insufficient resolution), or they could be caused by our imperfect toroidal low-pass filter. Irrespective of their origin, one purpose of this analysis is to check the spectral width of the islands. The poloidal mode number $m$ can be directly obtained from Fig.\ref{fig:APPENchaos1} by counting the number of O-points in an island chain. The corresponding toroidal mode number $n$ can be obtained via multiplication with the local value of the safety factor, $n = m q$. We then find that magnetic islands with $n>8$ are present in the $\rho_{eq0}<0.67$ region; for instance, $m/n=10/9$, $11/10$, and $12/11$.  The existence of these $n>8$ islands confirms the expectation that our low-pass filter does have the intended effect on perturbations relative to the HC equilibrium, though it is not clear whether the filter gave rise to these islands or failed to eliminate them.  A similar issue has also been reported in EUTERPE, a global gyrokinetic code\cite{kleiber2024euterpe}.  

In panels (c-d), we show the situation slightly after the emergence of magnetic chaos. The Poincar{\'e} plot in panel (d) shows that the chaotization (radial mixing of magnetic field lines) can be observed in the region $\rho_{eq0}<0.69$ ($q_\mathrm{min}< q \leq7/6$), and one can see that it is stronger in the region with $n>8$ magnetic islands.  Since the low-pass filter is not working correctly here, the degree of chaotization reported in Sections \ref{sec:level4}-\ref{sec:level5} and this Appendix may not be quantitatively accurate. Simply removing the low-pass filter does not solve the problem because the shorter wavelength modes can become unstable. At present, we cannot verify whether this low-pass filter suppresses or further amplifies the chaotization that arises from the overlaps of multi-helicity islands. In fact, we cannot even rule out the possibility that the islands (and the resulting chaos) are entirely products of the misaligned filter. 

\section*{References}
\bibliography{iopart-num}
\end{document}